\documentclass[lettersize,journal]{IEEEtran}
\usepackage{amsmath,amsfonts}
\usepackage{algorithmic}
\usepackage{algorithm}
\usepackage{array}
\usepackage[caption=false,font=normalsize,labelfont=sf,textfont=sf]{subfig}
\usepackage{textcomp}
\usepackage{stfloats}
\usepackage{url}
\usepackage{makecell}
\usepackage{verbatim}
\usepackage{graphicx}
\usepackage[backend=bibtex,%
    maxcitenames=3,
    style=numeric,%
    sorting=none,%
    giveninits=true,%
    eprint=false,%
    url=false,doi=true,isbn=false]{biblatex}
\usepackage{hyperref}
\usepackage{booktabs}
\usepackage{multirow}
\usepackage{gensymb}
\usepackage{tikz}
\usetikzlibrary{matrix, positioning,arrows.meta,calc}
\DeclareUnicodeCharacter{2009}{}

\begin{document}

\title{Pruning Sparse Tensor Neural Networks Enables Deep Learning for 3D Ultrasound Localization Microscopy}

\author{Brice Rauby,~\IEEEmembership{Member,~IEEE}, 
Paul Xing,~\IEEEmembership{Graduate Student Member,~IEEE},
Jonathan Porée,~\IEEEmembership{Member,~IEEE}, Maxime Gasse, Jean Provost, ~\IEEEmembership{Member,~IEEE}
\thanks{This work was supported in part by the Institute for Data Valorization (IVADO), in part by the Canada Foundation for Innovation under Grant 38095, in part by the Canadian Institutes of Health Research (CIHR) under Grant 452530, and in part by a Natural Sciences and Engineering Research Council of Canada (NSERC) discovery grant (RGPIN-2020-06786).
The work of Brice Rauby was supported in part by IVADO, and in part by the TransMedTech Institute, and in part by the Fonds de recherche du Québec—Nature et technologies.
The work of Paul Xing was supported by IVADO and in part by the TransMedTech Institute.
The work of Jonathan Porée was supported in part by IVADO, in part by the TransMedTech Institute, and in part by the Canada First Research Excellence Fund (Apogée/CFREF).
This research was enabled in part by support provided by Calcul Québec (\href{https://www.calculquebec.ca/}{calculquebec.ca}) and the Digital Research Alliance of Canada (\href{https://alliancecan.ca/en}{alliancecan.ca})} (Correspondingauthor:JeanProvost.)
\thanks{B. Rauby is the Department of Engineering Physics, Polytechnique Montréal, Montréal, QC H3T 1J4, Canada, and also with Mila-Quebec artificial intelligence institute, Montréal, QC H2S 3H1, Canada (email: brice.rauby@polymtl.ca)}
\thanks{P. Xing, and J. Porée are with the Department of Engineering Physics, Polytechnique Montréal, Montréal, QC H3T 1J4, Canada (e-mail: firstname.lastname@polymtl.ca)}
\thanks{M. Gasse is with ServiceNow, Montréal, QC H2S 3G9, Canada, also with the Department of Computer Engineering and Software Engineering, Polytechnique Montréal, Montréal, QC H3T 1J4, Canada, and also with Mila-Quebec artificial intelligence institute, Montréal, QC H2S 3H1, Canada (e-mail: maxime.gasse@servicenow.com)}
\thanks{J. Provost is with the Department of Engineering Physics, Polytechnique Montréal, Montréal, QC H3T 1J4, Canada, and also with the Montreal Heart Institute, Montréal, QC H1T 1C8, Canada (email: jean.provost@polymtl.ca)}

\thanks{Manuscript received XXXX XX, 202X; revised XXXX XX, 202X.}}

\markboth{Journal of \LaTeX\ Class Files,~Vol.~14, No.~8, August~2021}%
{Shell \MakeLowercase{\textit{et al.}}: A Sample Article Using IEEEtran.cls for IEEE Journals}

\maketitle
\begin{abstract}
Ultrasound Localization Microscopy (ULM) is a non-invasive technique that allows for the imaging of micro-vessels \textit{in vivo}, at depth and with a resolution on the order of ten microns. ULM is based on the sub-resolution localization of individual microbubbles injected in the bloodstream. Mapping the whole angioarchitecture requires the accumulation of microbubbles trajectories from thousands of frames, typically acquired over a few minutes. ULM acquisition times can be reduced by increasing the microbubble concentration, but requires more advanced algorithms to detect them individually. Several deep learning approaches have been proposed for this task, but they remain limited to 2D imaging, in part due to the associated large memory requirements. Herein, we propose the use of sparse tensor neural networks to enable deep learning-based 3D ULM by improving memory scalability with increased dimensionality. We study several approaches to efficiently convert ultrasound data into a sparse format and study the impact of the associated loss of information. When applied in 2D, the sparse formulation reduces the memory requirements by a factor 2 at the cost of a small reduction of performance when compared against dense networks. In 3D, the proposed approach reduces memory requirements by two order of magnitude while largely outperforming conventional ULM in high concentration settings. We show that Sparse Tensor Neural Networks in 3D ULM allow for the same benefits as dense deep learning based method in 2D ULM i.e. the use of higher concentration \textit{in silico} and reduced acquisition time.
\end{abstract}


\section{Introduction}
Ultrasound Localization Microscopy (ULM) is an imaging method that non-invasively maps the vascular tree and blood velocities at depth \textit{in vivo}. By localizing and tracking individual microbubbles injected into the blood flow \cite{erricoUltrafastUltrasoundLocalization2015,christensen-jeffriesVivoAcousticSuperResolution2015}, ULM achieves an imaging resolution approximately equal to one tenth of the diffraction-limited resolution. More recently, Dynamic Ultrasound Localization Microscopy (DULM) \cite{bourquinVivoPulsatilityMeasurement2022,cormierDynamicMyocardialUltrasound2021} has extended the capabilities of ULM by enabling the generation of retrospectively-gated, super-resolved movies of the blood flow dynamics, with applications in pulsatility mapping \cite{bourquinVivoPulsatilityMeasurement2022}, functional imaging of the brain \cite{renaudinFunctionalUltrasoundLocalization2022}, and cardiac imaging \cite{cormierDynamicMyocardialUltrasound2021}. 
ULM has been applied in pathological mouse models, such as to classify ischemic and hemorrhagic stroke ~\cite{chavignon3DTranscranialUltrasound2022} and to observe vascular impairments in Alzheimer's Disease ~\cite{lowerisonSuperResolutionUltrasoundReveals2024}.
Studies on patients have demonstrated the feasibility of ULM in various human organs, including the brain ~\cite{demeneTranscranialUltrafastUltrasound2021}, breast ~\cite{opacicMotionModelUltrasound2018,porteUltrasoundLocalizationMicroscopy2024}, kidney ~\cite{huangSuperresolutionUltrasoundLocalization2021,denisSensingUltrasoundLocalization2023,bodardUltrasoundLocalizationMicroscopy2023}, prostate ~\cite{solomonExploitingFlowDynamics2019}, lower limb muscle ~\cite{harputTwoStageMotionCorrection2018}, liver ~\cite{huangSuperresolutionUltrasoundLocalization2021}, pancreas ~\cite{huangSuperresolutionUltrasoundLocalization2021}, vasa vasorum of the carotid wall ~\cite{goudotAssessmentTakayasuArteritis2023}, testicular micro-circulation ~\cite{liSuperresolutionUltrasoundLocalization2024}, and lymph node metastatic cancer ~\cite{zhuSuperResolutionUltrasoundLocalization2022}.
Existing works have used fully addressed \cite{heilesUltrafast3DUltrasound2019,demeulenaereVivoWholeBrain2022,heilesVolumetricUltrasoundLocalization2022}, multiplexed array \cite{bourquinQuantitativePulsatilityMeasurements2024, chavignon3DTranscranialUltrasound2021, lokThreeDimensionalUltrasoundLocalization2022}, and row-column array probes \cite{jensenAnatomicFunctionalImaging2022,hansen-shearerUltrafast3DTranscutaneous2024,wu3DTranscranialDynamic2024} to extend ULM and DULM to 3D imaging. In addition to the inherent advantages from 3D imaging, such as reduced user dependence in imaging plane selection and robustness to out-of-plane motion, 3D ULM is less sensitive to detection errors caused by out-of-plane trajectories and velocity estimation bias from elevation direction projection, which likely enhances clinical relevance and reliability.
Both localization and velocity estimation can be improved by rejecting microbubbles that do not appear across several frames \cite{erricoUltrafastUltrasoundLocalization2015}. Such tracking can be performed, e.g., using the Nearest Neighbor algorithm \cite{bourquinVivoPulsatilityMeasurement2022, erricoUltrafastUltrasoundLocalization2015} or the Hungarian method \cite{heilesPerformanceBenchmarkingMicrobubblelocalization2022, songEffectsSpatialSampling2018}. Some approaches have also incorporated Kalman filtering to refine the position estimations of a track \cite{bourquinQuantitativePulsatilityMeasurements2024, lokThreeDimensionalUltrasoundLocalization2022,taghaviUltrasoundSuperresolutionImaging2022, tangKalmanFilterBasedMicrobubble2020}. 
The acquisition time necessary to construct a complete vascular map is mainly dependent on the required time to perfuse all vessels and, thus, on microbubble concentration \cite{hingotMicrovascularFlowDictates2019}. However, a trade-off exists between the microbubble concentration and the localization precision and accuracy \cite{belgharbiAnatomicallyRealisticSimulation2023}, which can be partially lifted using, e.g., methods based on the compressed sensing theory \cite{bar-zionSUSHISparsityBasedUltrasound2018}, on the division of the k-space in several subregions \cite{huangShortAcquisitionTime2020}, or tracking the microbubble signals prior to sub-pixel localization  \cite{leconteTrackingPriorLocalization2024}. Deep learning-based methods have also investigated frame-by-frame, spatial-only approaches \cite{liuDeepLearningUltrasound2020, vanslounSuperResolutionUltrasoundLocalization2021,chenDeepLearningBasedMicrobubble2022}, and, more recently, the spatio-temporal context through convolution \cite{mileckiDeepLearningFramework2021, shinContextAwareDeepLearning2023} or sequential modeling \cite{chenLocalizationFreeSuperresolution2023}. 

Despite promising results with increased microbubble concentrations both \textit{in silico} \cite{mileckiDeepLearningFramework2021} and \textit{in vivo} \cite{mileckiDeepLearningFramework2021,shinContextAwareDeepLearning2023}, deep-learning based approaches have been limited to 2D imaging. Indeed, the addition of a third spatial dimension considerably increases the size of intermediate feature maps and  highly-resolved outputs. For example, a straightforward implementation of Deep-stULM \cite{mileckiDeepLearningFramework2021} in 3D would require at least two orders of magnitude more memory than its 2D counterpart. Thus, the development of deep learning based models for 3D ULM is conditioned on successfully addressing their memory complexity.

\begin{figure}
 \centering
\includegraphics[width=0.45\textwidth]{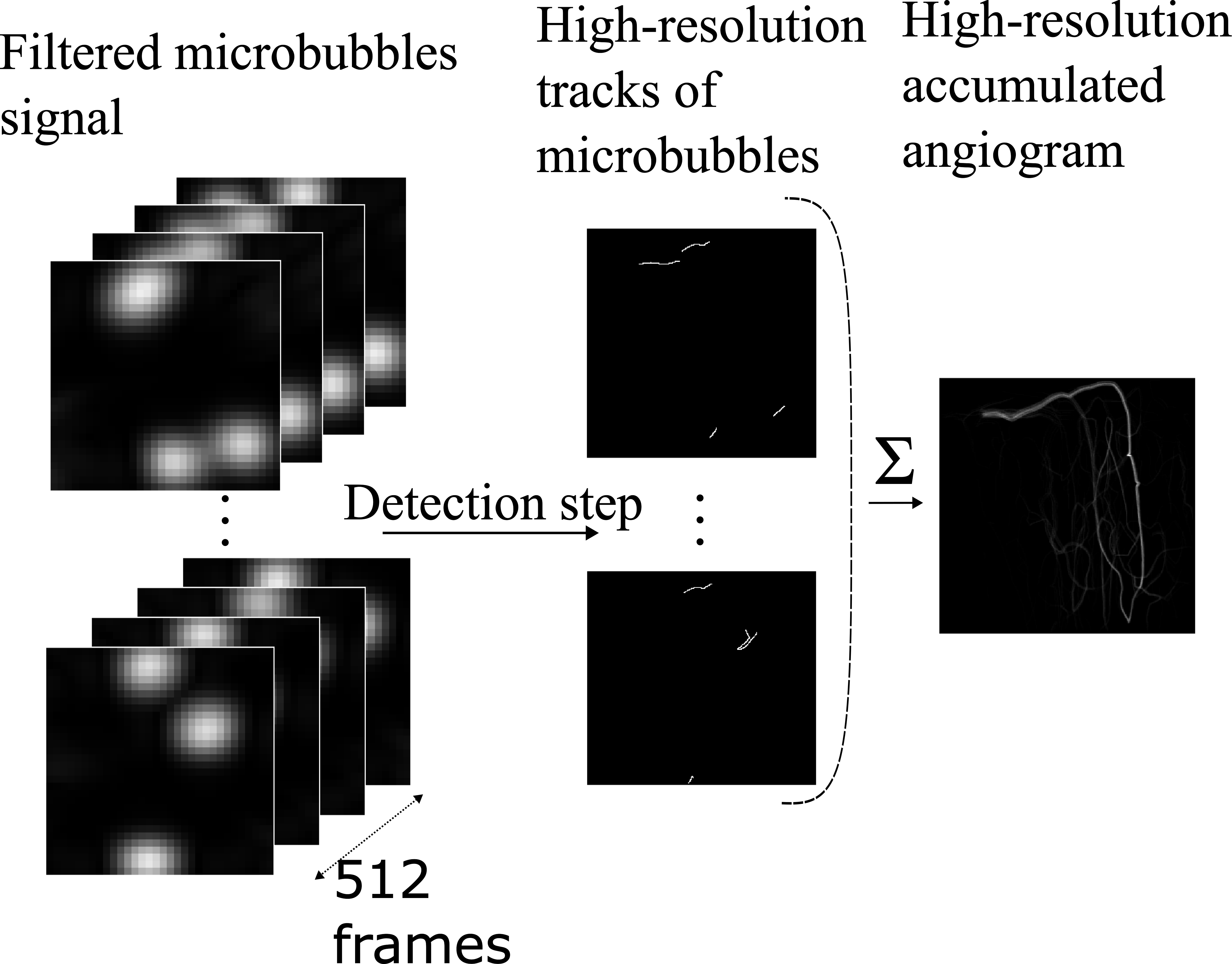}
\caption{The left column represents the filtered microbubble signal (i.e, the input of the network), the center column represents the corresponding microbubble tracks (i.e., the desired output of the network) and the right column represent the final result after summation of all the predictions from a dataset (i.e., the vascular structure imaged) }
\label{ulm_explained}   
\end{figure}

\begin{figure*}
 \centering
\includegraphics[width=\textwidth]{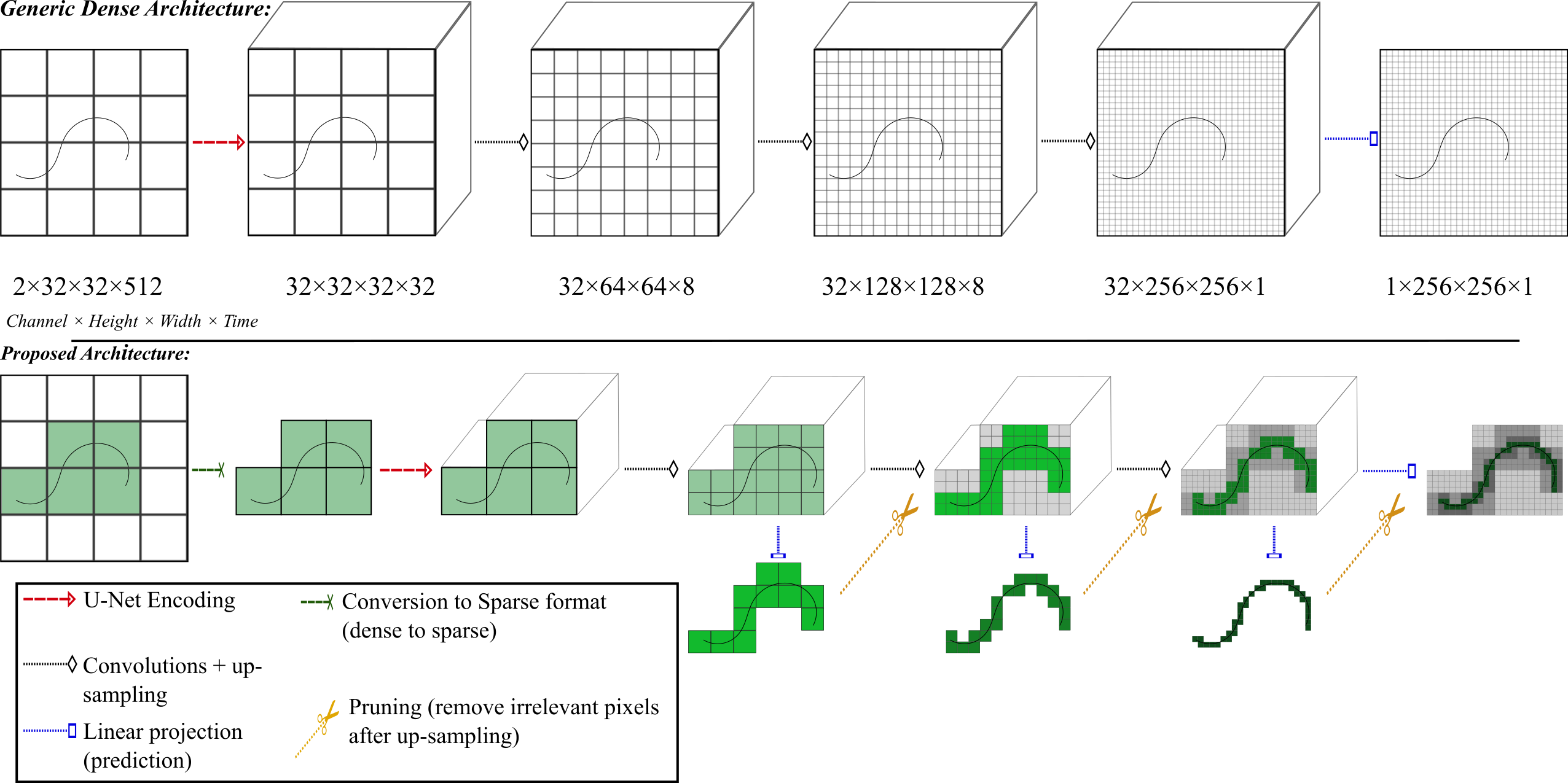}
\caption{The top row shows a dense representation of a trajectory in Deep-stULM as well as the intermediate map dimension. The bottom row illustrates how sparse formulations could reduce the memory cost: the green pixels represent the pixels of interest at each resolution, and the gray pixels represent the pixels removed through pruning based on intermediate prediction}
\label{method_figure}   
\end{figure*}

To improve the scaling of memory complexity of deep learning approaches in ULM and enable deep learning application for 3D ULM, we propose to leverage the recently introduced Sparse Tensor Neural Networks \cite{choy4DSpatioTemporalConvNets2019}.  Sparse Tensor Neural Networks store only non-zero pixels (referred to as active sites), along with their spatial coordinates and corresponding pixel values (often described as COO format in the literature). This enables the handling of arbitrarily shaped tensors.
During training and inference, convolutions and subsequent operations are only applied to these active sites. 
While ultrasound images are typically dense data that cannot be stored directly as sparse tensors efficiently, filtered microbubbles responses are sparsely distributed in space and time as seen in Figure \ref{ulm_explained}.
To leverage the sparsity of microbubbles response, one must thus design a filter that extracts microbubble responses from ultrasound images and is sufficiently restrictive that it minimizes memory requirements without discarding information enables the neural network to outperform conventional approaches. 
Hereafter, filtering out most of the input signal from dense tensors before conversion to sparse format while keeping the signal of interest is designated as the \textit{dense-to-sparse} operation (represented in dashed green in Figure \ref{method_figure}).

Our contributions can be summarized as follows:
\begin{itemize}
    \item A sparse formulation of Deep-stULM enabling deep learning-based 3D ULM.
    \item A comparative study \textit{in silico} between ULM and the proposed approach under varying concentrations in 3D.
    \item A 2-D \textit{in silico} comparison of performance and memory usage with deep learning baselines and conventional ULM
\end{itemize}
We show that Sparse Tensor Neural Networks reduce memory cost and scale better with added input dimensions, which allows for the first deep learning application to 3D ULM and outperforms existing 3D ULM in high concentration. 
We also provide the code and the dataset needed to reproduce the results at \href{https://github.com/provostultrasoundlab/SparseTensorULM} {https://github.com/provostultrasoundlab/SparseTensorULM}.

\section{Theory}
After image formation, ULM data is typically filtered to remove the signal from the tissue while retaining microbubble responses. In this study, we focus on the subsequent step, which is to find the sub-resolution positions of the microbubbles based on their response. As illustrated in Figure \ref{ulm_explained}, this detection step is performed on a group of frames and is followed by an accumulation across several sets of frames, where the detected positions are later added to form the image or the volume of the vascular network.
Several deep learning approaches \cite{liuDeepLearningUltrasound2020,vanslounSuperResolutionUltrasoundLocalization2021} use convolutional architectures where the microbubble position is projected onto a grid with a finer resolution than the input signal. The upscaling factor $r$ between the input dimension and the output dimension typically ranges from $4$ to $8$. 
Therefore, $c_\text{dense}$, the memory complexity for storing the output grid of such architectures in dense format scales with: 
\begin{equation}
    c_{dense}(r,d, D)=(r\times D)^d 
\end{equation}
where $d$ represents the dimensionality of the output grid ($d=2$ for 2D ULM, and $d=3$ for 3D ULM), and $D$ is a typical dimension of the input in pixels. ULM assumes a sparse distribution of microbubbles, thus an upper bound on the number of microbubbles that can be detected in a certain volume is given by:
\begin{equation}\label{upperBoundN}
    N<(\frac{\rho D}{\alpha})^d
\end{equation} where $N$ is the number of microbubbles, $\rho$ is the size of a pixel in wavelength, and $\alpha$ approximately describes the size of the point spread function in wavelengths.
 In sparse format, $c_\text{sparse}$, the memory complexity of storing the microbubble positions scales with $d\times N$ and can be written as:  
\begin{equation}
    c_\text{sparse}(N,d) \simeq \eta \times d\times N_d
\end{equation} 
with $\eta$ being a constant depending on the implementation of the sparse tensors.
 Consequently, $\gamma_d$, the ratio between the sparse and the dense representation of the outputs of the networks in dimension $d$ scales with: 
\begin{equation}
\label{eq:gamma_def}
    \gamma_d \simeq  \frac{\eta \times d\times N_d}{(r\times D)^d }
\end{equation}
When extending from 2D to 3D, the ratio between the sparse and the dense representation of an ideal outputs from such network is expected to be multiplied by a factor $\delta$: 
\begin{equation}
    \delta = \frac{\gamma_\text{3D}}{\gamma_\text{2D}}   
\end{equation}
Using eq. \ref{eq:gamma_def}, we obtain
\begin{equation}
    \delta = \frac{N_\text{3D}}{N_\text{2D}} \times \frac{3}{2 (r\times D)}
\end{equation}
This scaling law makes the sparse representation very attractive for the extension in 3D. Indeed, assuming that N reaches its upper bound from eq. \ref{upperBoundN}, the ratio between sparse and dense representation becomes: 
\begin{equation}
    \delta =\frac{3}{2} \frac{\rho}{\alpha r}
\end{equation} 
Using typical values, $\alpha = 3$, $\rho = \frac{1}{2}$, and $r=8$, we obtain:
\begin{equation}
\label{eq:scaling_factor}
    \delta =  \frac{1}{32}
\end{equation} 
This numerical application illustrates the fact that in 3D the use of sparse formulation is from one to two orders of magnitude more efficient than in 2D when compared to the dense representation of ULM output images.
 However, practical factors such as the variability of the upscaling factor $r$ within the network architecture, the reduced sparsity of intermediate representations, the non-uniform distribution of microbubbles in space, or temporal context considerations, can alter this scaling law. 

Time complexity follows a similar scaling law to memory complexity, as convolutions are only applied to active sites (i.e., non-zero pixels) in sparse representation. In contrast, convolutions are applied to every site in dense representation. In practice, convolution implementations on dense tensors are heavily optimized in deep learning libraries and efficiently leverage GPUs and parallel computations \cite{paszkePyTorchImperativeStyle2019}. Sparse convolutions require storing active sites, which can be sped up using unordered maps and caching \cite{choy4DSpatioTemporalConvNets2019}. Different optimization levels of convolution implementations may outweigh the theoretical gains in computational costs of the proposed architectures. For these reasons, we proposed inference time comparisons reported in Table \ref{tab:results_time_memory}.

\section{Method}
First, we detail the approach used to simulate the 2D and 3D datasets used for training and evaluation of the different methods. Then we describe the model architecture, training parameters, and evaluation metrics. Additional studies on the impact of the \textit{dense-to-sparse} operations and on further architecture modifications such as pruning \cite{gwakGenerativeSparseDetection2020} (represented in gold dotted line and scissors in Figure \ref{method_figure}) and deep-supervision \cite{leeDeeplySupervisedNets2015} are presented.

\subsection{Simulations}
\subsubsection{2D dataset}
To compare Sparse Tensor Neural Networks with their dense counterpart, we based our study on a previously introduced dense method \cite{mileckiDeepLearningFramework2021} and used the same 2D dataset based on the previously published simulation pipeline.
Microbubble flow was simulated using a realistic model \cite{belgharbiAnatomicallyRealisticSimulation2023} based on \textit{ex vivo} mice brains obtained with two-photon imaging.
Four portions of different mice brains were used to generate the training set, one other portion was used for model selection and validation and the last one was kept as a test set to assess the performance. 
This approach ensures that the test set comprises microbubble trajectories from a completely unseen mouse brain, allowing us to evaluate the model inter-individual generalization.
Since each region covers a volume of only $ 500 \times 500 \times 500 \mu\text{m}^{3}$, we dilated the vascular network by a factor of 2 to fill a $1000 \times 1000 \times 1000 \mu\text{m}^{3}$ area, as done previously \cite{mileckiDeepLearningFramework2021}.
The ultrasound signal corresponding to the microbubble position was simulated using an in-house GPU implementation of SIMUS \cite{garciaSIMUSOpensourceSimulator2022} with parameters corresponding to an L22-14 probe (Vermon, Tours).
Three $15.625$-MHz, tilted plane waves with angles of $-1\degree$, $0\degree$, and $1\degree$ were simulated with a frame rate of $1$ kHz.
The simulated signal was subsampled to match the $100\% $ bandwidth IQ signal, mimicking the Verasonics Vantage system. 
Finally, the IQ data were beamformed with a GPU implementation of the delay and sum algorithm on a grid of $32 \times 32$ pixels with a resolution of $\frac{\lambda}{4}$ (i.e.,$25\mu \text{m}$), in groups of $512$ frames.
The point spread function (PSF) of the system was simulated at the center of the grid and used to compute the local correlation between the beamformed IQ data with the PSF. The obtained correlation maps were used as the input for the different deep-learning models.
In total, $2250$ movies were generated for training, $250$ for validation, and $500$ for testing. The concentration of microbubbles simulated for the training and validation sets was set to $5$ microbubbles per field of view (FOV), as done previously \cite{mileckiDeepLearningFramework2021}.
Several test sets were simulated based on the trajectories from the test angiogram with varying concentrations ($1$, $5$, $10$, and $20$ microbubbles per FOV) also matching the test set from the previous study \cite{mileckiDeepLearningFramework2021}.

\subsubsection{3D Dataset}
The $3$D dataset was obtained similarly but since 3D convolution filters have more than their 2D counterpart, additional  microbubble trajectories were included to prevent 3D models overfitting.
We divided the generated spatio-temporal samples in three groups: $3500$ samples for training, $500$ for the validation, and $2000$ for testing.
We dilated the vascular network by a factor of $8$ to account for the larger wavelength and the coarser beamforming grid ($\frac{\lambda}{2}$).
We simulated a $750$-fps imaging sequence containing 5 angles ($\{-2\degree,0\degree\},\ \{2\degree,0\degree\},\ \{-1\degree,0\degree\},\ \{1\degree,0\degree\},\ \{0\degree,0\degree\}$) emitted with a 7.8125-MHz center frequency using a matrix array with parameters matching a commercially available 8-MHz 2D matrix probe (Verasonics, WA, USA).
The concentration of microbubbles simulated for the training, validation, and test sets was increased to 30 microbubbles (compared to 5 for the 2D-case) per field of view (FOV) given the additional dimension.

\subsubsection{Additional test sets}
To complement the evaluation on anatomically realistic dataset, we generated random trajectory datasets and simulated corresponding ultrasound signal based on similar imaging sequences. 
500 movies were generated for each of the following concentrations $1$, $5$, $10$, and $20$ microbubbles per FOV in 2D. As there is no vascular network to reconstruct, the Dice was computed for each frame and averaged across the whole dataset. Since this metric measures the overlap between few microbubble trajectories, it is hereafter referred to as \textit{trajectory Dice} and is expected to yield lower value than typical Dice between vascular networks.

We assessed the out-of-distribution generalization and robustness of the proposed method and baselines to additive Gaussian noise.  At test time, random noise was added to the real and imaginary parts of the correlation map of the angiogram-based test set at $5$ MB/FOV. The noise had a mean of $0$, and the standard deviation was increased from $0.1$ to $0.25$ in steps of $0.05$.

\subsection{Model training and evaluation}
\subsubsection{Sparse Tensor Neural Network and 4D convolutions}
After the \textit{dense-to-sparse} operation, the sparse tensor containing the low-resolution signal was given as input to a Sparse Tensor Neural Network implemented using the Python library MinkowskiEngine \cite{choy4DSpatioTemporalConvNets2019}.
For each intermediate layer, Sparse Tensor Neural Networks only apply their convolutions and activations on non-zero values, yielding another sparse tensor.
Conventional operations used in CNNs are implemented in MinkowskiEngine, leading to a relatively straightforward translation of the model from dense to sparse format.
To assess the benefits of sparse formulation, we converted the dense Deep-stULM architecture to a sparse formulation without additional change, this approach is designated as \textit{Sparse Deep-stULM} hereafter. However, such dense architecture might not take most of the sparse tensor implementation. Pruning or cascaded learning could further improve the memory efficiency of the sparse formulation. Both of this additional modifications require intermediate supervision and are detailed in the following sections.
The resulting models are also extended directly to 3D imaging with 4D convolutions to handle 3D+T tensors.
4D convolutions are directly implemented in the Python library MinkowskiEngine \cite{choy4DSpatioTemporalConvNets2019}.

\subsubsection{Training procedure}
For the 2D models based on Deep-stULM, the hyperparameters were set to the same value as in the original study \cite{mileckiDeepLearningFramework2021}:  the optimizer used was Adam \cite{kingmaAdamMethodStochastic2017} and the training was divided into two parts. During the first 150 epochs, the ground truth microbubble trajectories were dilated with a radius of 2 to stabilize the training. The initial learning rate was set to 0.1 and then decayed by a factor of 10 at the epochs 15, 45, 75, and 100. During the last 150, the ground truths were no longer dilated, and the learning rate was set to 0.001 at epoch 150 and then decayed by a factor of 10 at epochs 160, 200 and 250. We did not optimize the hyperparameters for the sparse formulation of Deep-stULM and used the same as the original study. The batch size was set to 8 for all the runs in 2D.
For the 3D models, we also used the Adam optimizer with an initial learning set of 0.1. We trained the 3D networks for 20 epochs in total, and the learning rate was decayed by a factor of 10 at epochs 15 and 17. For the first epoch, the batch size was set to 2 to allow every configuration to fit in memory, then for the remaining epochs, the batch size was increased to 4. 
Deep-stULM was trained using Dice loss in both is sparse and dense formulation. 
\subsubsection{Performance comparison with ULM and Deep-stULM}

\paragraph{Evaluation metric}
To compare our results with the previously established method \cite{mileckiDeepLearningFramework2021} and conventional ULM, we measure the overlap between the network prediction and ground truth using the Dice coefficient :
\begin{equation*}
    \text{Dice} = \frac{2 \times | \text{GT} \cup \text{Pred}|}{| \text{GT}| + | \text{Pred}|}
\end{equation*}
with GT being the projection of all the trajectories from the ground truth to the super-resolved grid and Pred the prediction of the network. Similarly to the previous study \cite{mileckiDeepLearningFramework2021}, the Dice values displayed use a Dice computed between the binarized angiograms (i.e., between the logical summation of all the microfilms from the test set).

\paragraph{Conventional ULM}
We also provide the results of a standard, non-deep-learning ULM method, described in \cite{bourquinVivoPulsatilityMeasurement2022}. Briefly, Gaussian fitting was used to localize microbubbles and the Hungarian method \cite{kuhnHungarianMethodAssignment1955} was used for the tracking step. The number of detections in each frame was set to the optimal value based on the number of microbubbles simulated in the FOV (i.e., for the $5$MB/FOV concentration, the number of detection would be set to $5$). Note that this setting is ideal and may favor the conventional ULM. Indeed, in real applications, the exact number of microbubbles in the FOV is unknown.

\paragraph{Deep learning baseline, mSPCN-ULM}
We retrained the mSPCN-ULM architecture \cite{liuDeepLearningUltrasound2020} on our datasets using original training parameters. We ensured that the loss had converged at the end of the training (60 epochs), and the learning rate was decayed by ten after 30 epochs. The output of the mSCPN-ULM was interpolated to match Deep-stULM output resolution. Since training with a Dice loss was unstable, mSCPN-ULM was trained using the original training loss \cite{liuDeepLearningUltrasound2020} based on L1 distance between the target and the prediction, both convolved with a Gaussian kernel.

\paragraph{Memory monitoring}
We monitored the memory usage of the training using CometML and took the maximum value reached during the training of each method. As there is some stochasticity involved both in training and in the measurement of the memory, we used the average across 3 different runs and provided the standard deviations between each run for deep learning approaches. For the 3D dense formulation of Deep-stULM, it was not possible to train the model due to practical memory limitation. Therefore, we provide only an estimate of the memory usage. This estimate was based on scaling the 2D memory usage based on the increase of memory for the intermediate maps due to the addition of a spatial dimension. For mSPCN-ULM, the 3D memory usage was linearly extrapolated from memory usage with smaller batch size.

\paragraph{Computation time measurements}
\label{paragraph:Computation_time_method}
Processing times for one group of 512 frames were measured on NVIDIA V100 Volta (32G HBM2 memory) GPUs. To account for I/O latency on computation servers, processing times were measured and averaged across $5$ redundant forward paths. Measured times for multiple samples were averaged to provide robust estimates. For 3D inference times, mSPCN-ULM could be implemented by simply translating PyTorch 2D operations to 3D operations. However, the whole group of $512$ could not fit in memory. We measured and linearly extrapolated the processing times for smaller subsets of $16$, $32$, and $64$ frames to estimate the processing times of $512$ frames. 3D Dense formulation of DeepST-ULM was not implemented as it would have required 4D convolutions not supported in PyTorch.

\subsection{Additional studies}
\subsubsection{\textit{dense-to-sparse} strategies}
 To assess the loss of information and its impact on the performance caused by the \textit{dense-to-sparse} filtering, we compared the performance of the sparse model for two simple \textit{dense-to-sparse} strategy referred as Top-k and thresholding strategy with varying value for their respective parameters. To provide better intuition on the performance that one can expect with more sophisticated filtering,  we developed a deep learning based solution.
To compare between each method, we computed the average number of non-zero pixels in the test set movies to compare across methods and plotted it in Figure \ref{fig:performance_curve_dense_to_sparse}. 
In addition, to differentiate between the performance loss induced by the \textit{dense-to-sparse} strategy from the effect of the sparse implementation, we also evaluated a dense model with inputs filtered according to the \textit{dense-to-sparse} strategy (each model was retrained on the filtered data).
 
 \paragraph{Top-k strategy} Typical ULM approaches use regional maxima for microbubble detection before localizing them with high precision \cite{heilesPerformanceBenchmarkingMicrobubblelocalization2022}.
 Based on the same underlying assumption that the microbubble signals are located near local maxima, it is reasonable to consider only the k-largest pixel of each input tensor. This operation is designated as top-k operation later on. In practice, due to the smoothness of the input, this approach is very similar to the use of local maxima value while providing better control of the memory usage of the input tensor. 
 The explored values used for the top-k approaches range between $5000$ to $50 000$ pixels.  
 \paragraph{Thresholding strategy} Previously published deep-learning methods \cite{mileckiDeepLearningFramework2021} used a threshold based on the value of the local correlation between the signal and the point spread function of the imaging system  to remove residual after clutter filtering \textit{in vivo}. We applied this same approach directly on our simulated datasets.
For the thresholding approaches, the threshold values were set to $\{0.01, 0.05, 0.10, 0.25\}$. As the thresholding strategy with a threshold set at $0.10$ in 2D offered a good trade-off between performance and sparsity, we used it for all the experiments where the \textit{dense-to-sparse} strategy was not specified. The threshold was heuristically set to $0.25$ for the 3D experiments. 

\paragraph{Deep learning based strategy} 
We trained a dense CNN to localize microbubbles at low resolution. To do so, it is trained to predict the presence of microbubbles in every pixel of the beamforming grid (low resolution). 
The dense network used to localize  microbubbles at low resolution is fully convolutional both in space and time direction and takes as input a tensor of shape $2 \times H \times W \times T$ in $2D$. The inputs channels encode the real and imaginary parts of the input signal. The input signal is composed of the local correlation of the beamformed IQ data with the simulated PSF of the imaging system. The spatial resolution was kept unchanged throughout the whole network. However, the temporal dimension was reduced by a factor of 2. The output of this network was then interpolated to match the correlation map dimension. The resulting mask is used to convert the correlation map to a sparse format, where only the pixel values with microbubbles are stored along with their coordinates. This filtering network was trained using the Dice loss between its predictions and the ideal mask at low resolution obtained from the simulated microbubble position.

\subsubsection{Architecture modifications}
Herein, we describe further experiments to refine the sparsity using pruning on the intermediate representation of the network along with deep-supervision and cascaded learning.
\paragraph{Deep Supervision and pruning}
Similarly to other architectures\cite{chenDeepLearningBasedMicrobubble2022,liuDeepLearningUltrasound2020, vanslounSuperResolutionUltrasoundLocalization2021}, Deep-stULM \cite{mileckiDeepLearningFramework2021} uses upsampling layers that preserve the sparsity of the upsampled tensor. This is sub-optimal as at finer resolution the sparsity of the trajectory is increased as depicted in Figure \ref{method_figure}. To mitigate these issues, we used the previously introduced pruning operations \cite{gwakGenerativeSparseDetection2020}, that aim to gradually remove the pixels where no microbubbles are detected (green pixels versus gray pixels in Figure \ref{method_figure}). The feature maps are masked based on the output of an intermediate classifier. Consequently, the removed pixels are no longer considered in the following operation and their coordinates are not stored in memory. 
As depicted in Figure \ref{method_figure}, we implemented pruning at every resolution level based on intermediate prediction. Since pruning requires the training of intermediate classifiers at each level, we trained them in a supervised fashion. These intermediate classifiers are trained using the same loss as the final loss and consist of pointwise convolution directly applied to the intermediate representation of the network. This form of supervision is similar to deep supervision \cite{leeDeeplySupervisedNets2015} and can also serve as regularization and improve the performance of the network. These intermediate classifiers are required to perform pruning, as they provide the mask used to remove the less relevant pixels from the following operations.
\paragraph{Cascaded learning}
In the case of super-resolution, deep supervision also makes possible a certain form of cascaded learning inspired by \cite{marquezDeepCascadeLearning2018}.
Indeed, the intermediate classifiers can be trained sequentially: during the first phase of the training, only the first classifier is trained to predict the presence of microbubbles on a grid at the input resolution. Then, during the following phase, the intermediate classifiers corresponding to higher resolution levels are sequentially added to the global loss.  When applied, the cascaded learning strategy used one epoch for each intermediate level, and the number of epochs for the last resolution level was the same as in standard training.

\section{Results}

\begin{table*}
        \caption{Comparison of the memory usage, inference time and angiogram reconstruction performance. * Value estimated. 
   }

    \centering
    \begin{tabular}{ccccccccc}
        &  \multicolumn{3}{c}{2D ($5$MB/FOV)} & & \multicolumn{3}{c}{3D ($30$MB/FOV)}\\
        \cmidrule{2-4} \cmidrule{6-8}
            &\makecell{GPU \\ Inference \\time (ms)} & \makecell{ Memory \\ requirements
        \\ (GB)}  &  Dice (\%) & &\makecell{GPU \\ Inference \\time (ms)} & \makecell{ Memory \\ requirements
        \\ (GB)}  & Dice (\%)  \\
        \hline
        Sparse Deep-stULM & \boldmath{$19 \pm 1.3$} & \boldmath{$6.8 \pm 0.2$} & $73.93 \pm 1.96$ & & \boldmath{$70\pm 13.8$} & \boldmath{$9.9\pm 0.1$} & \boldmath{$49.97 \pm 1.79$}\\
        \hline
        Deep-stULM  & $34 \pm 0.8$ & $12.6$ & \boldmath{$80.01 \pm 1.76$} & & N.A & $694$* & N.A\\
        \hline
        mSPCN-ULM & $77 \pm 0.9$ & $22.7$ & $62.54  \pm 0.30$& &   $8016 \pm 533.2$* & $366$* & N.A\\
        \hline
        ULM & N.A & N.A & $60.80$ & & N.A & N.A &  $12.34$ \\

    \end{tabular}
    \label{tab:results_time_memory}
    
\end{table*}

\begin{figure*}[h!]
\begin{tikzpicture}[
    arrow/.style={arrows={Triangle[length=3mm, width=2mm]-}}
]
  \matrix (figs) [matrix of nodes, inner sep=0, column sep=1mm, row sep=1mm, nodes={inner sep=0pt}] {
    \includegraphics[width=.2\linewidth]{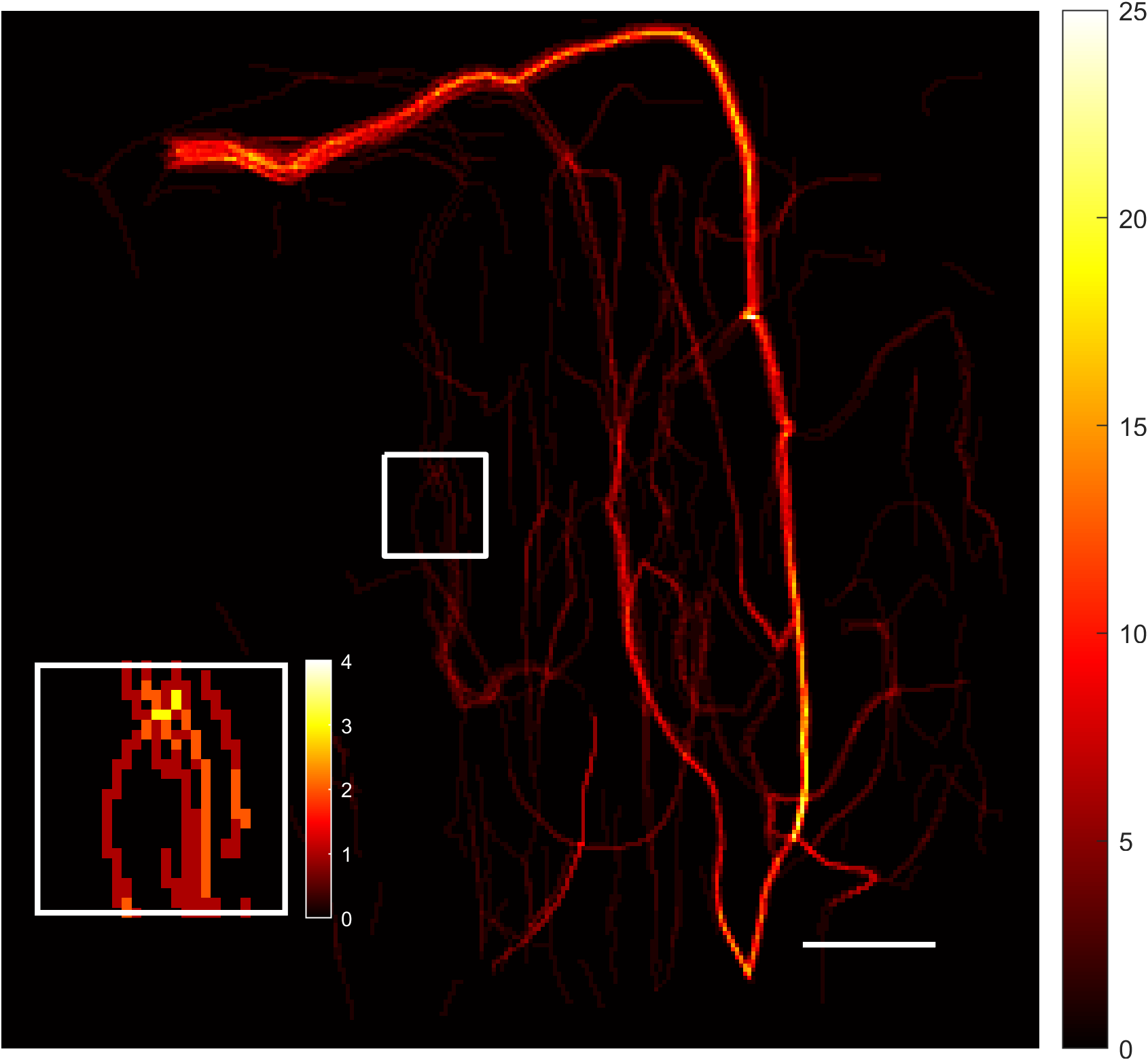} &
    \includegraphics[width=.2\linewidth]{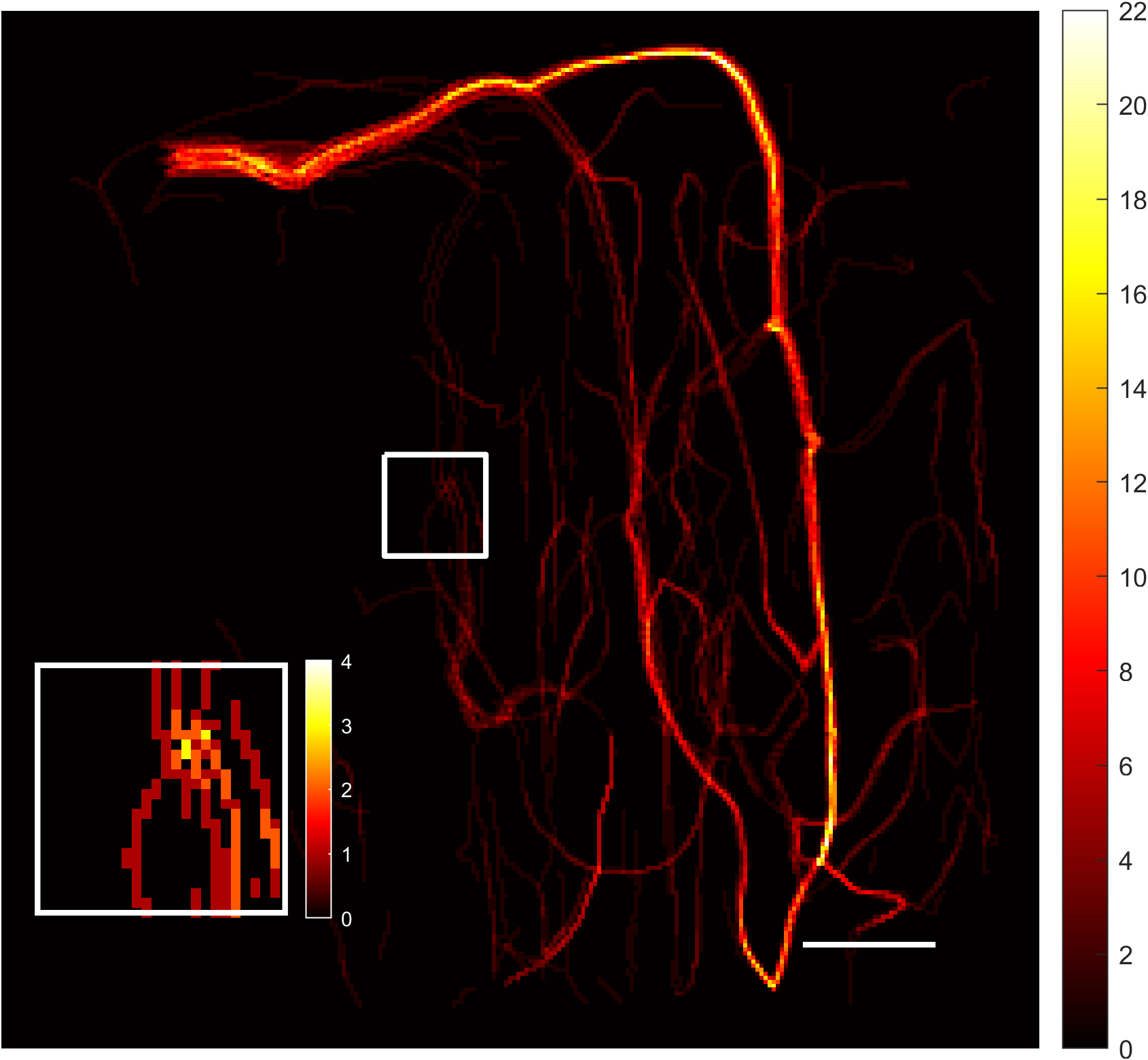} &
    \includegraphics[width=.2\linewidth]{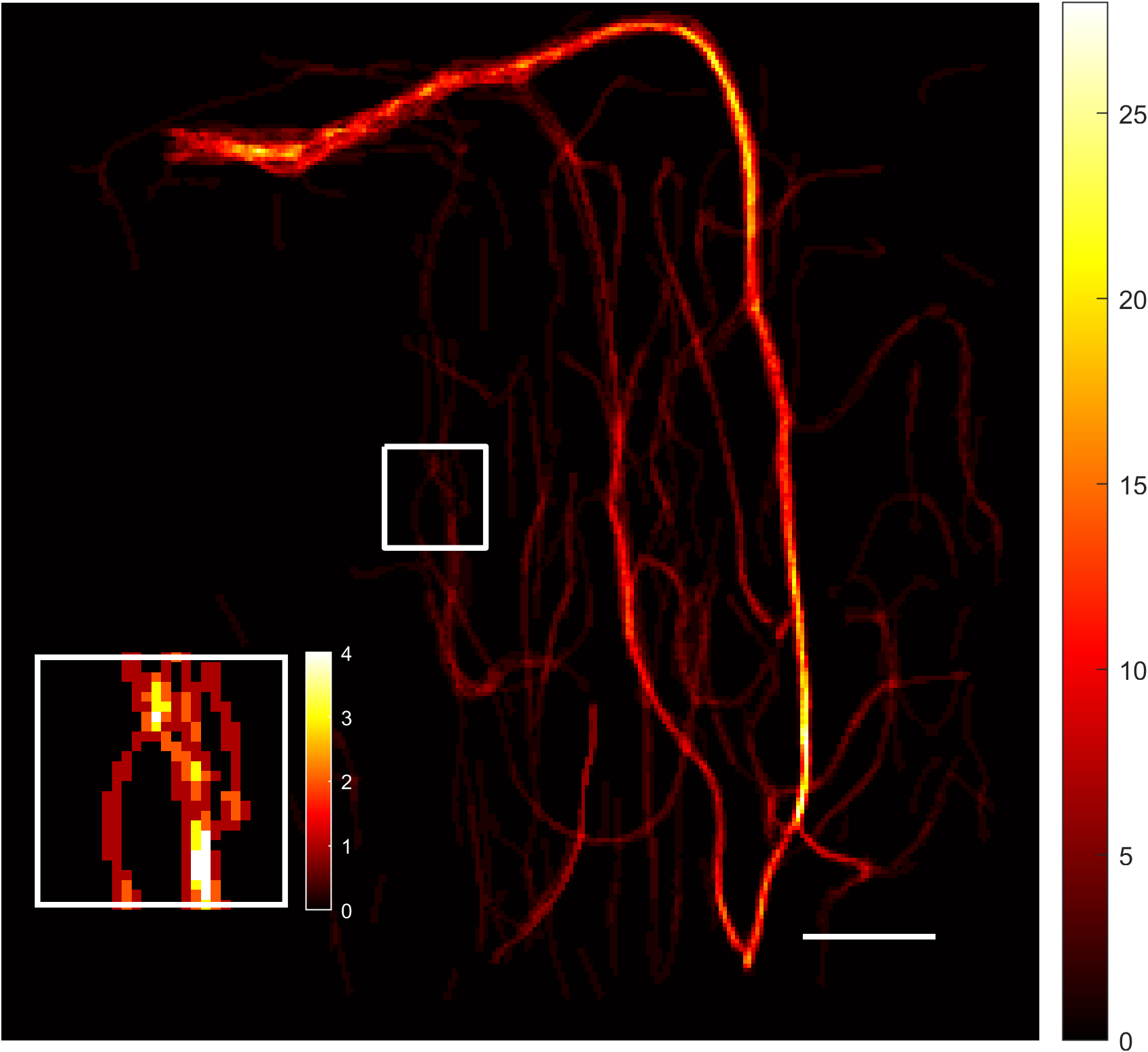} &
    \includegraphics[width=.2\linewidth]{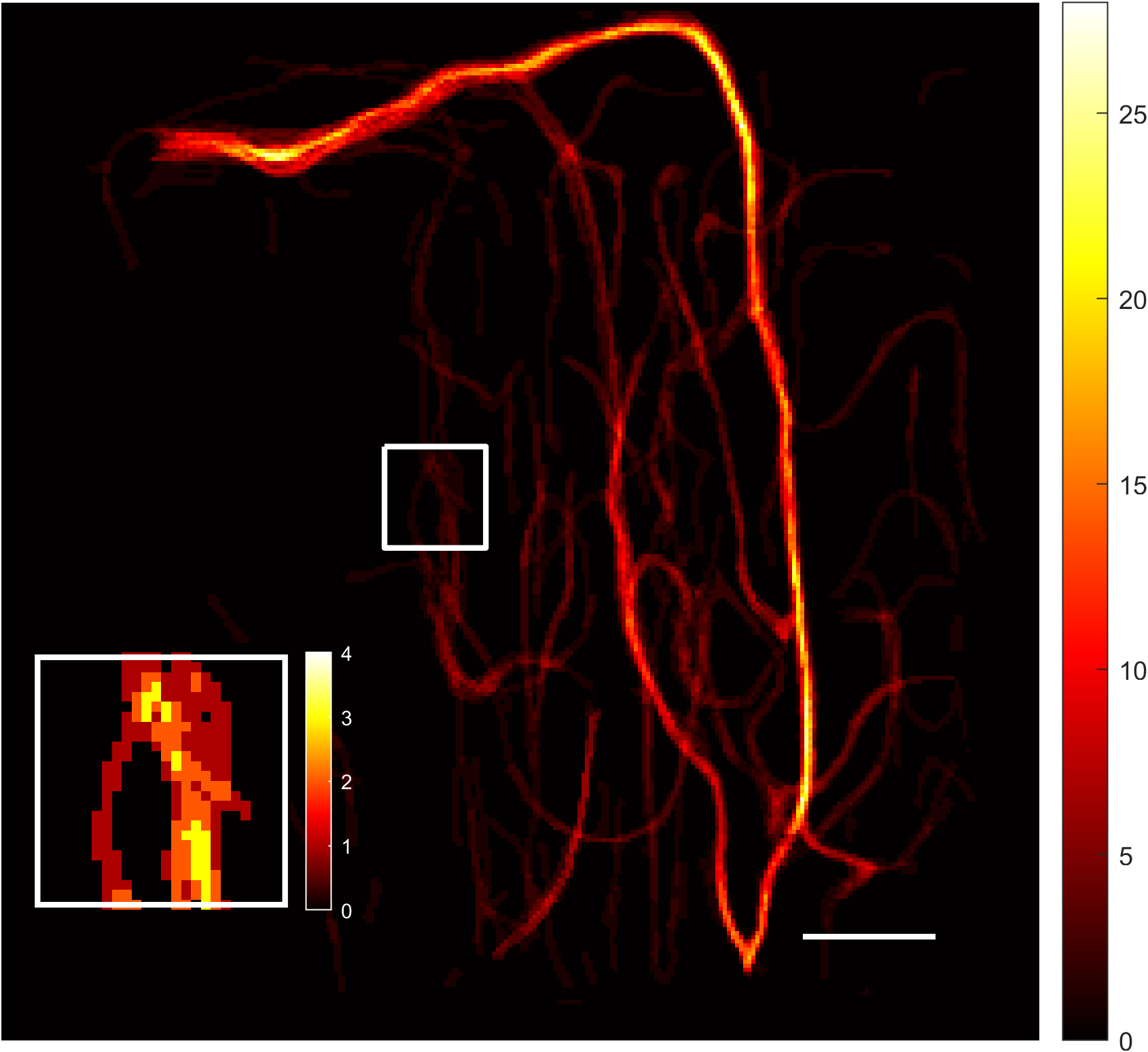} \\
    \includegraphics[width=.2\linewidth]{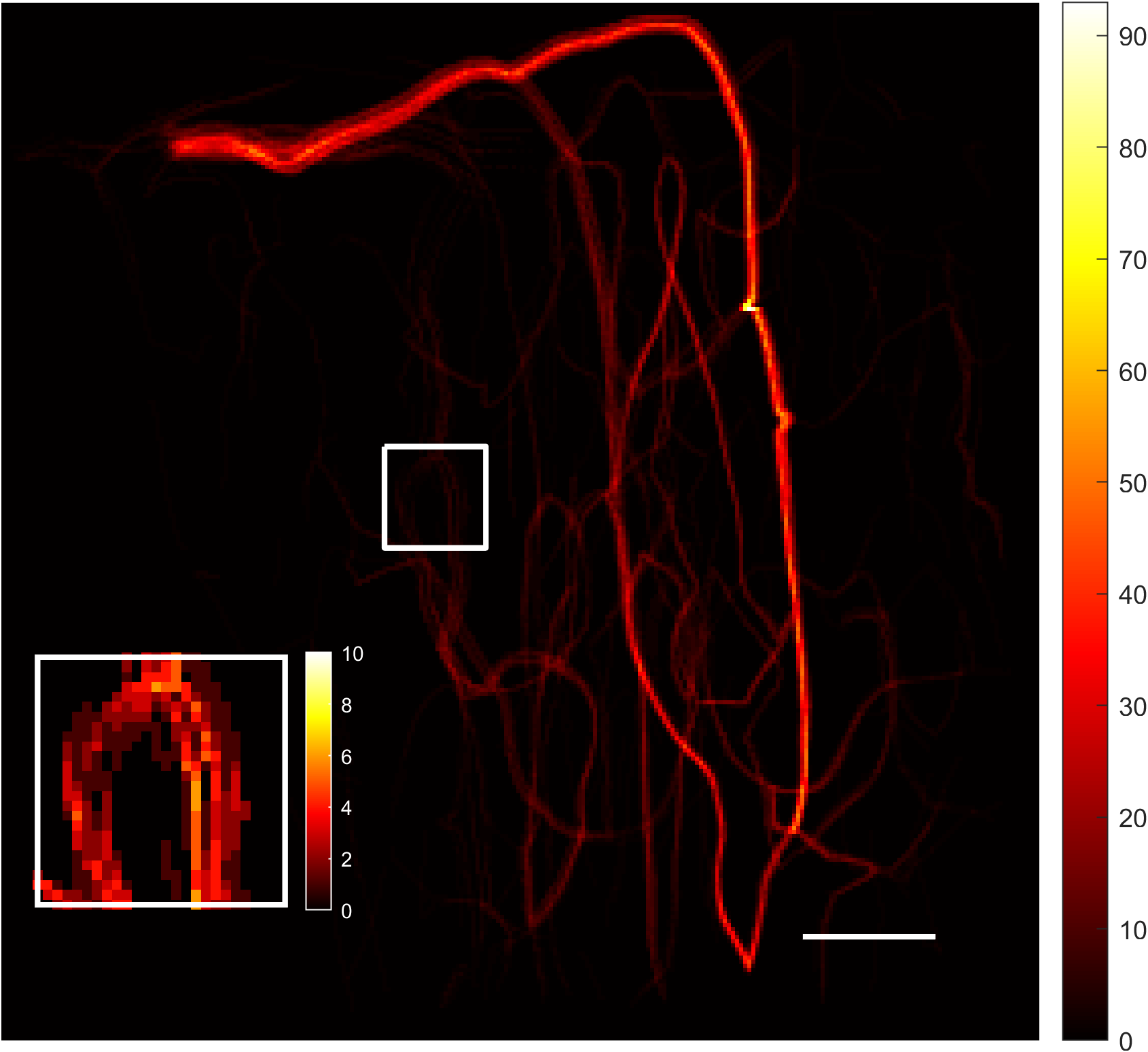} &
    \includegraphics[width=.2\linewidth]{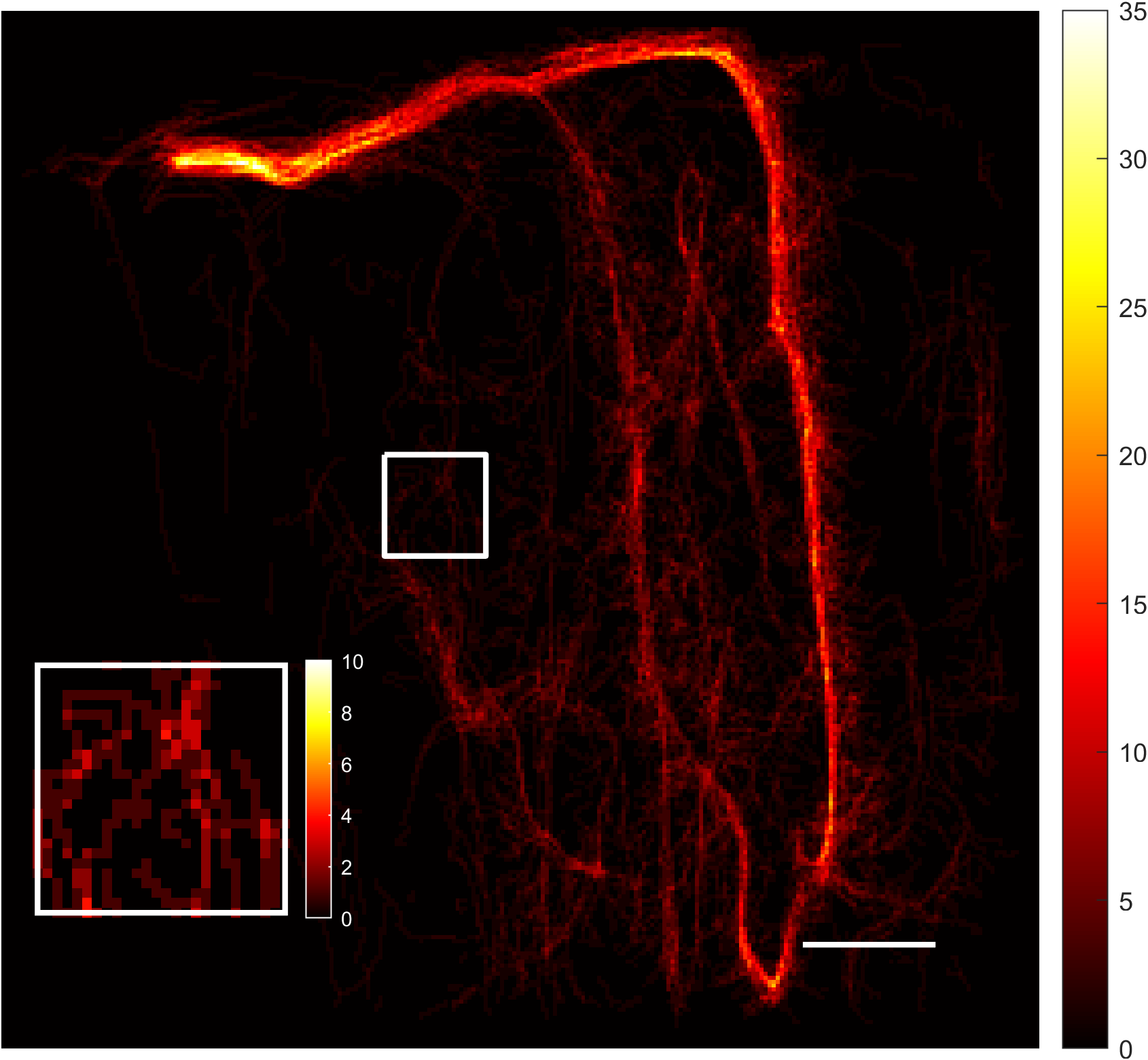} &
    \includegraphics[width=.2\linewidth]{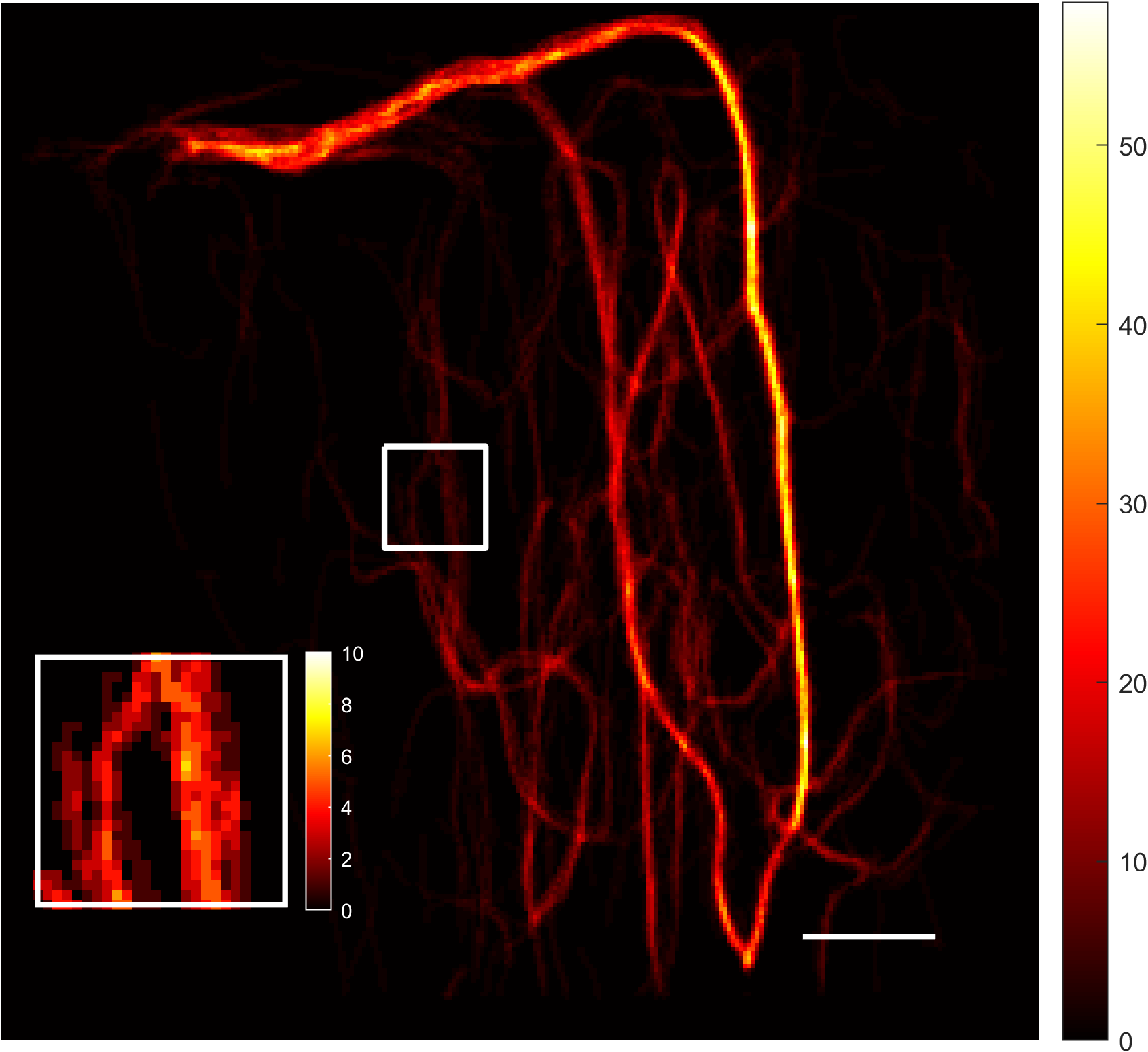} &
    \includegraphics[width=.2\linewidth]{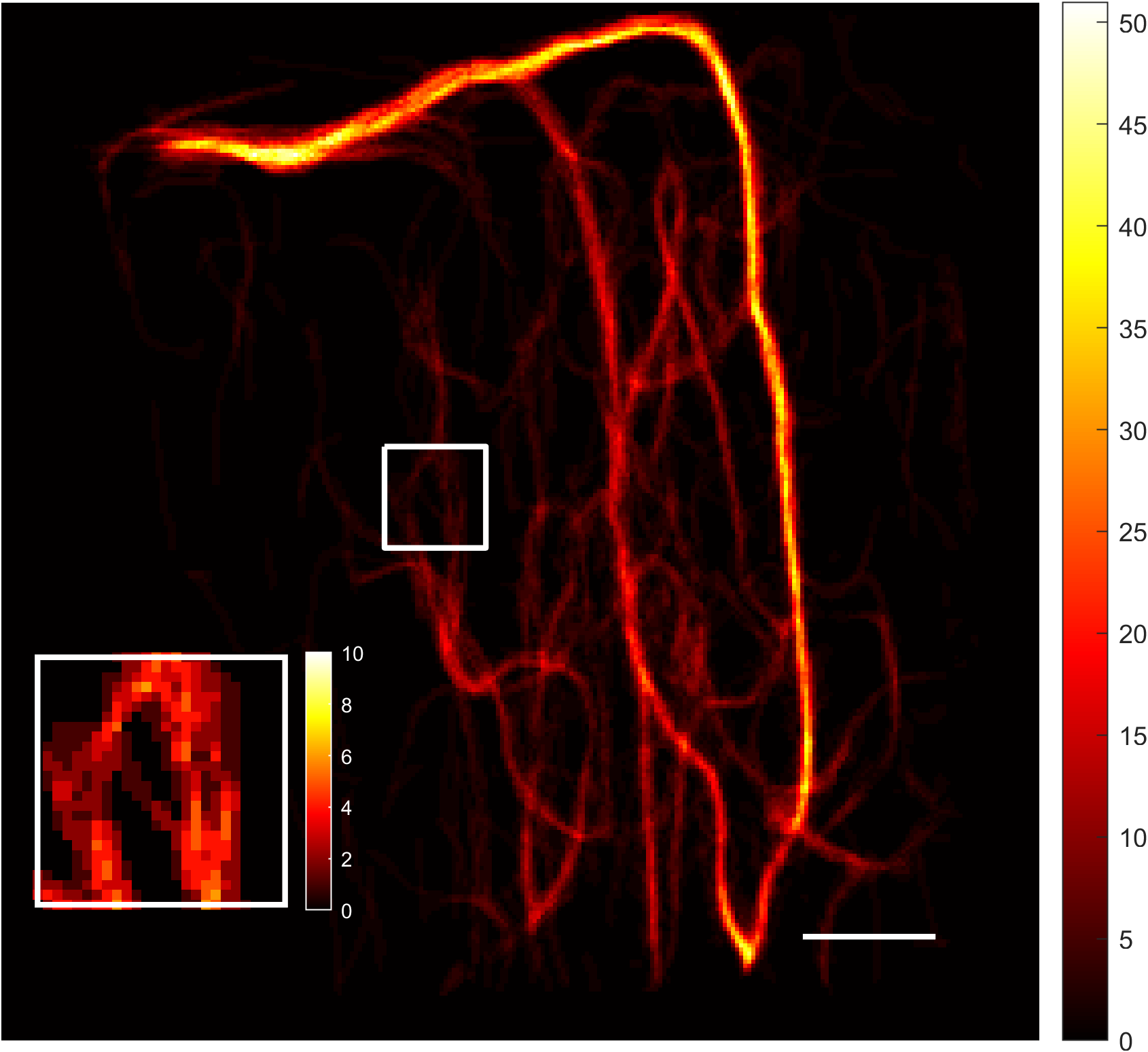} \\
    
    \includegraphics[width=.2\linewidth]{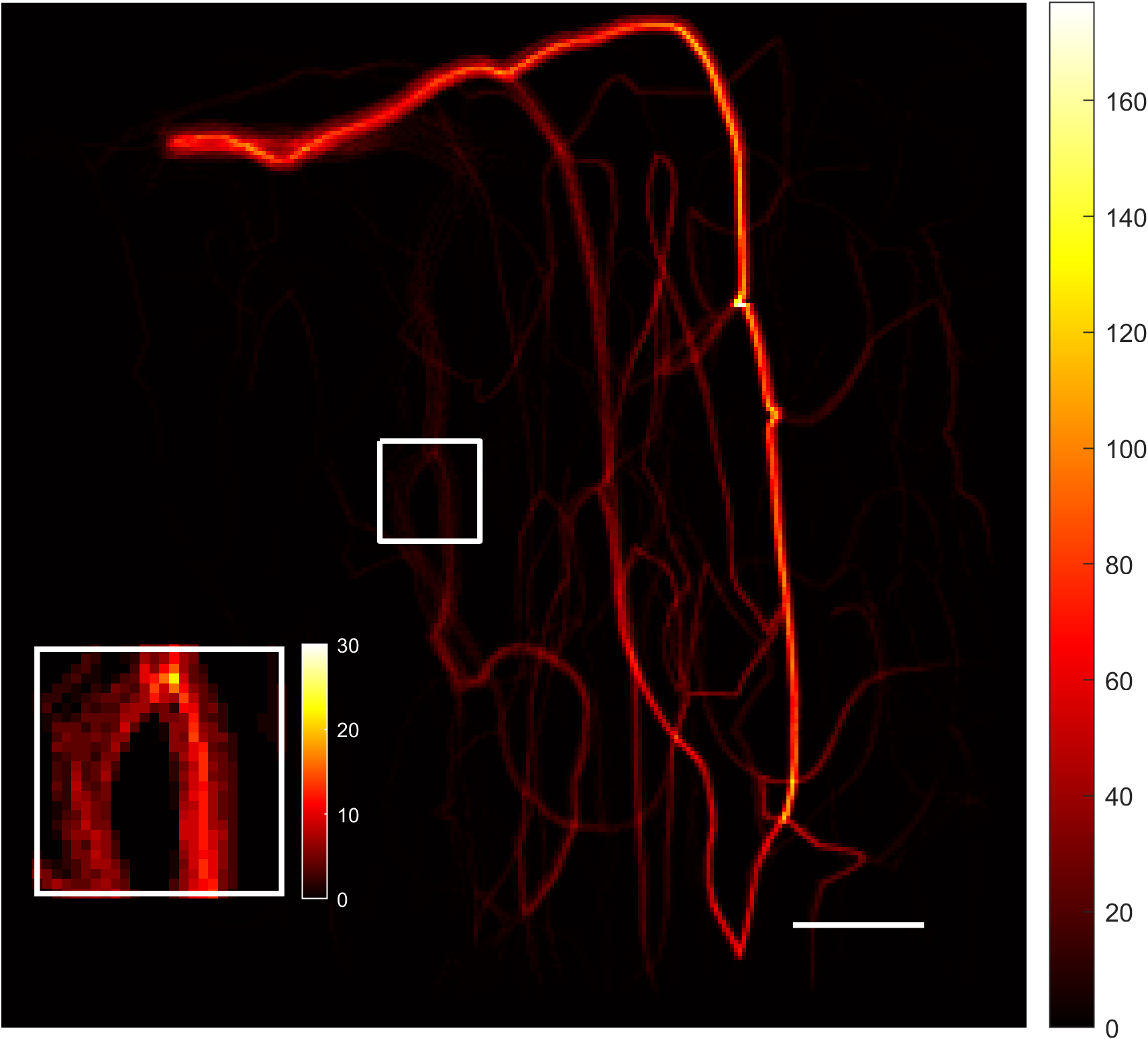} &
    \includegraphics[width=.2\linewidth]{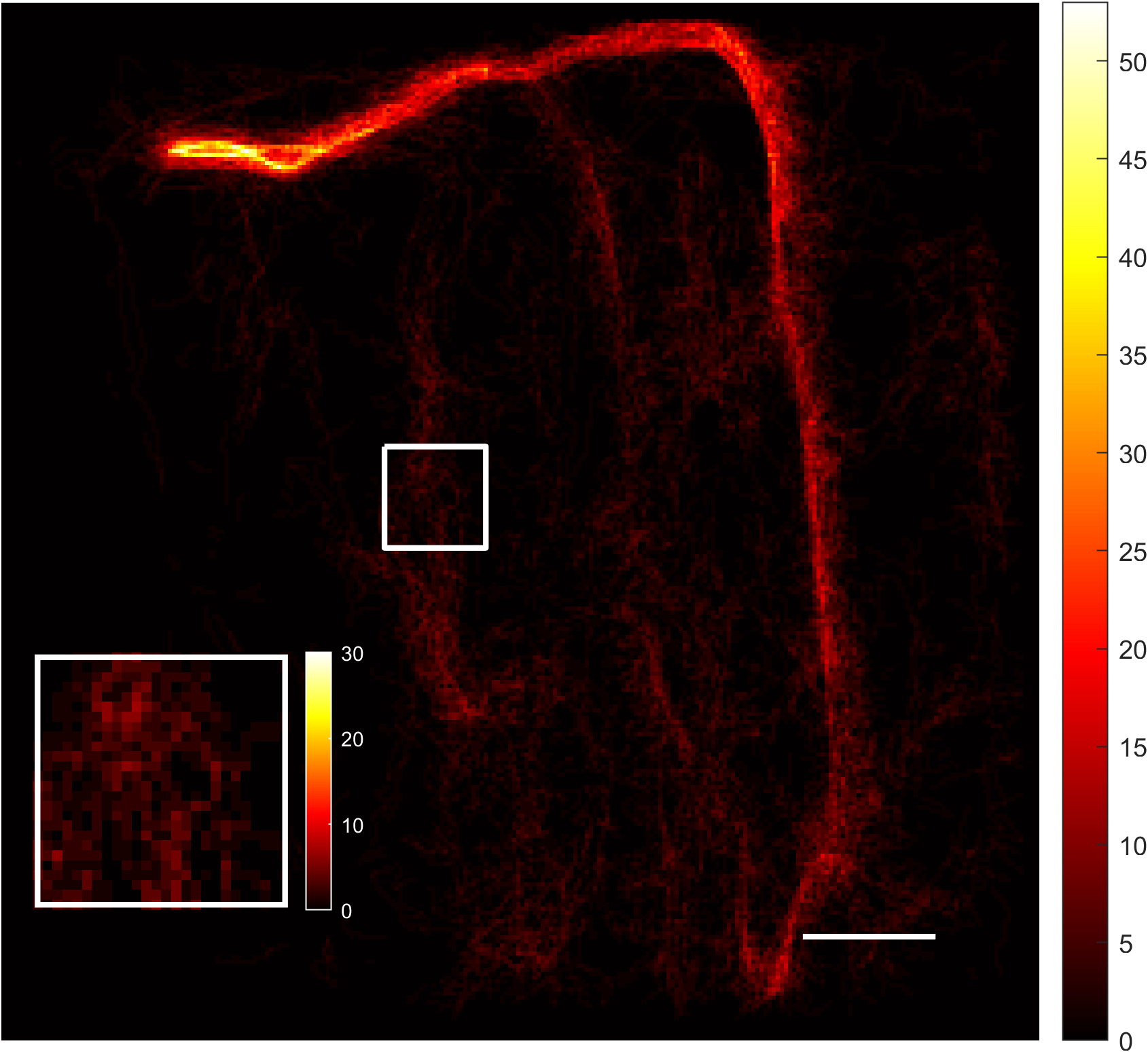} &
    \includegraphics[width=.2\linewidth]{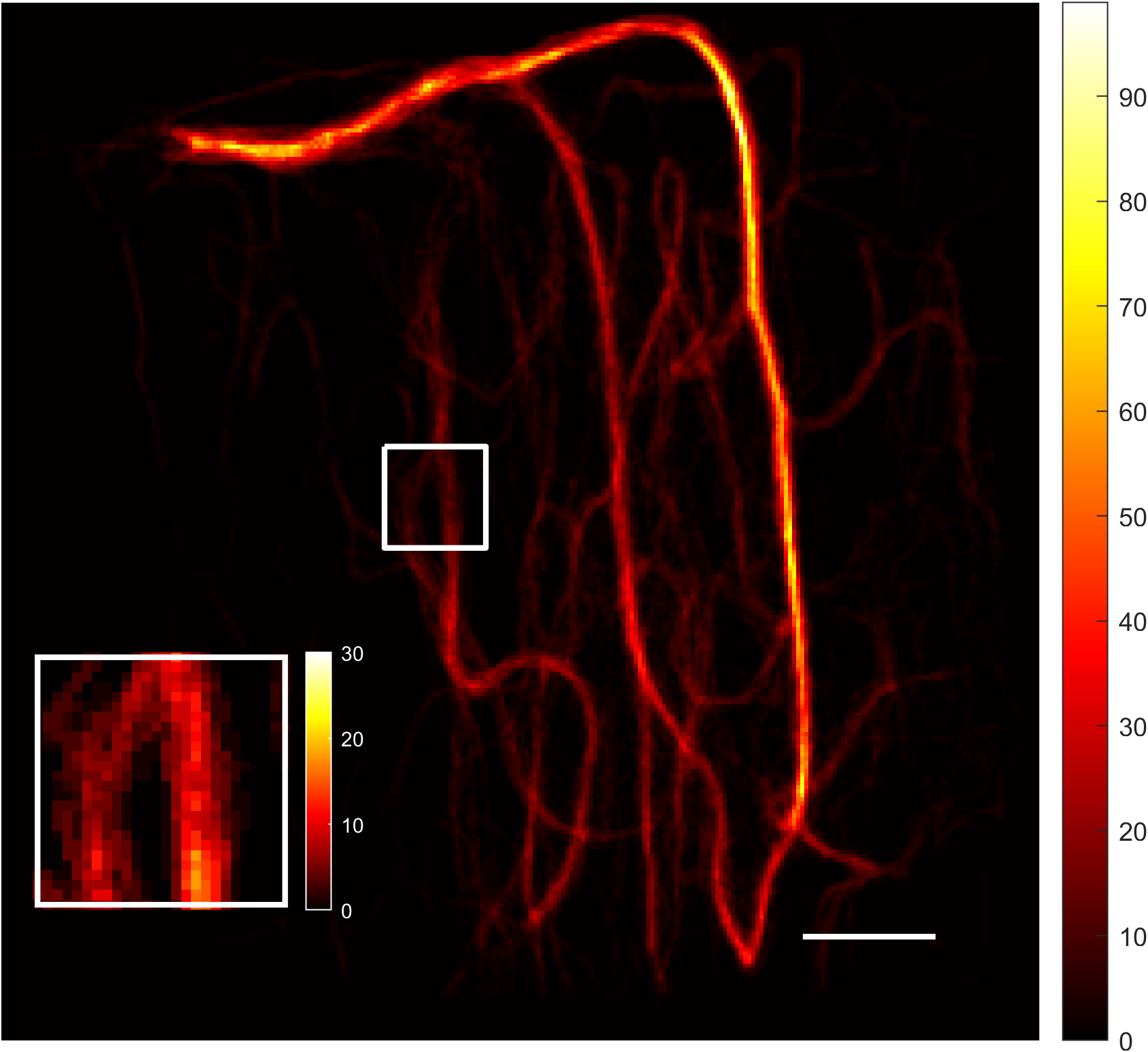} &
    \includegraphics[width=.2\linewidth]{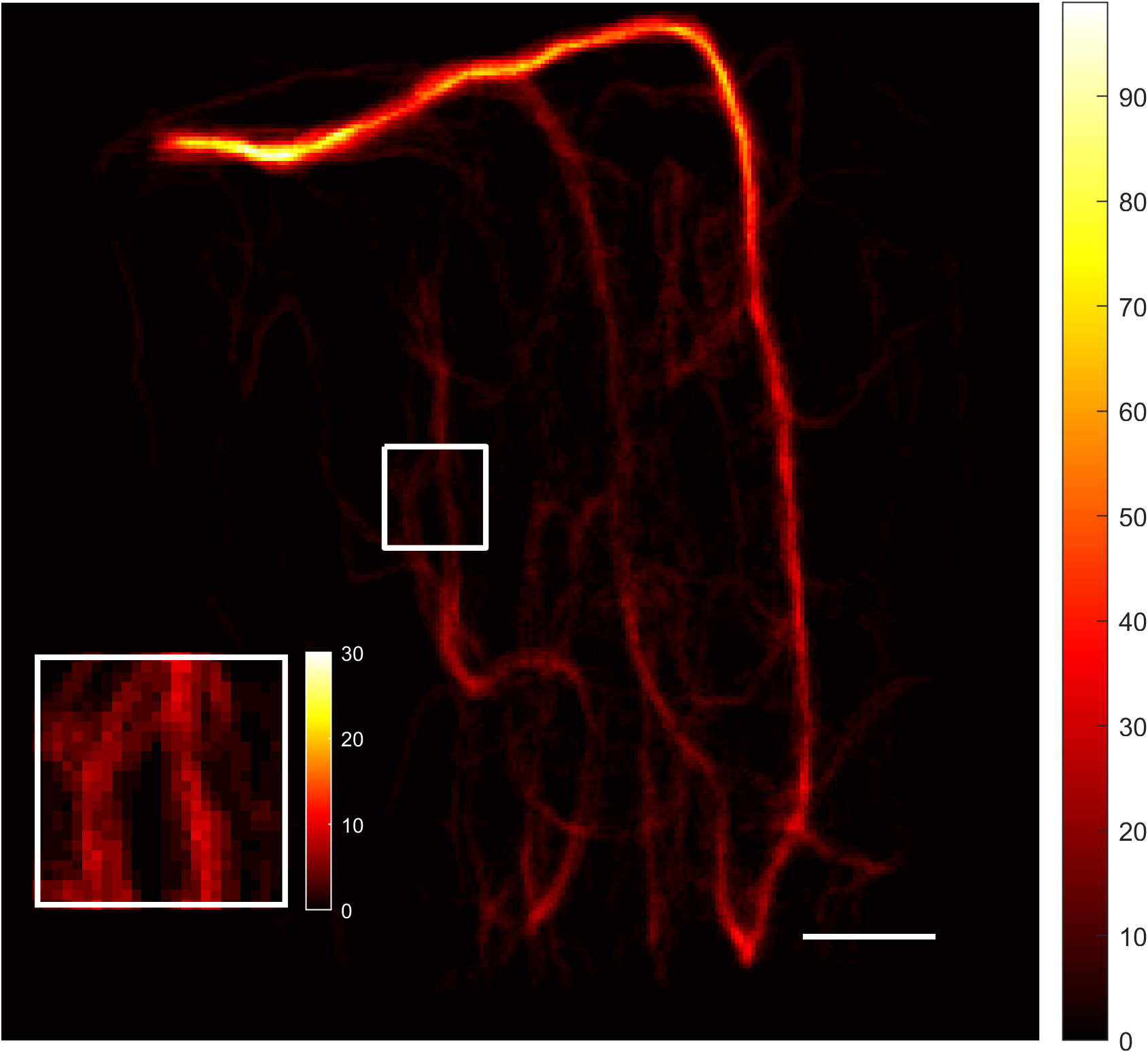} \\
    
    \includegraphics[width=.2\linewidth]{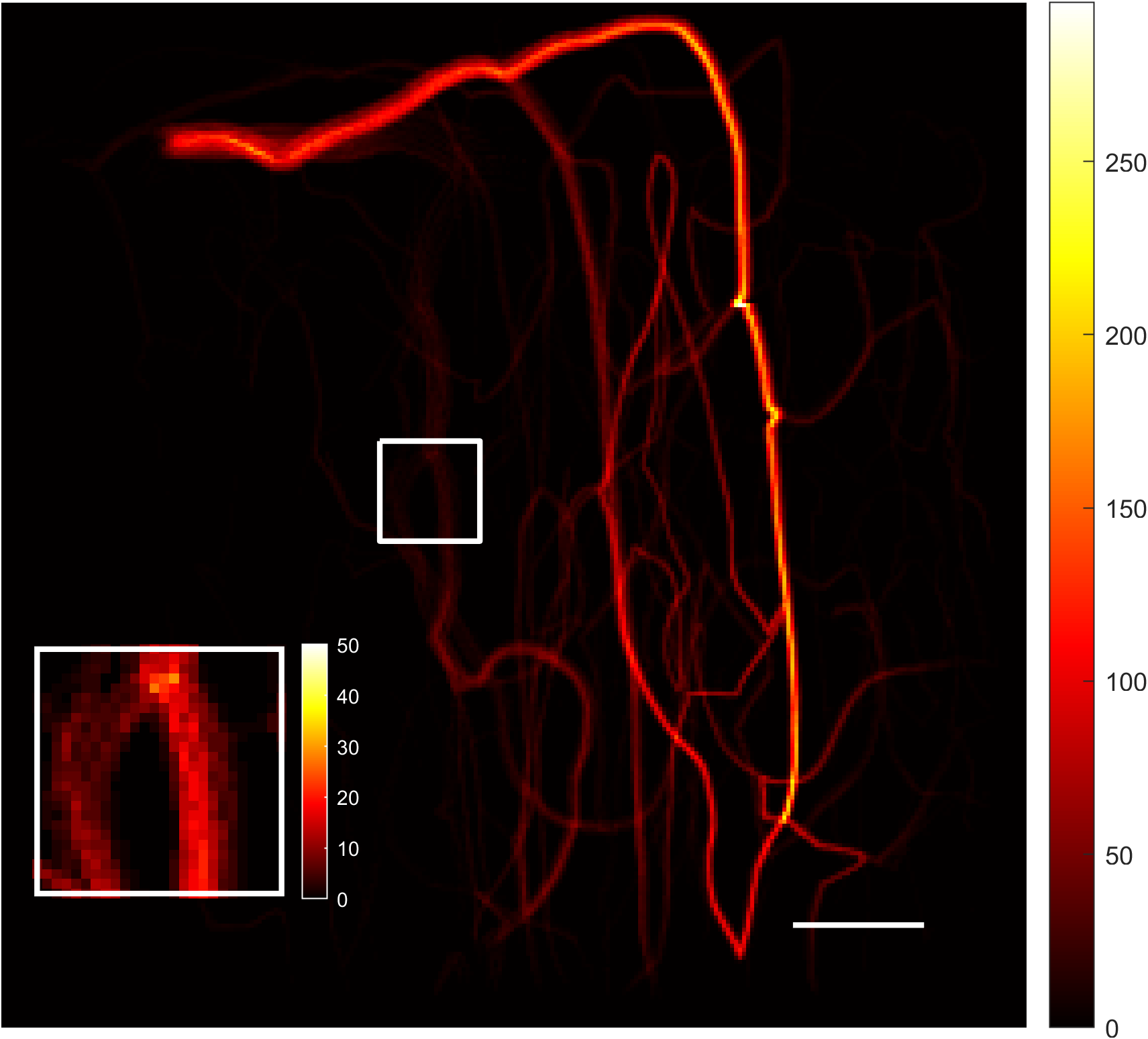} &
    \includegraphics[width=.2\linewidth]{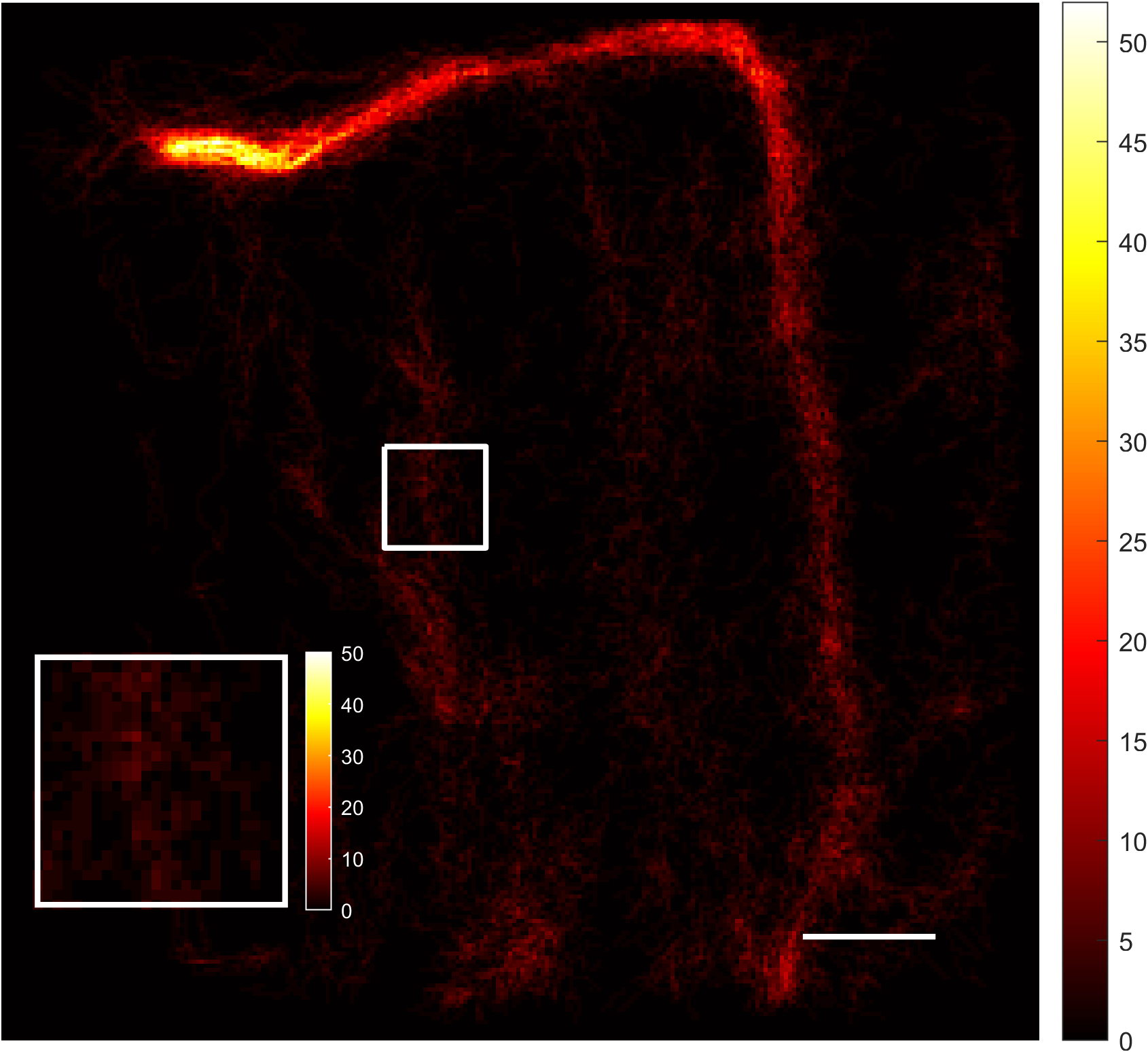} &
    \includegraphics[width=.2\linewidth]{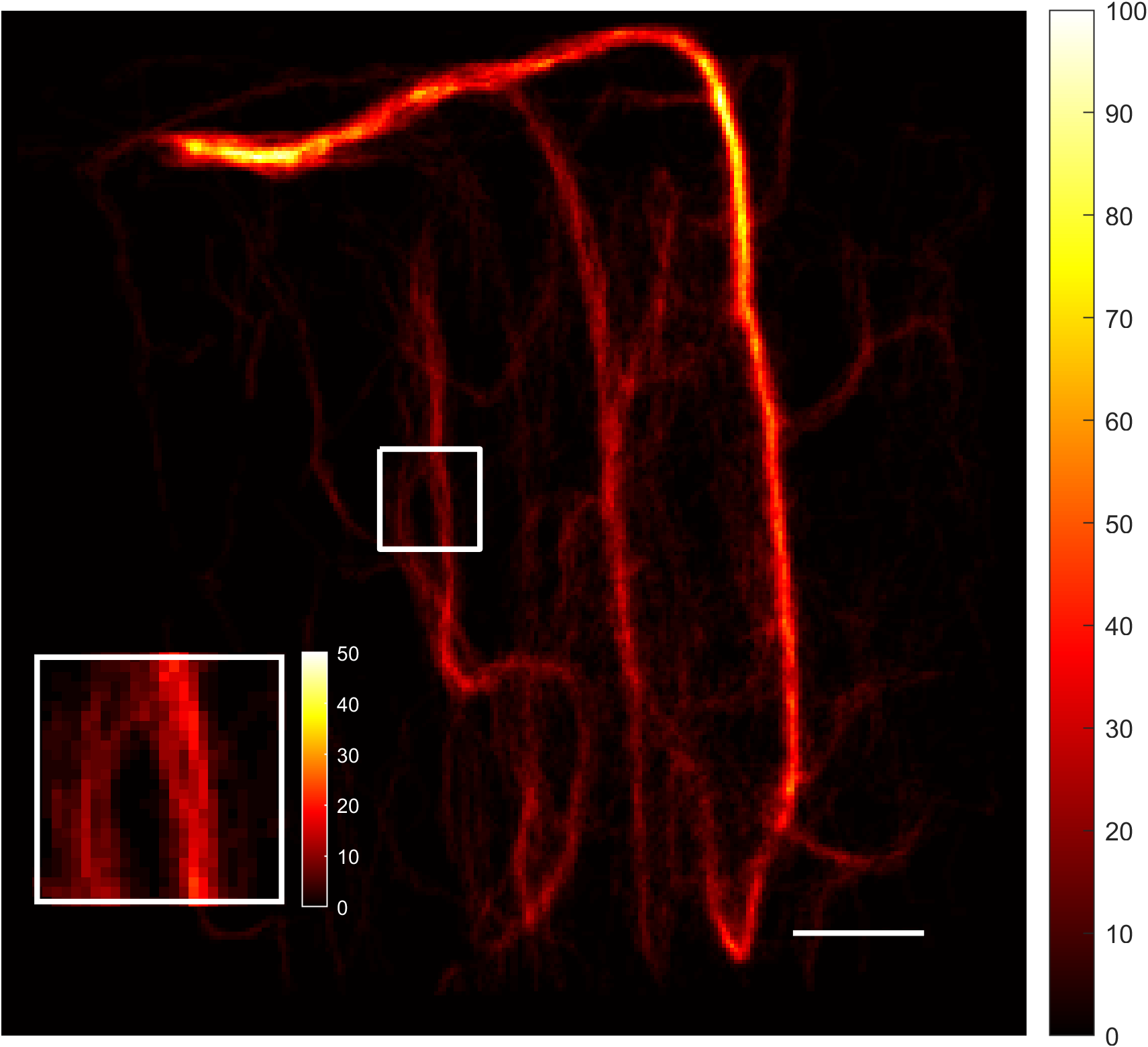} &
    \includegraphics[width=.2\linewidth]{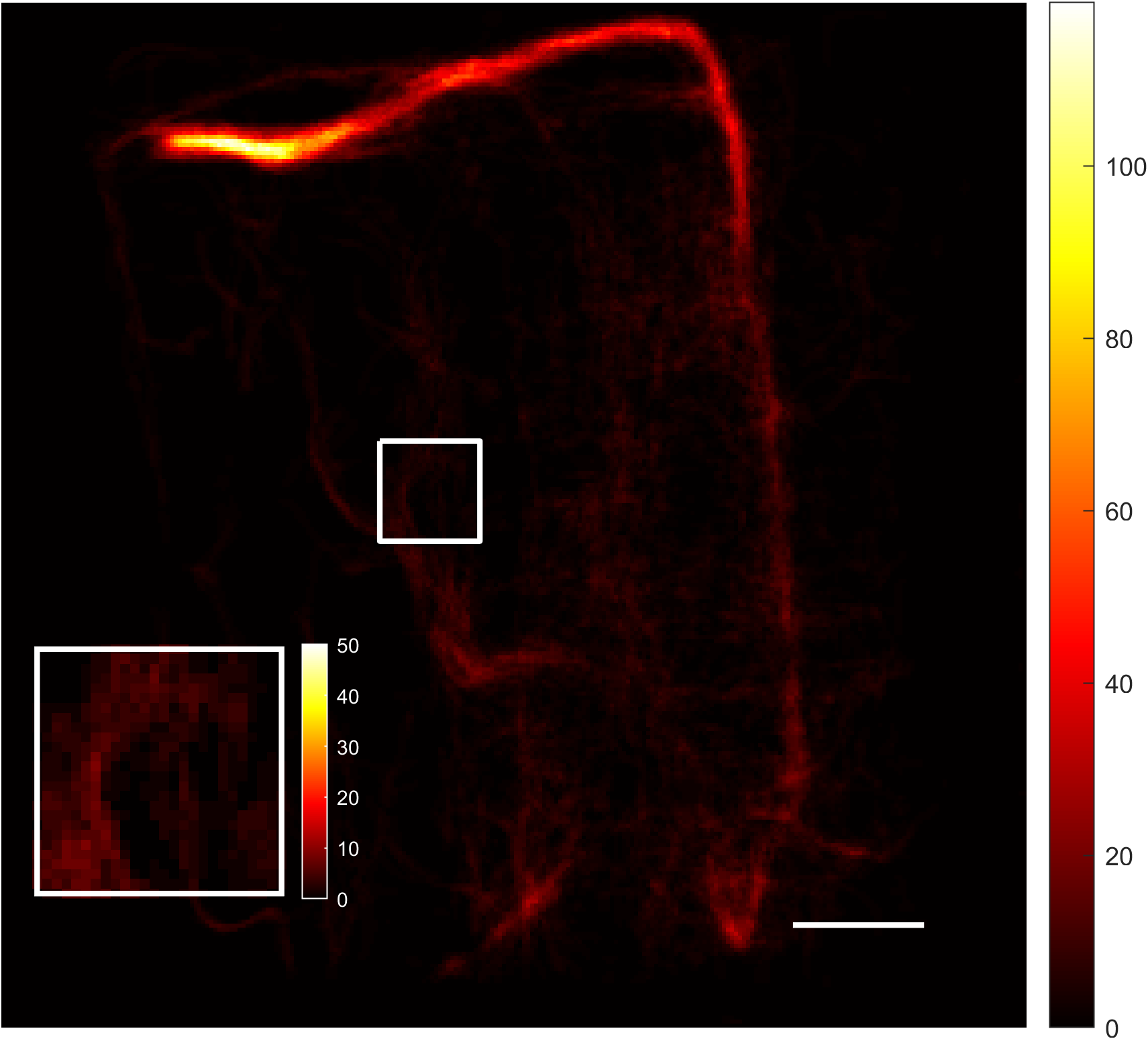} \\
  };
  
    \node [left=4mm of figs-1-1] {\scriptsize 1 MB};
    \node [left=4mm of figs-2-1] {\scriptsize 5 MB};
    \node [left=4mm of figs-3-1] {\scriptsize 10 MB};
    \node [left=4mm of figs-4-1] {\scriptsize 20 MB};
    \node [above=1mm of figs-1-1,text width=.2\linewidth, align=center] {\scriptsize Ground Truth};
    \node [above=1mm of figs-1-2,text width=.2\linewidth, align=center] {\scriptsize Conventional ULM};
    \node [above=1mm of figs-1-3,text width=.2\linewidth, align=center] {\scriptsize Dense \\ Deep-stULM};
    \node [above=1mm of figs-1-4,text width=.2\linewidth, align=center] {\scriptsize Sparse\\ Deep-stULM};
  
\draw[arrow] ([xshift=-1mm]figs-4-1.west) -- ([xshift=-1mm]figs-1-1.west);
  \node[rotate=90, left=3mm of figs, anchor=center] {Concentration};
\end{tikzpicture}
\caption{Comparison of performance under increasing concentration for conventional ULM (center left column), Deep-stULM dense formulation (center right column) and its sparse formulation (right column). Ground truth is given for comparison (left column). The scale bar is 98µ and corresponds to the wavelength of the simulated pulse. Concentration increases from 1 (top row), 5, 10, and 20 (bottom row) microbubbles per field of view.}
\label{fig:angio2D}

\end{figure*}

\subsection{Processing time, memory reduction and performance comparison in 2D}
\begin{figure}
\centering
\includegraphics[height=4cm]{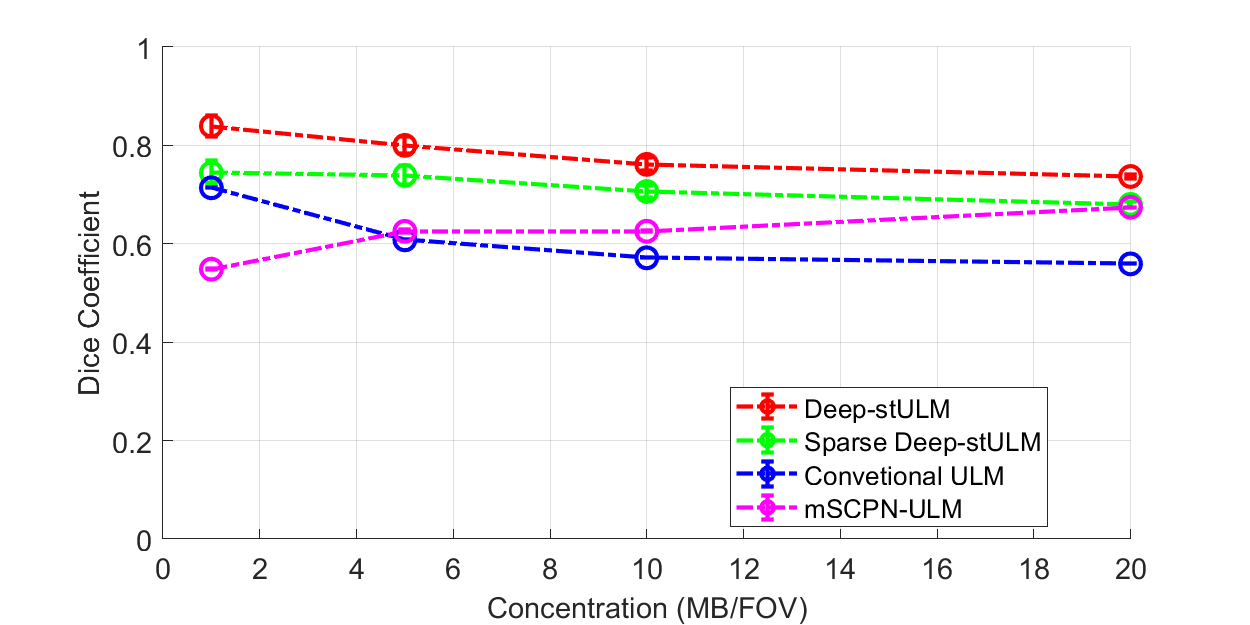}  
     \caption{Evolution of the angiogram reconstruction performance on vascular network datasets with increasing concentration}
     \label{fig:performance_curve_concentrationAngiogram}
\end{figure}

\begin{figure}
         \centering
        \includegraphics[height=4cm]{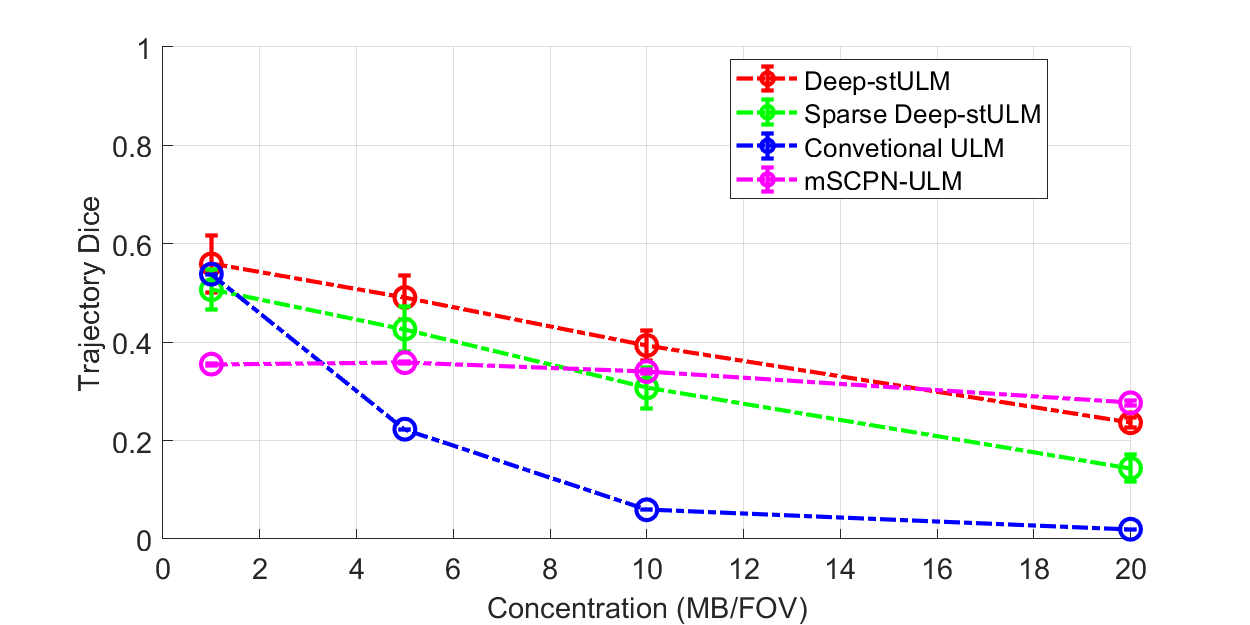}
         \caption{Evolution of the trajectory detection on random trajectory datasets with increasing concentration}
         \label{fig:performance_curve_concentration_random}
\end{figure}

The memory usage results and performance at 5 MB/FOV are reported in Table \ref{tab:results_time_memory} for Conventional ULM, Deep-stULM, mSPCN-ULM and Sparse Deep-stULM.
The threshold strategy was used for the sparse method to convert the input to sparse formulation. 
Sparse formulation reduced the memory usage of Deep-stULM from $12.6$ GB to $6.8$ GB during training, the processing time from $34$ ms to $19$ ms, while the Dice decreased from $80.1\%$ to $73.9\%$.
In comparison, mSPCN-ULM, with $4$ times more processing time and with nearly four-fold higher memory usage in training,  obtained a lower Dice of $62.54\%$, yet slightly outperforming conventional ULM (Dice of $60.8\%$).
The results of the different methods in 2D under varying concentrations are shown in Figure \ref{fig:angio2D}, and the evolution of the Dice value computed is reported in Figure \ref{fig:performance_curve_concentrationAngiogram} along with the results on random trajectory datasets with varying concentrations in Figure \ref{fig:performance_curve_concentration_random}. Qualitatively, the standard ULM performances degraded as the concentration increased with degradation in resolution starting from 5 microbubbles per field of view (MB/FOV).
 Quantitatively, this diminution of performance was highlighted by a drop in Dice coefficient from $71.4\%$ with 1 MB/FOV to $60.8\%$ for 5 MB/FOV on angiogram reconstruction.
 The performance continued to decrease with the concentration until it reached a Dice of $55.9\%$ at 20 MB/FOV, with only the biggest vessels being visible. 
In contrast, the dense Deep-stULM approach exhibited a smaller performance degradation from $83.9\%$ for 1 MB/FOV to $73.7\%$ for 20 MB/FOV. The sparse formulation of Deep-stULM showed robustness to increased concentration and reached performance levels at high concentration (with a Dice coefficient of $70.6\%$ (resp. $68.0\%$) for 10 (resp. 20) MB per FOV that were very close to its performance level at low concentration ($74.4\%$ for 1MB per FOV). However, the performances at low concentration (1MB per FOV) were lower than Deep-stULM ($83.9\%$) but were comparable to conventional ULM ($71.4\%$). In opposition, mSPCN-ULM performance increased with the concentration from $54.9$ at 1MB/FOV to  $67.4$ at 20MB/FOV, reaching a level of performance similar to sparse and dense Deep-stULM and outperforming conventional ULM.
Evaluation on random trajectory datasets in Figure. \ref{fig:performance_curve_concentration_random} showed a similar trend for Conventional ULM and sparse and dense Deep-stULM. However, for mSPCN-ULM the \textit{trajectory Dice} slightly decreased from $35.5\%$ at 1MB/FOV to $27.7\%$ at 20MB/FOV outperforming the other methods.

\begin{figure}
    \centering
        \includegraphics[height=4cm]{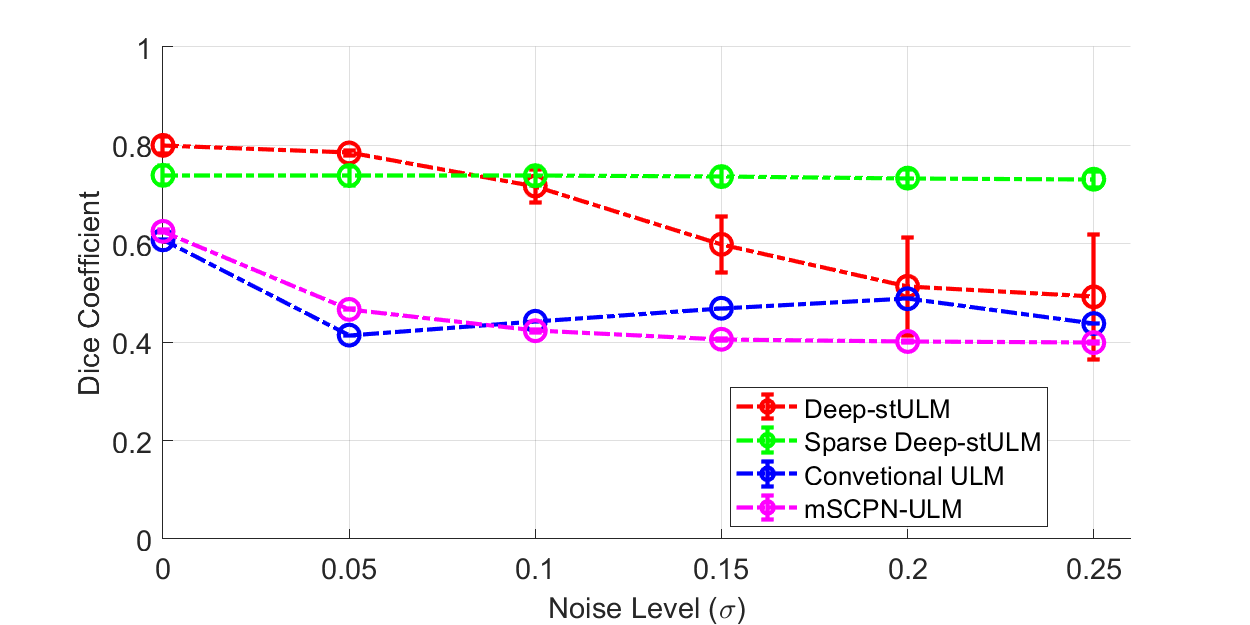}
    \caption{Evolution of the performance as a function of additive noise standard deviation}
    \label{fig:performance_curve_Noise}
\end{figure}

As seen in Figure \ref{fig:performance_curve_Noise}, evaluation under additive noise at test time showed a decrease in performance for all the methods but the sparse formulation of Deep-stULM, which performed similarly at all noise levels (from $73.9$ for $\sigma = 0.1$ to $73.1$ for $\sigma = 0.25$). 

\subsection{3D feasibility study}
\begin{figure*}[h!]
\begin{tikzpicture}[
    arrow/.style={arrows={Triangle[length=3mm, width=2mm]-}}
]
  \matrix (figs) [matrix of nodes, inner sep=0, column sep=1mm, row sep=1mm, nodes={inner sep=0pt}] {
    \includegraphics[width=.3\linewidth, trim=85mm 55mm 85mm 55mm, clip]{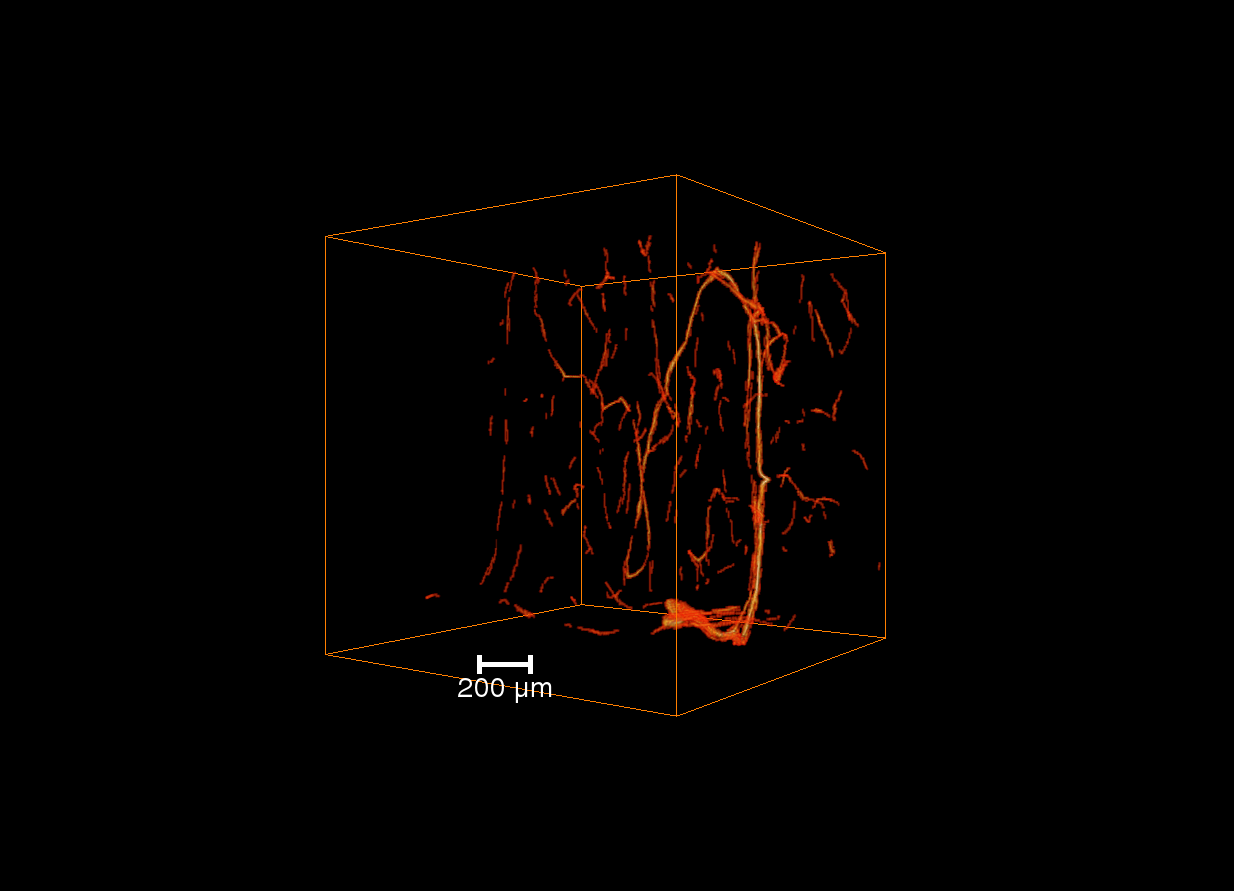} &
    \includegraphics[width=.3\linewidth, trim=85mm 55mm 85mm 55mm, clip]{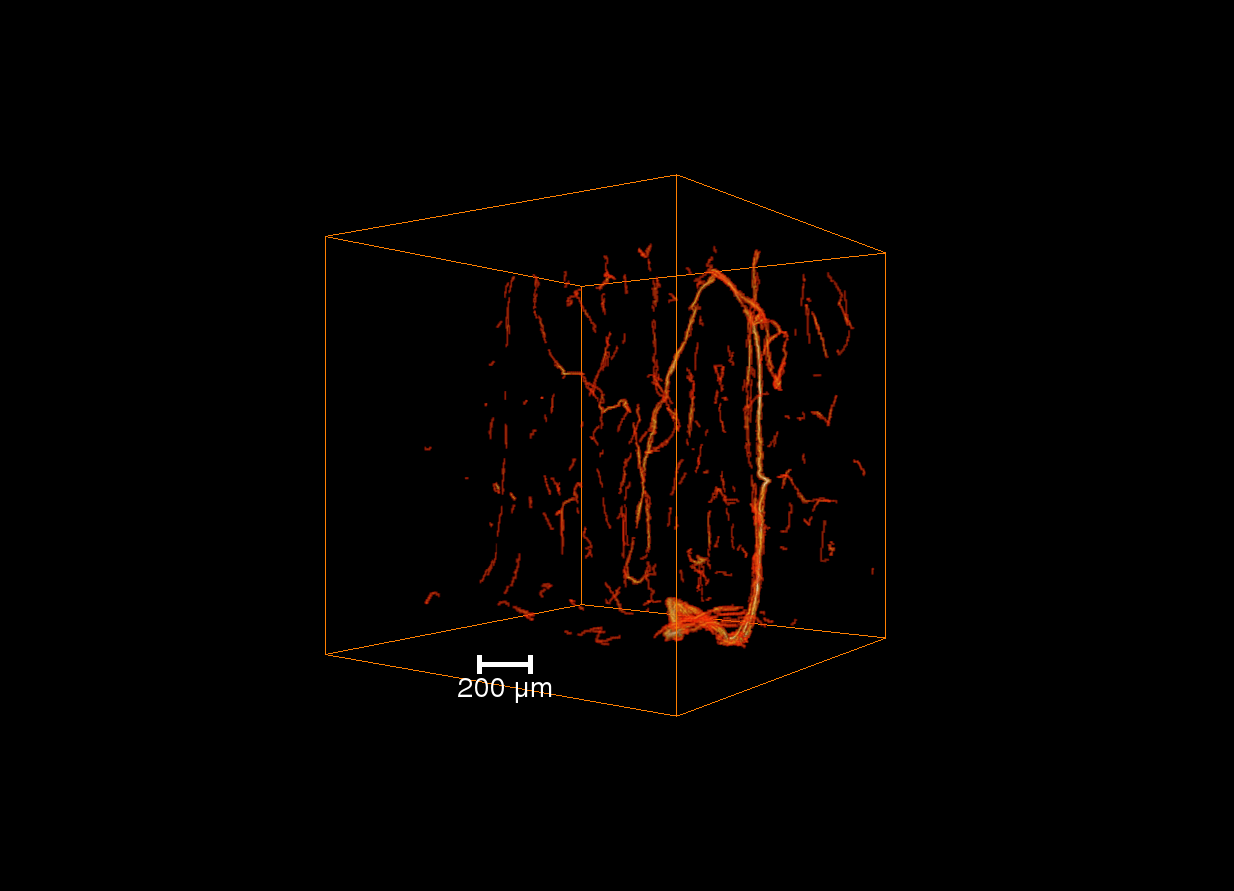} &
    \includegraphics[width=.3\linewidth, trim=85mm 55mm 85mm 55mm, clip]{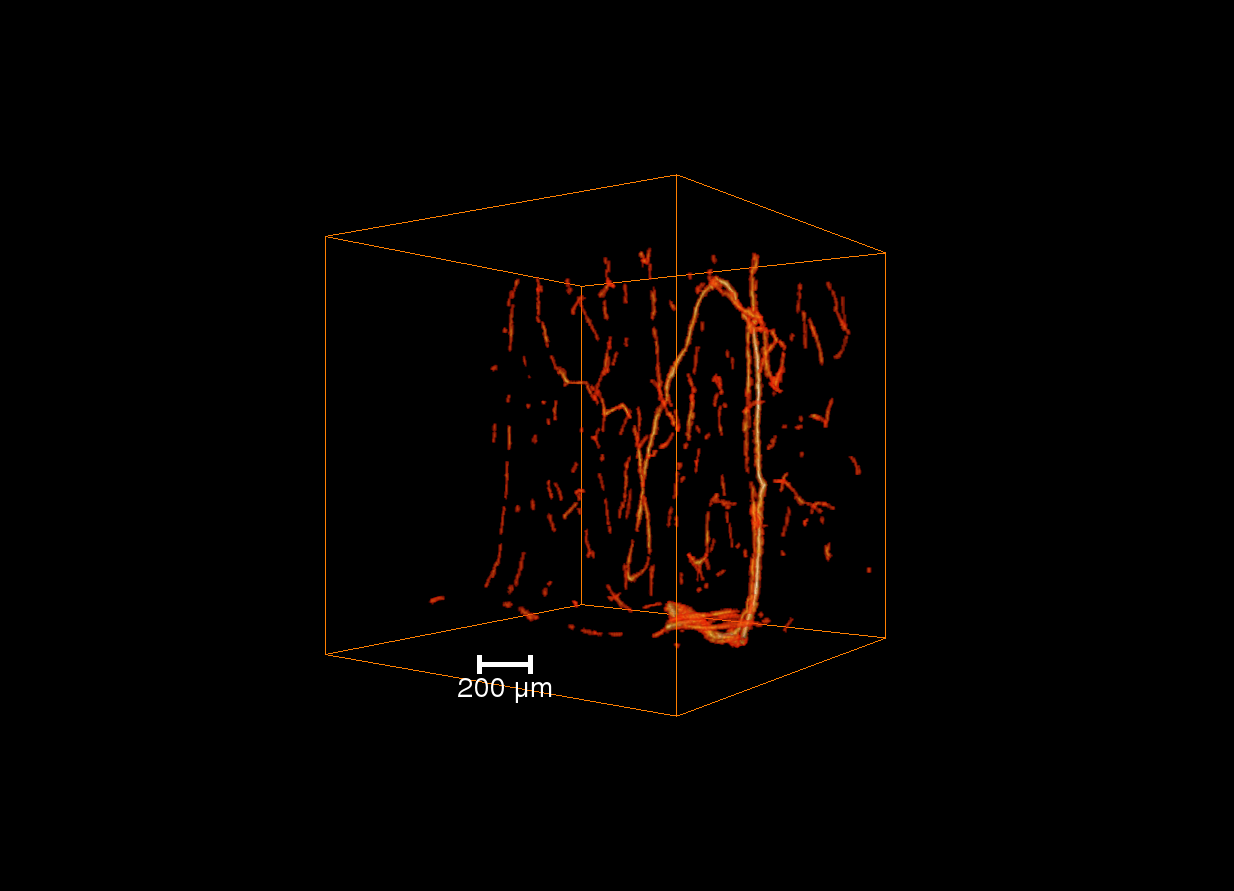} \\
    \includegraphics[width=.3\linewidth, trim=85mm 55mm 85mm 55mm, clip]{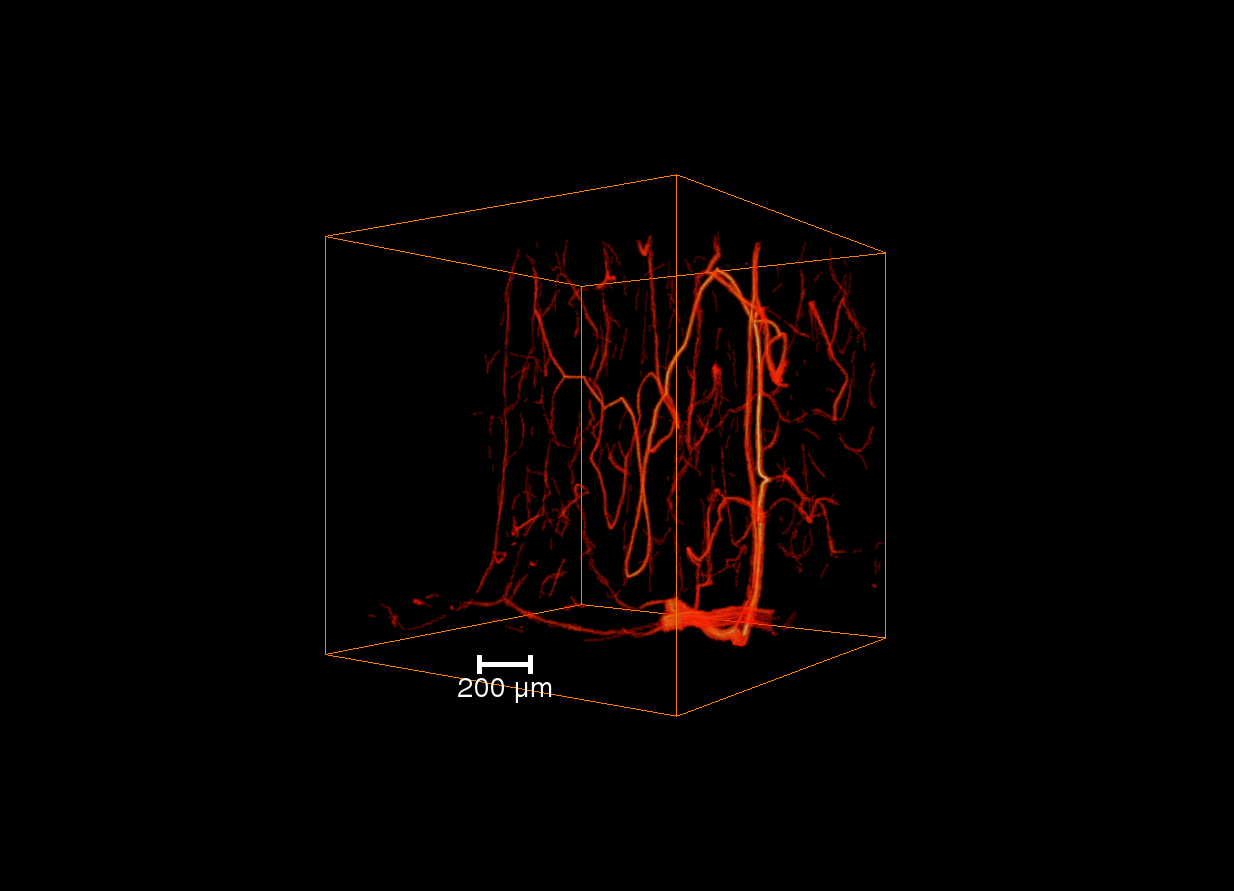} &
    \includegraphics[width=.3\linewidth, trim=85mm 55mm 85mm 55mm, clip]{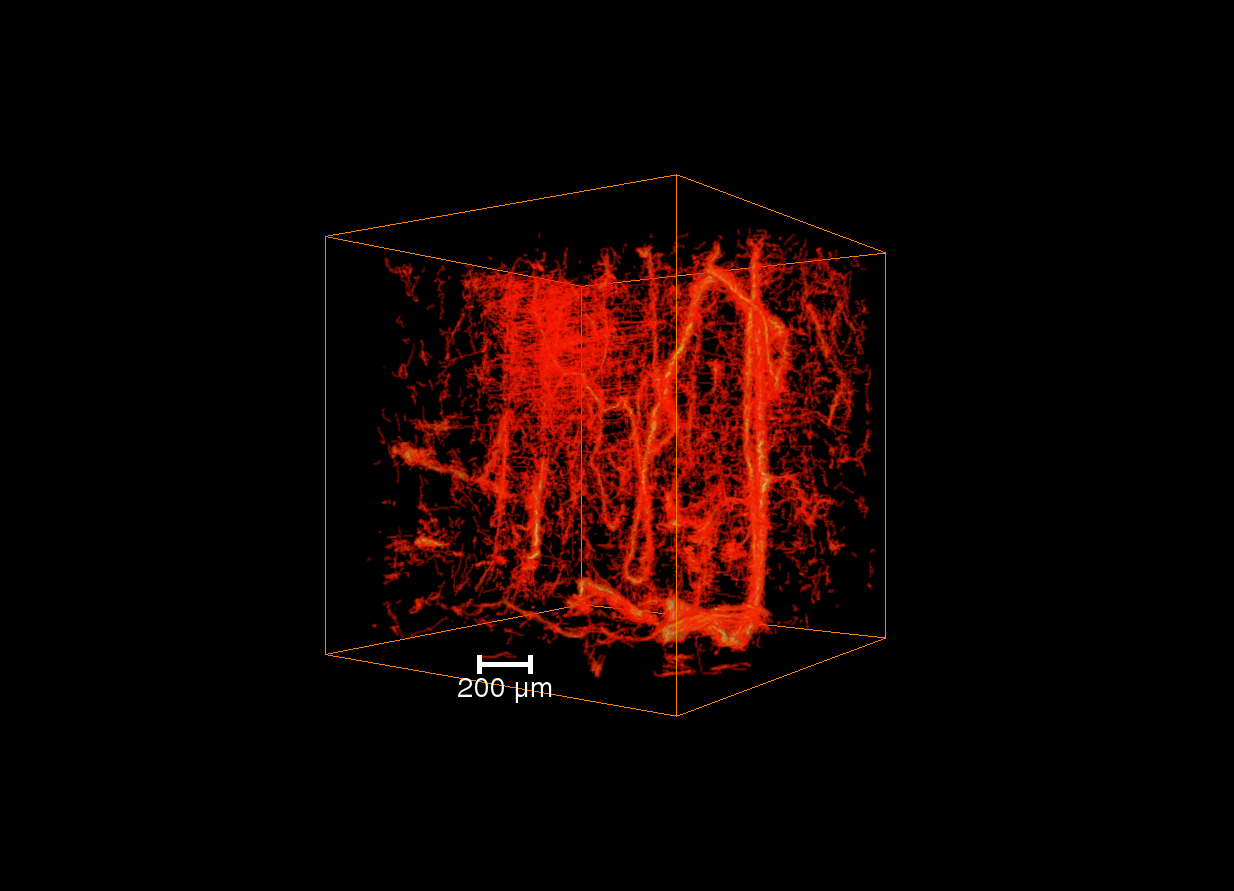} &
    \includegraphics[width=.3\linewidth, trim=85mm 55mm 85mm 55mm, clip]{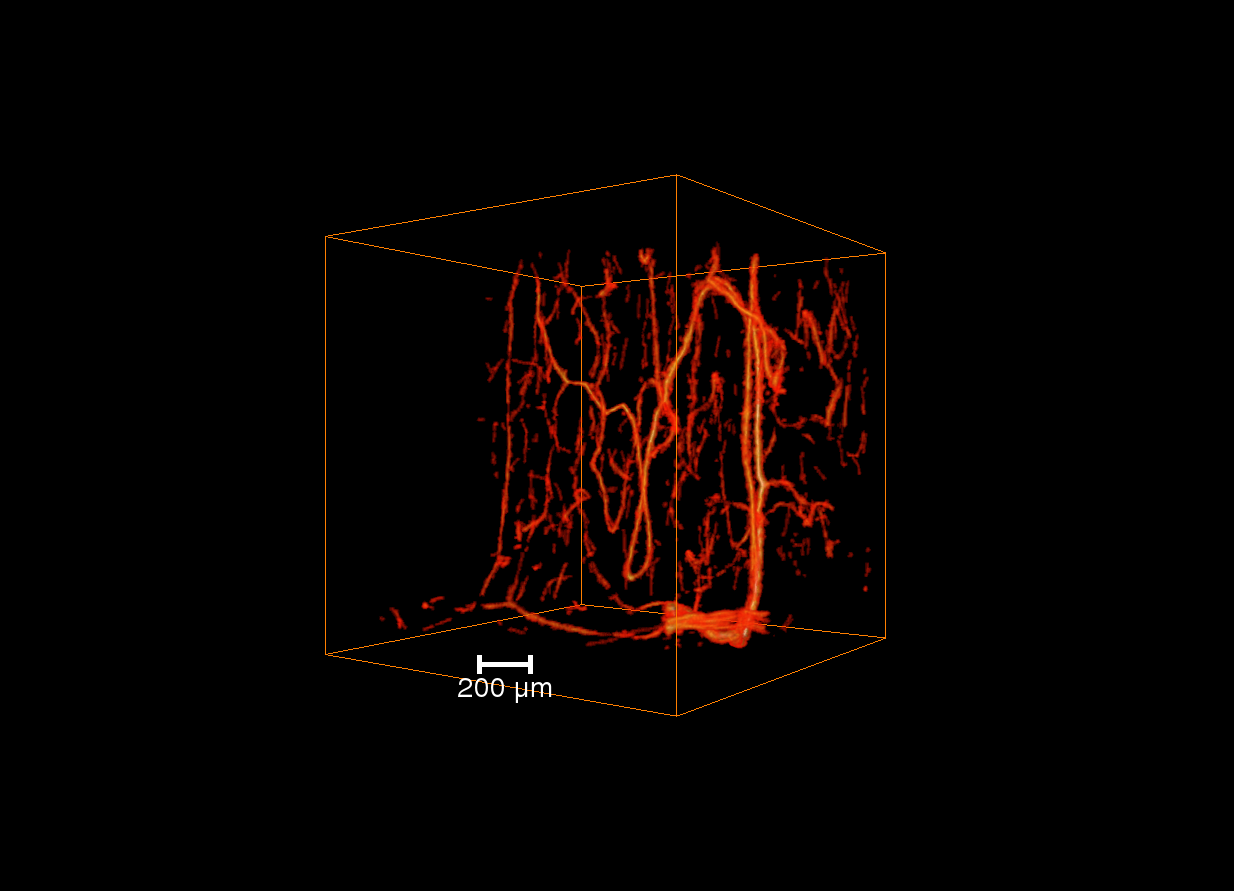} \\
    \includegraphics[width=.3\linewidth, trim=85mm 55mm 85mm 55mm, clip]{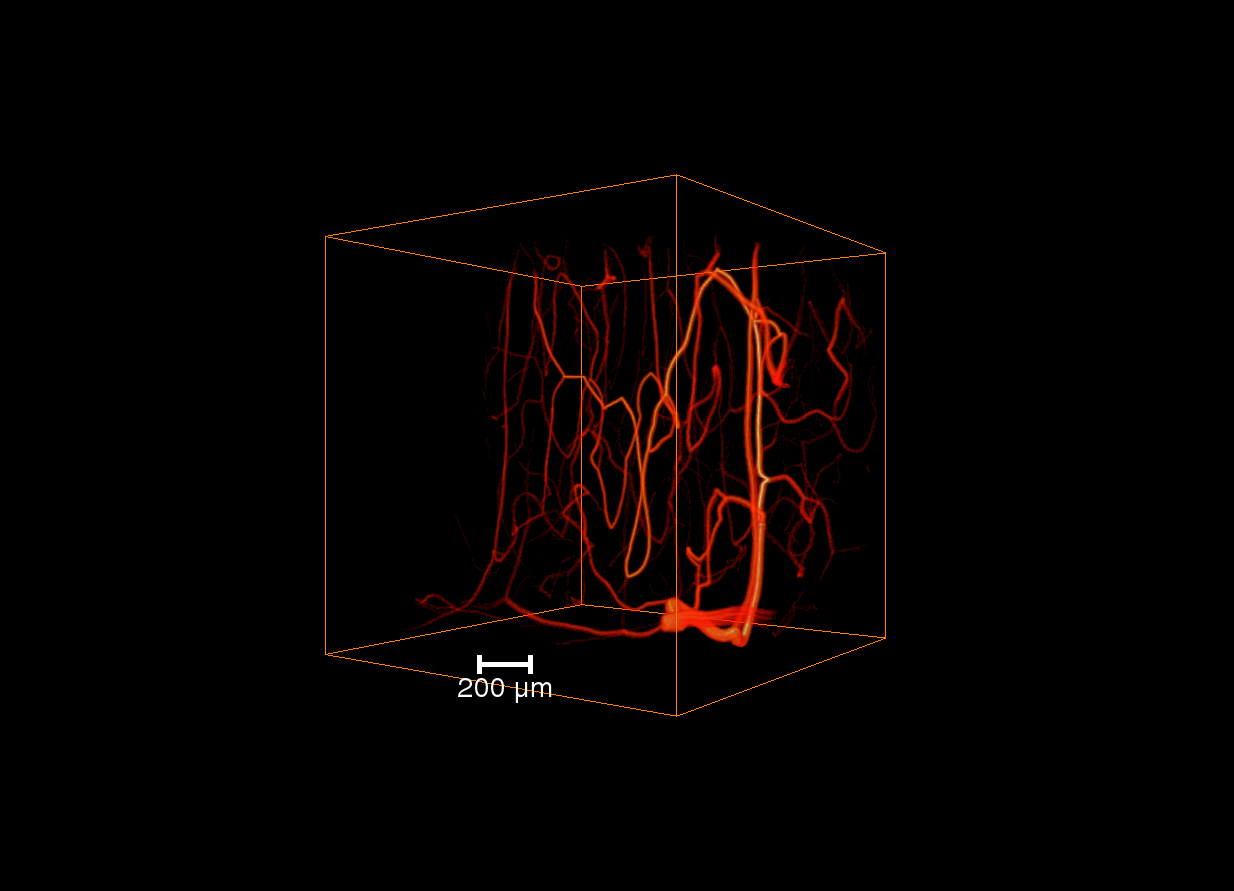} &
    \includegraphics[width=.3\linewidth, trim=85mm 55mm 85mm 55mm, clip]{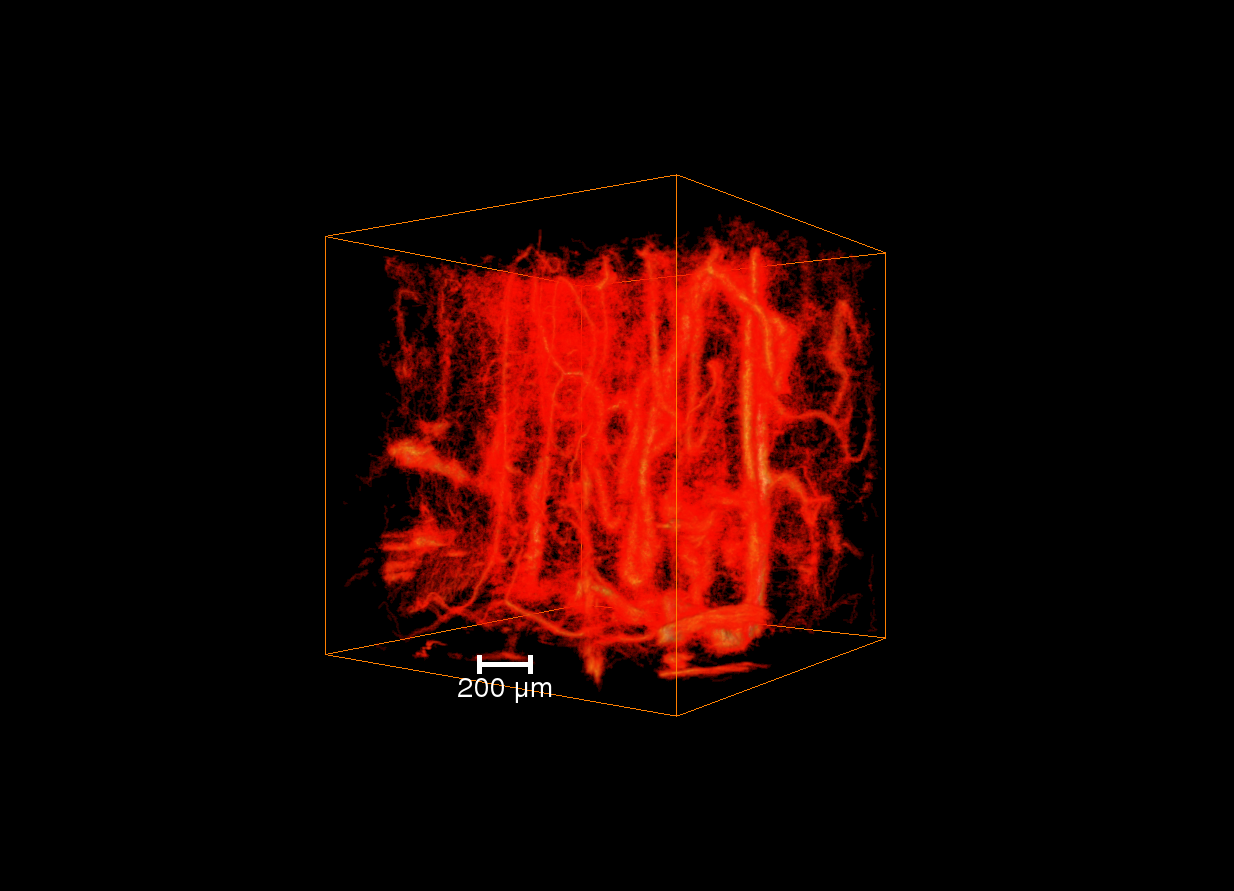} &
    \includegraphics[width=.3\linewidth, trim=85mm 55mm 85mm 55mm, clip]{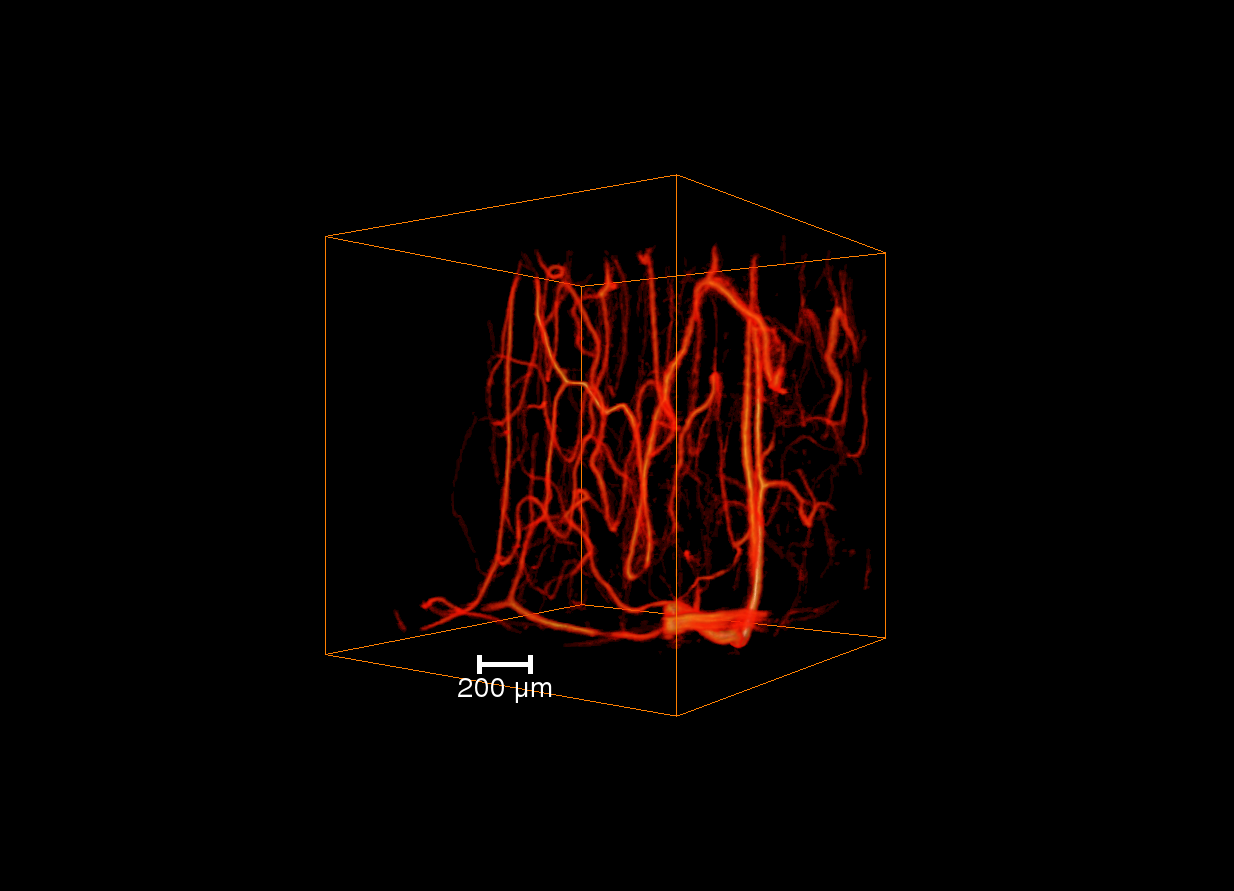} \\
  };
  
    \node [left=4mm of figs-1-1] {\scriptsize 1 MB};
    \node [left=4mm of figs-2-1] {\scriptsize 10 MB};
    \node [left=4mm of figs-3-1] {\scriptsize 30 MB};
    \node [above=1mm of figs-1-1,text width=.2\linewidth, align=center] {\scriptsize Ground Truth};
    \node [above=1mm of figs-1-2,text width=.2\linewidth, align=center] {\scriptsize Conventional ULM};
    \node [above=1mm of figs-1-3,text width=.2\linewidth, align=center] {\scriptsize Sparse\\ Deep-stULM};
  
\draw[arrow] ([xshift=-1mm]figs-3-1.west) -- ([xshift=-1mm]figs-1-1.west);
  \node[rotate=90, left=3mm of figs, anchor=center] {Concentration};
\end{tikzpicture}
\caption{3D Comparison of performance under increasing concentration for conventional ULM (center column) and Deep-stULM sparse formulation (right column). Ground truth is given for comparison (left column). Scale bar is 200 $\mu$m and corresponds to the wavelength of the simulated pulse. Concentration increases from 1 (top row), 10 and 30 (bottom row) microbubbles per field of view.}
\label{fig:3D_angio}

\end{figure*}

In Table \ref{tab:results_time_memory}, we display the results and the memory usage for Sparse Deep-stULM and conventional ULM as well as estimated memory usage in 3D for Deep-stULM, mSPCN-ULM. We also provided inference times for Sparse Deep-stULM and mSPCN-ULM, showing that Sparse Deep-stULM was more than two orders of magnitude faster than mSPCN-ULM. The reported Dice are computed in 3D, which means that the angiograms are sparser than in 2D and leads to  values for the Dice in 3D that are typically lower. To provide comparative insights on the 3D Dice values, we projected to 2D the ground truth and reconstructed 3D angiograms and measured their Dice in 2D. The 2D projected Dice values were typically higher and comparable to the 2D results. Indeed, conventional ULM reached a projected Dice of $67.56\%$ for 1MB/FOV (close to the $71.4\%$ obtained in 2D with the same concentration) and $54.58$ for 30MB/FOV (close to the $55.9$ for 20MB/FOV in 2D). Projected Dice values for Sparse Deep-stULM were $77.71 \%$ for 1MB/FOV and $77.81\%$ for 30MB/FOV.
In addition, since the simulation parameters between the 2D and 3D datasets are different to match realistic imaging sequences, the 3D PSF is larger and has important side lobes, which makes the localization process more challenging.
It is important to note that just using the sparse formulation allowed us to train the network with less than $11$GB of GPU memory while outperforming conventional ULM ($50.0\%$ versus $12.3\%$). 
For qualitative analysis, the reconstructed angiograms from the test set are displayed in Figure \ref{fig:3D_angio} with concentration increasing from 1 MB/FOV to 30 MB/FOV. At high concentration (10 MB/FOV and 30 MB/FOV), the sparse model accurately reconstructed the angiogram when conventional ULM failed to do so. Indeed, the conventional ULM produced many false detections that were not present in the sparse model reconstruction. At low concentration (1 MB/FOV), both conventional ULM and sparse model reconstructed the angiogram with fidelity.

\subsection{Additional studies}
    \begin{figure}
    \centering
    \includegraphics[height=5cm]{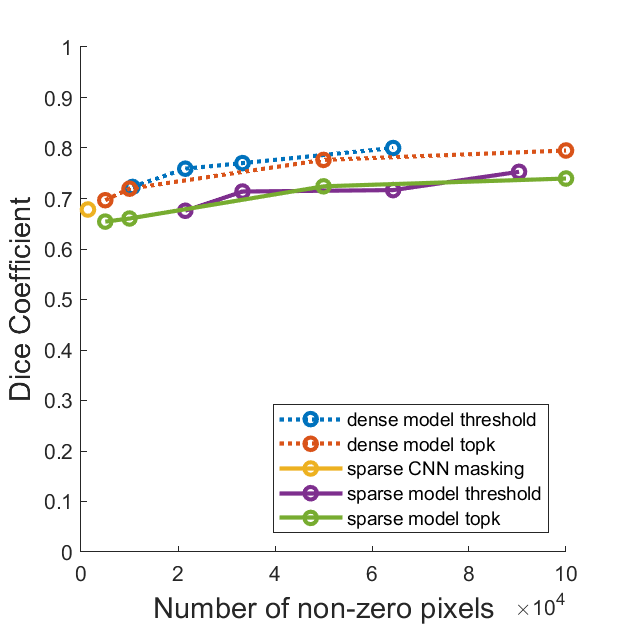}
    \caption{Evolution of performance as a function of the sparsity achieved for different \textit{dense-to-sparse} operations and comparison with dense network with masked input as well as CNN-based dense to sparse operations.}
    \label{fig:performance_curve_dense_to_sparse}
    \end{figure}

\begin{figure}[h!]
\begin{tikzpicture}[
    arrow/.style={arrows={Triangle[length=3mm, width=2mm]-}}
]
  \matrix (figs) [matrix of nodes, inner sep=0, column sep=.005\linewidth, row sep=.005\linewidth, nodes={inner sep=0pt}] {
    \includegraphics[width=.245\linewidth]{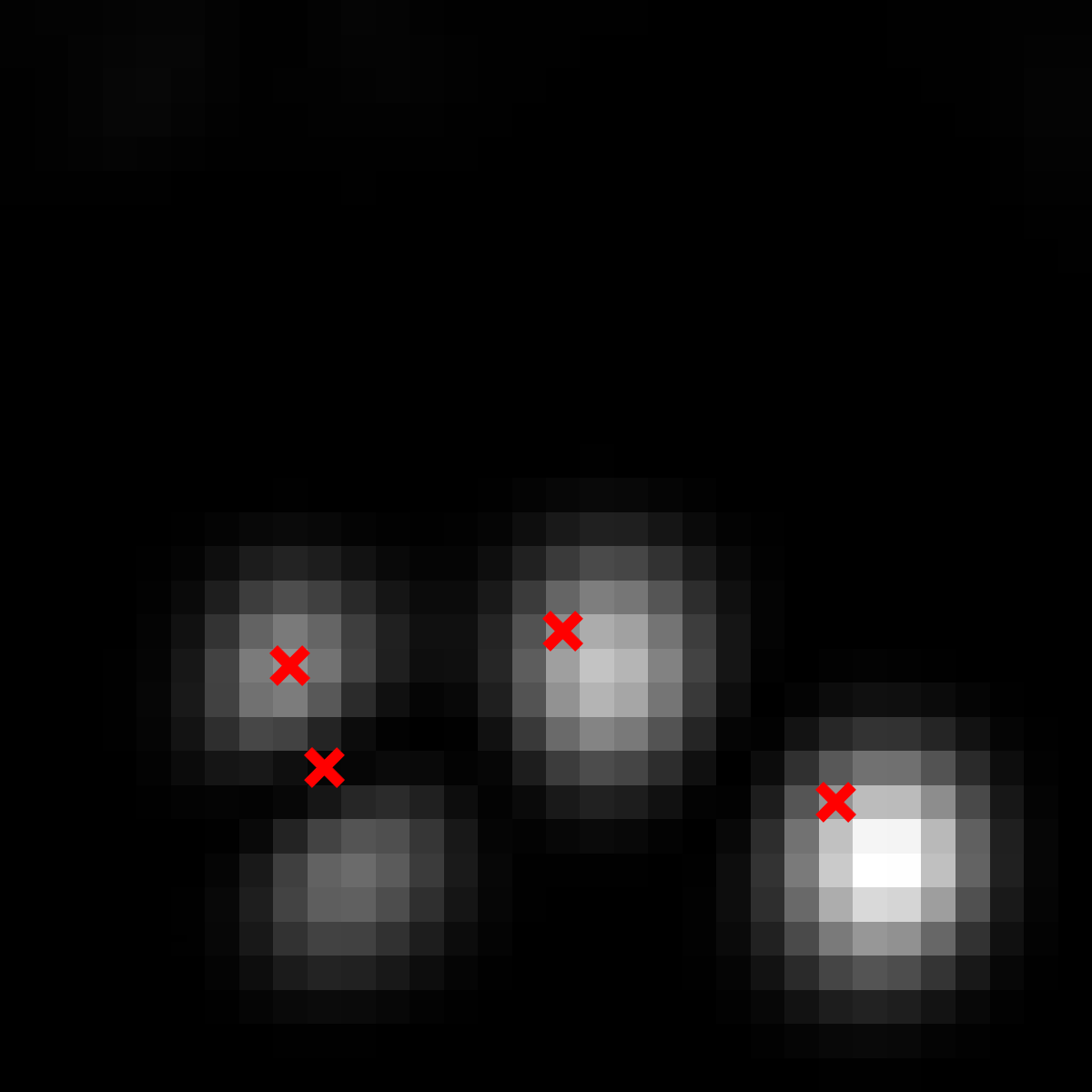} &
    \includegraphics[width=.245\linewidth]{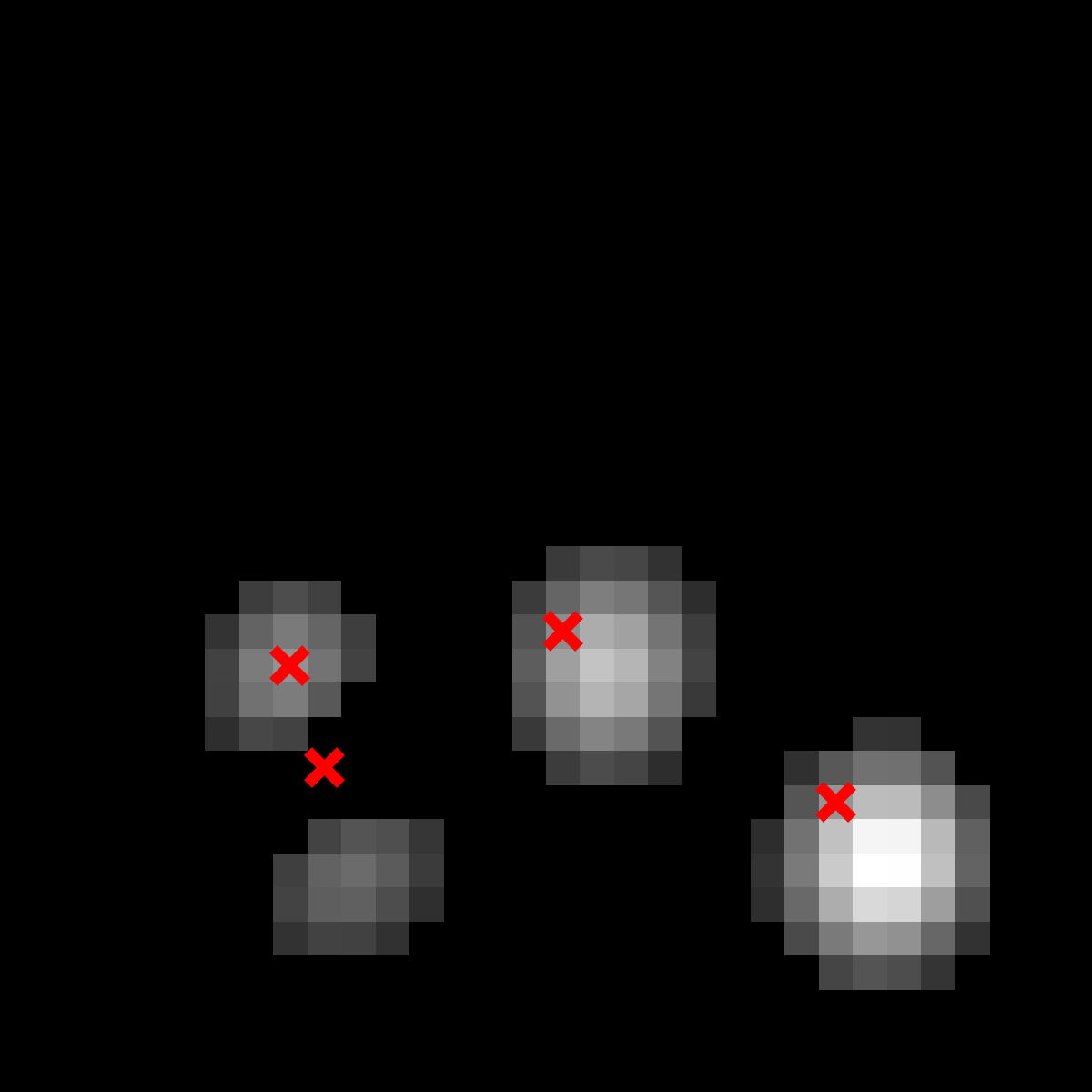} &
    \includegraphics[width=.245\linewidth]{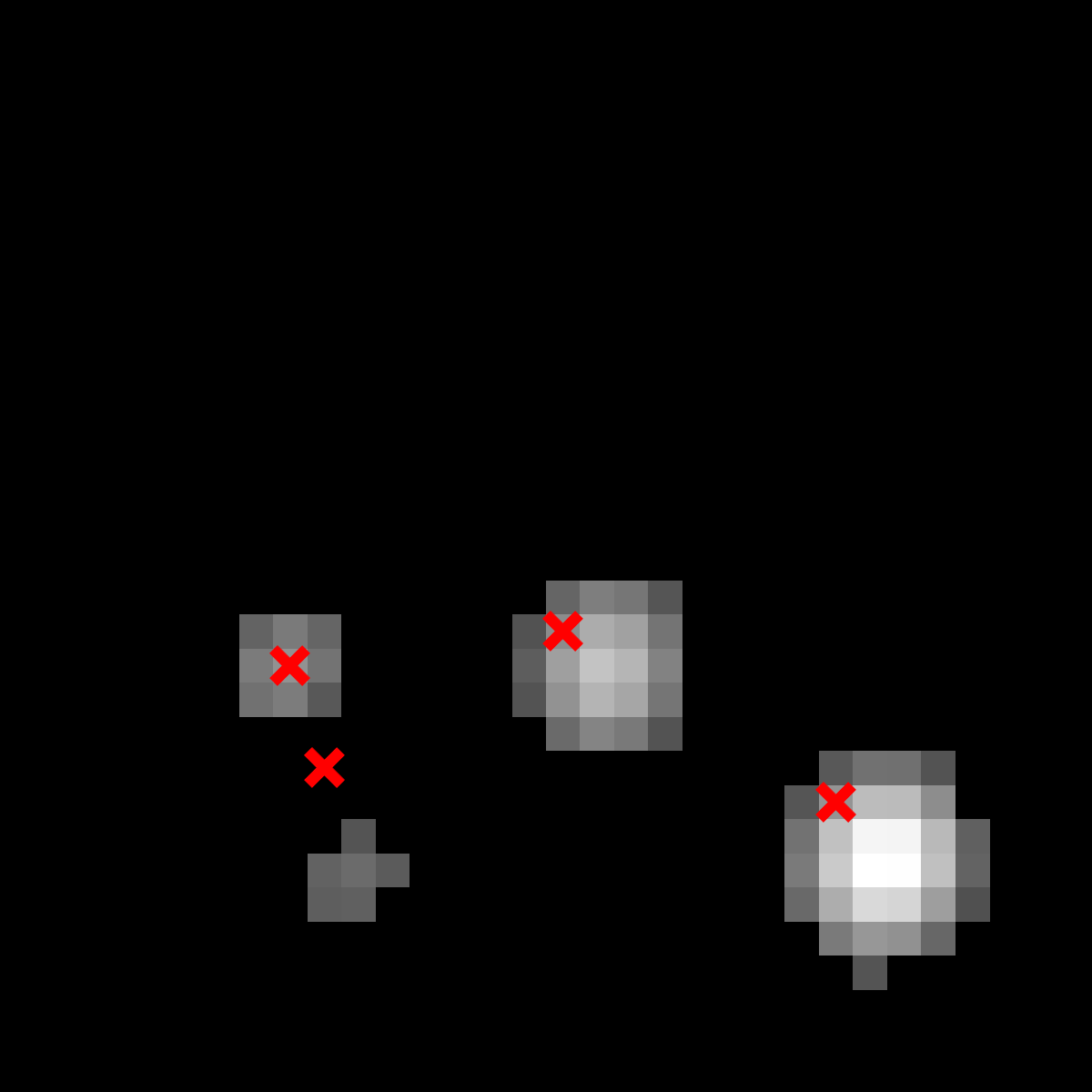} &
    \includegraphics[width=.245\linewidth]{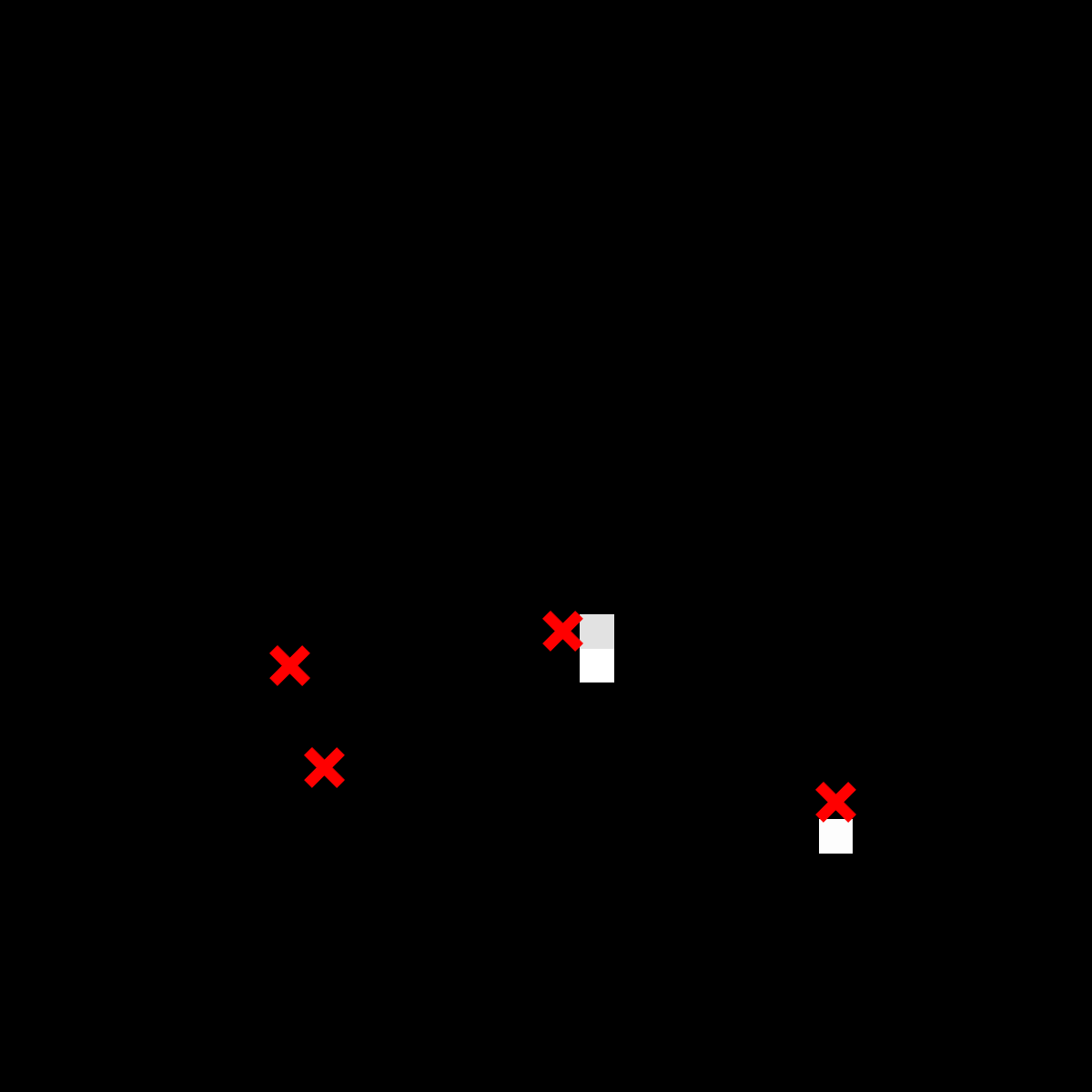} \\
    \includegraphics[width=.245\linewidth]{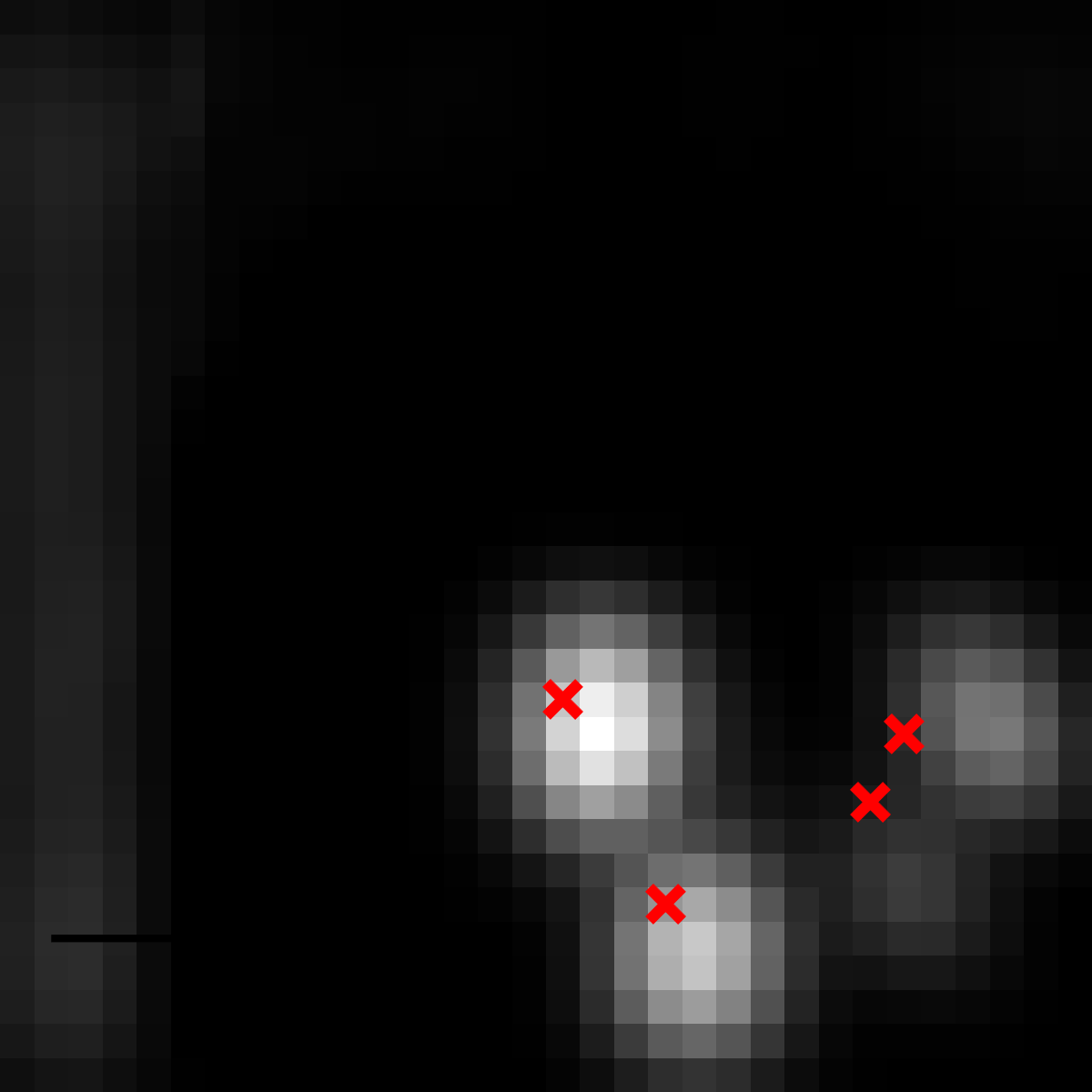} &
    \includegraphics[width=.245\linewidth]{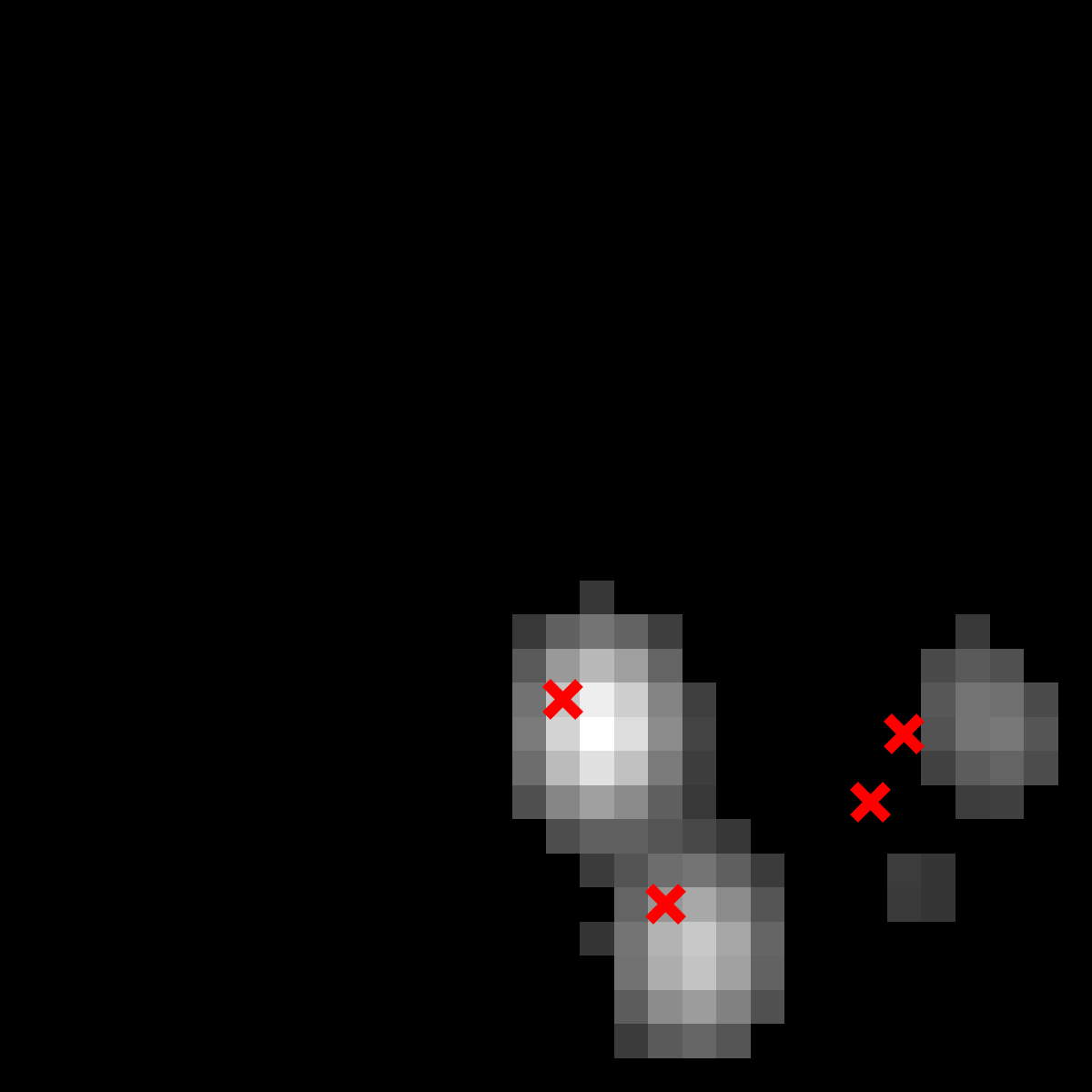} &
    \includegraphics[width=.245\linewidth]{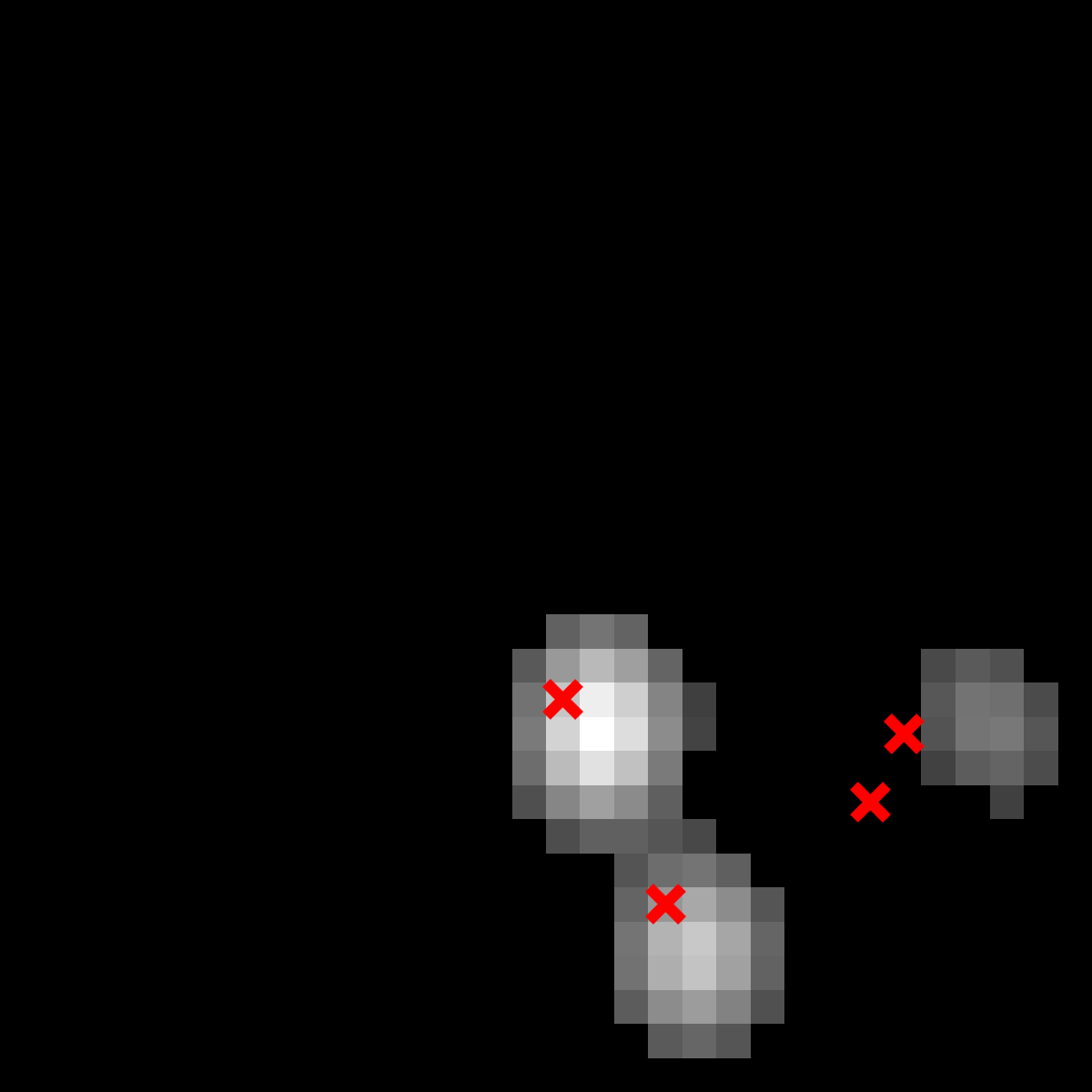} &
    \includegraphics[width=.245\linewidth]{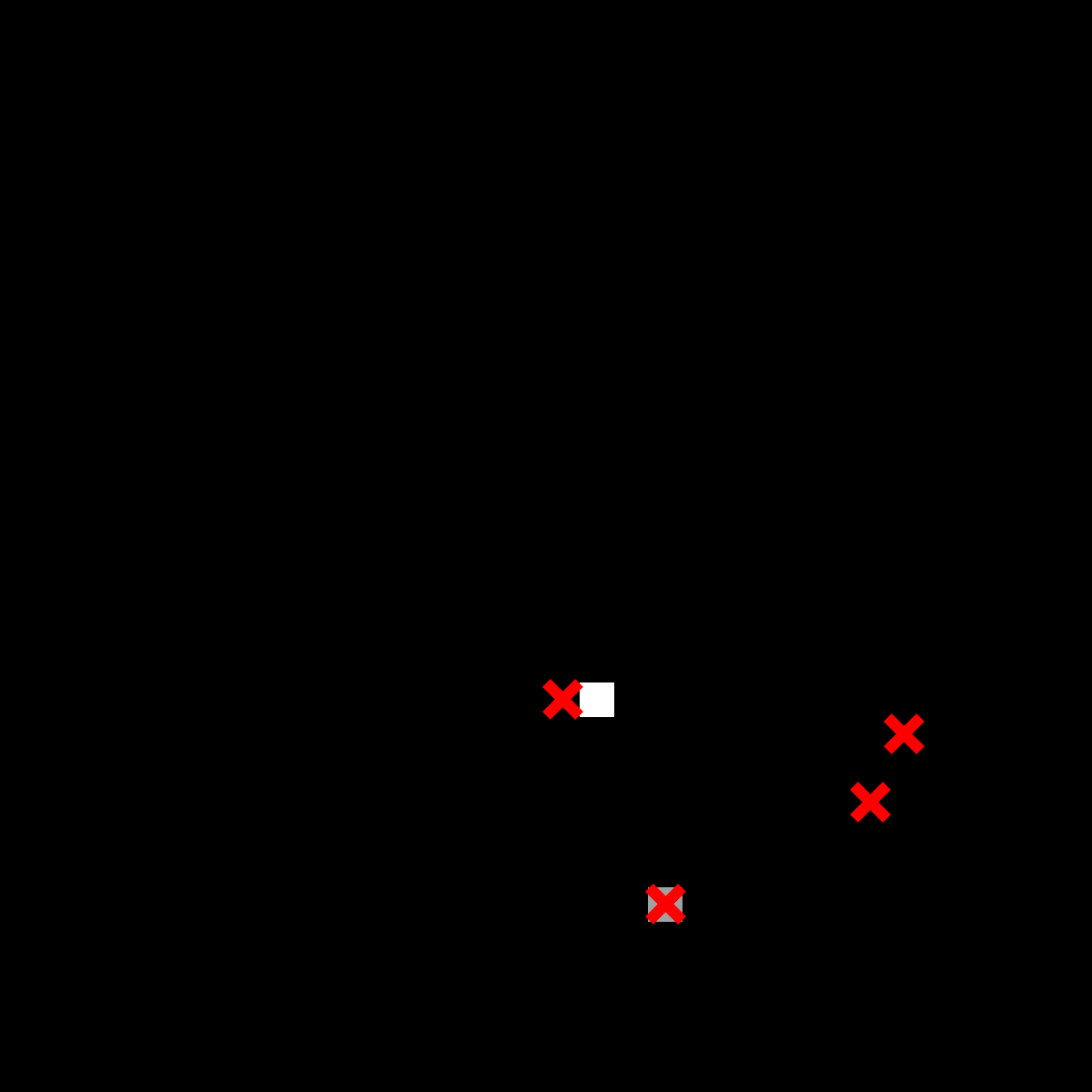} \\
    \includegraphics[width=.245\linewidth]{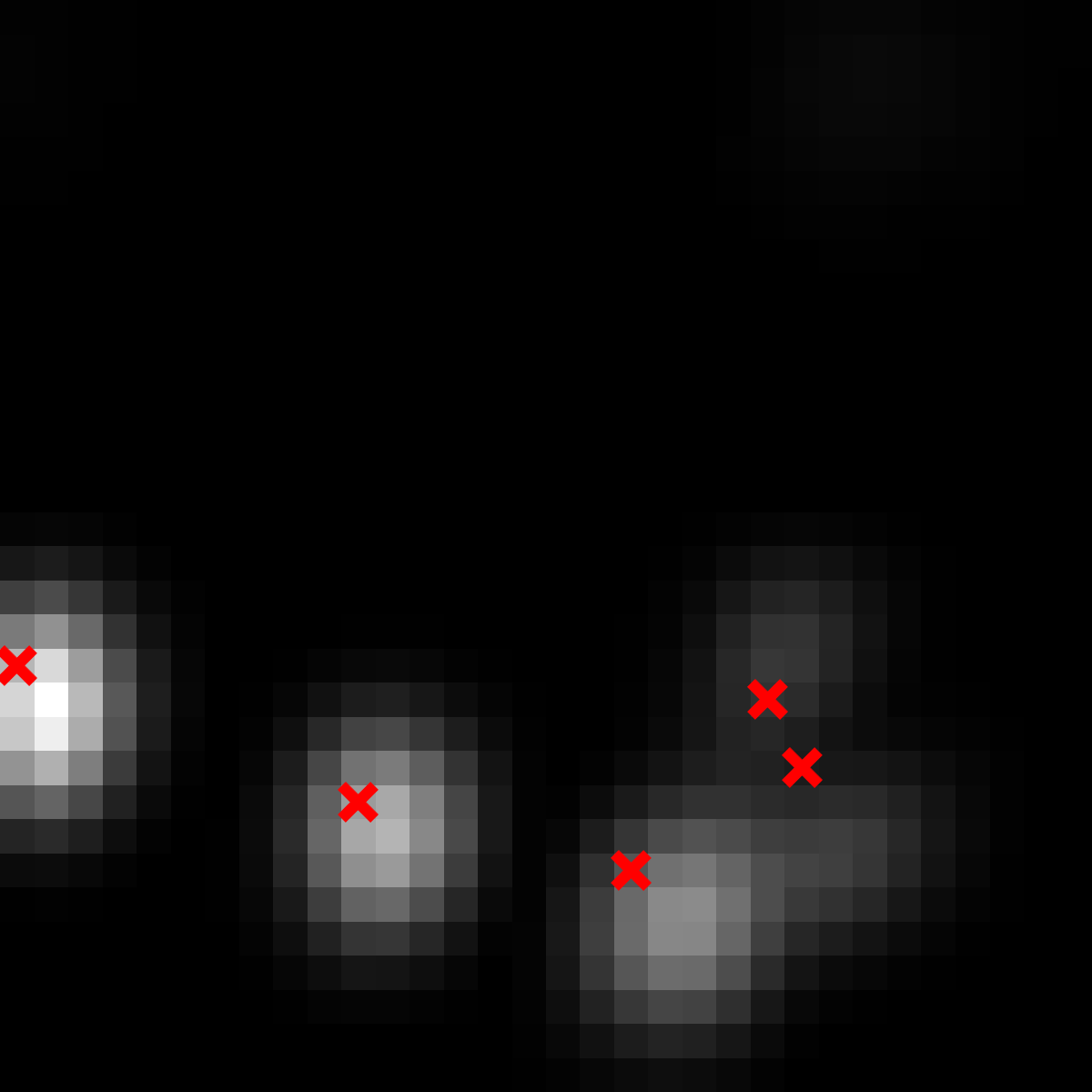} &
    \includegraphics[width=.245\linewidth]{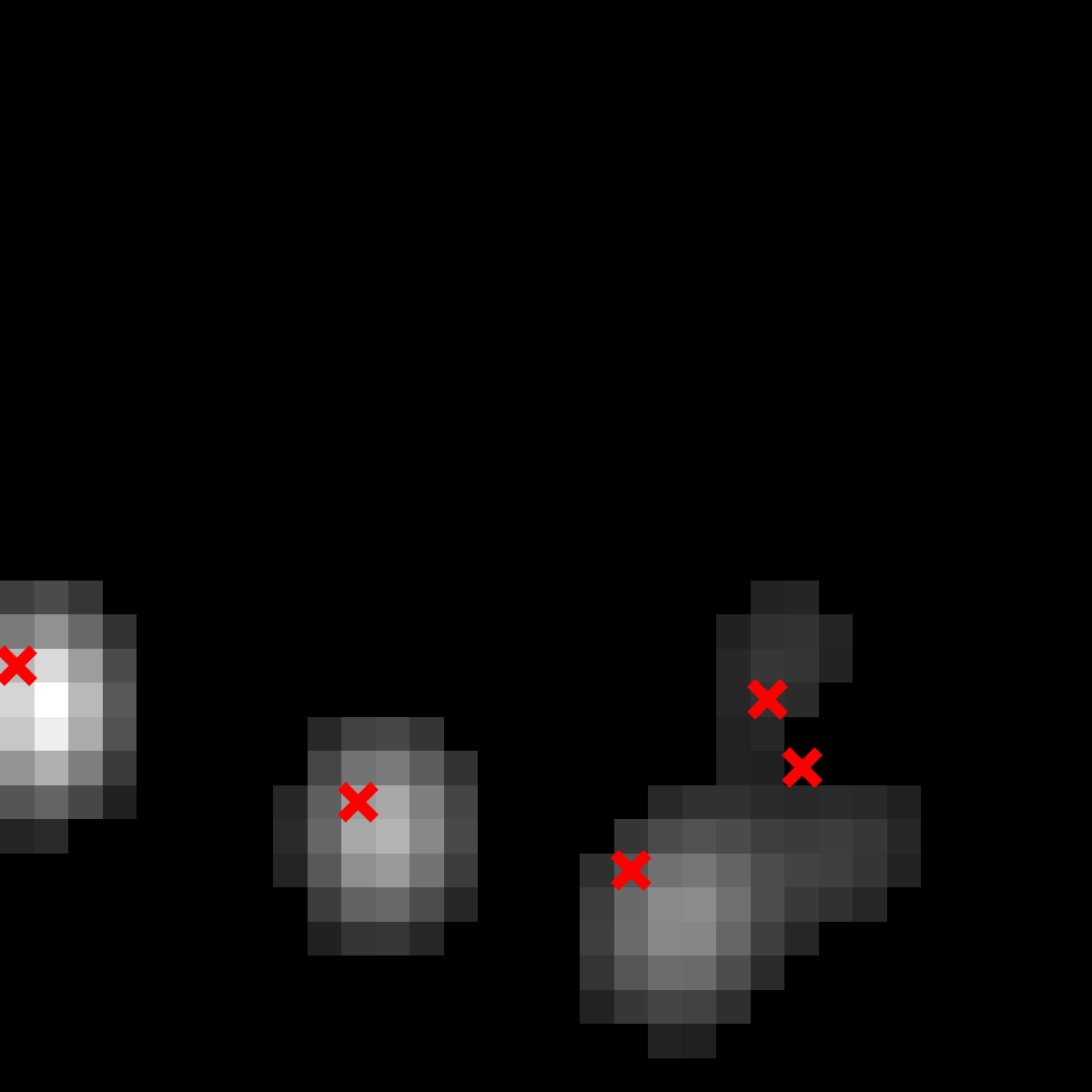} &
    \includegraphics[width=.245\linewidth]{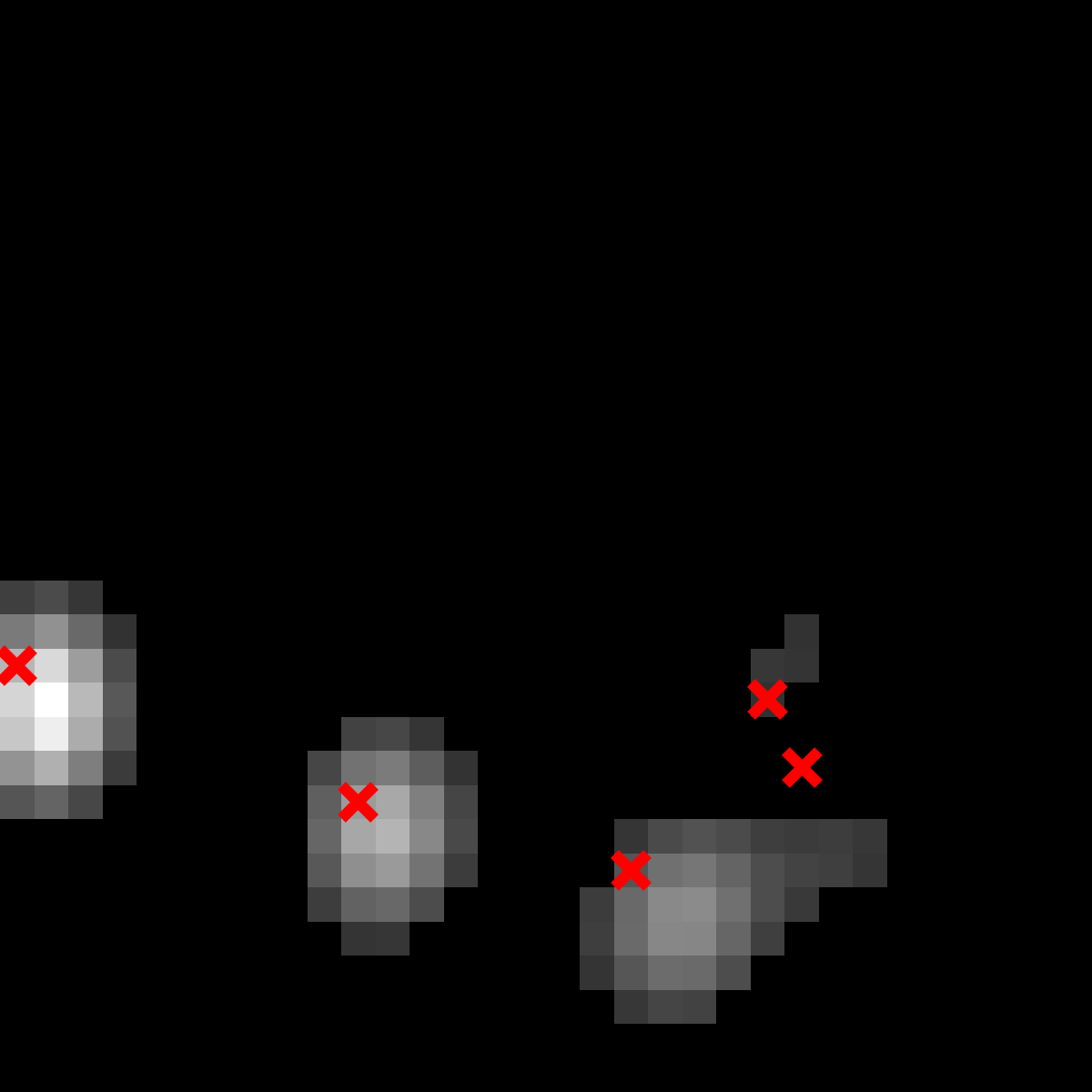} &
    \includegraphics[width=.245\linewidth]{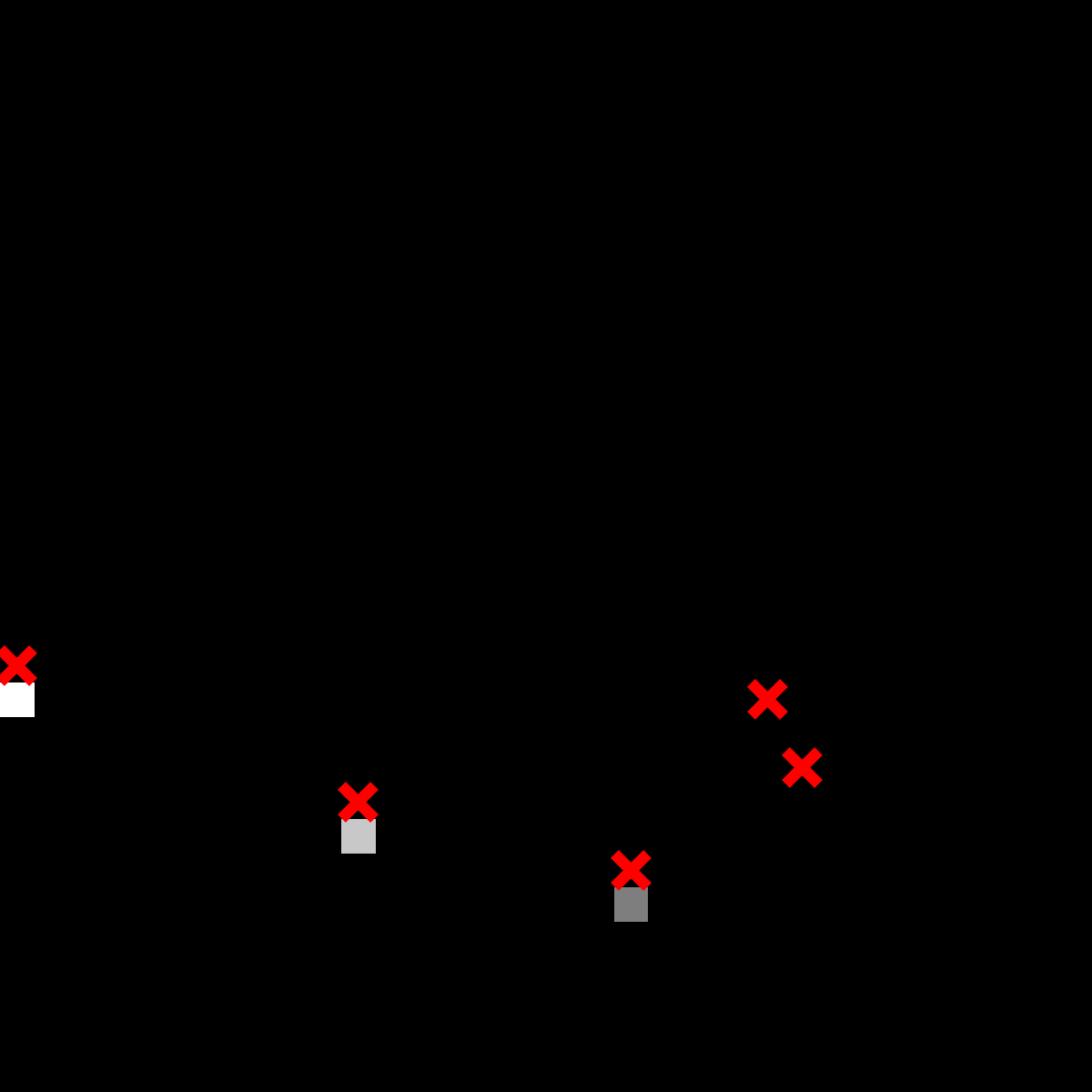} \\
  };
  
    \node [above=1mm of figs-1-1,text width=.24\linewidth, align=center] {\scriptsize Dense };
    \node [above=1mm of figs-1-2,text width=.24\linewidth, align=center] {\scriptsize Sparse Threshold};
    \node [above=1mm of figs-1-3,text width=.24\linewidth, align=center] {\scriptsize Sparse top-k};
    \node [above=1mm of figs-1-4,text width=.24\linewidth, align=center] {\scriptsize Sparse CNN};
  
\end{tikzpicture}
\caption{Failure cases of the \textit{dense-to-sparse} thresholding strategy with a 0.1 threshold (center left column) along with the dense correlation map (left column), the results of the top-k with $50000$ values (center right column), the results of the CNN based strategy  (right column)}
\label{fig:sparse_failure}

\end{figure}

\begin{table*}
        \caption{Comparison of the memory usage and performance of the different additions to Sparse DeepST-ULM architecture.
    }

    \centering
    \begin{tabular}{ccccccccc}
                  & \multirow{2}{*}{Sparse input} & \multirow{2}{*}{\makecell{Intermediate \\loss}} & \multirow{2}{*}{Pruning}&  
        \multirow{2}{*}{\makecell{Cascaded \\ learning}}   & \multicolumn{2}{c}{2D ($5$MB/FOV)}  & \multicolumn{2}{c}{3D ($30$MB/FOV)}  \\
        \cmidrule{6-7} \cmidrule{8-9}
        & & & & & \makecell{ Memory \\ requirements
        \\ (GB)}  &  Dice (\%) & \makecell{ Memory \\ requirements
        \\ (GB)}  & Dice (\%)  \\
        \hline
        \makecell{Sparse  Deep-stULM\\ + Pruning \\ + cascaded learning} & \checkmark & \checkmark & \checkmark & \checkmark & \boldmath{$5.6 \pm 0.2$} & $59.22 \pm 0.49$ & $3.7 \pm 0.4$ & $38.53 \pm 0.27$\\
        \hline
        \makecell{Sparse Deep-stULM\\ + Pruning}& \checkmark & \checkmark & \checkmark &  & $5.7 \pm 0.1$ &$62.36 \pm 1.57$  & \boldmath{$3.6 \pm 0.2$}  &  $38.65 \pm 0.58$\\
        \hline
        \makecell{Sparse  Deep-stULM\\ + int. loss} & \checkmark & \checkmark &  &  & $6.9 \pm 0.1$  &  $71.98 \pm 0.16$ & $10.6 \pm 0.3$ & $ 43.51 \pm 0.99$\\
        \hline
        Sparse Deep-stULM & \checkmark &  &  &  & $6.8 \pm 0.2$ & \boldmath{$73.93 \pm 1.96$} & $9.9 \pm 0.1$ & \boldmath{$49.97 \pm 1.79$}
    \end{tabular}

    \label{tab:ablatation_2D}
\end{table*}
\subsubsection{Evaluation of strategies to convert to sparse formulation}
In figure \ref{fig:performance_curve_dense_to_sparse}, it is observed that the thresholding and top-k \textit{dense-to-sparse} approaches reached a very similar trade-off between the sparsity obtained and the level of performance independently of the formulation of the model (dense or sparse). Indeed, for the sparse model with the thresholding \textit{dense-to-sparse} strategy the Dice varies from $67.5\%$ with around $21000$ pixels to $75.3\%$ with $90000$ pixels while it varies from $65.4\%$ with around $5000$ pixels to $73.9\%$ with $100000$ pixels for the top-k strategy. Similarly, for the dense model with the thresholding \textit{dense-to-sparse} strategy the Dice varies from $72.3\%$ with around $11000$ pixels to $79.9\%$ with $64000$ pixels while it varies from $69.6\%$ with around $5000$ pixels to $79.5\%$ with $100000$ pixels for the top-k strategy.  
Finally, the CNN \textit{dense-to-sparse} operation yielded a better trade-off than all the other approaches. Indeed, the CNN \textit{dense-to-sparse} operation reached a Dice of $67.8\%$ with only $1400$ pixels.

\textit{Dense-to-sparse} strategy failure cases are displayed in Figure \ref{fig:sparse_failure}. These failure cases were selected when the threshold strategy, with the threshold set to 0.1, removed at least one pixel containing a microbubble and accounted for between 5\% and 10\% of the total test set frames. For the threshold and top-k strategies, some low-intensity microbubbles in the vicinity of other microbubbles were filtered out, especially when the peak of the correlation map did not match the microbubble's position. These microbubbles were not detected by the CNN masking strategies, which also made localization errors, even for clearer microbubbles. 

\subsubsection{Impact of architecture modifications} 
In table \ref{tab:ablatation_2D}, we observed that using pruning jointly with sparse formulation led to a decrease in memory requirements in 2D ($6.9$ GB to $5.6$ GB) but also led to an important degradation of the performance ($9.6\%$ of Dice). It appeared that combining pruning and cascaded learning has a small impact on the memory ($5.6$ GB versus $5.7$ GB) while degrading the performance (more than $3\%$ of Dice). 
The addition of intermediate loss for deep supervision did not benefit the training performance and had a small memory cost.
In 3D, every variation of the model trained significantly outperformed the conventional ULM (between $38\%$ and $50\%$ for the sparse models and $12.3\%$ for the conventional ULM). In contrast to 2D, the use of pruning  reduced memory usage by a factor of approximately 4. However, similarly to the 2D case, pruning also degraded the performance ($50\%$ to $38.6.\%$).

\section{Discussion} 
In this work, we studied the impact of using a sparse formulation on the memory usage and performance of Deep-stULM. We also investigated more multiple methods to further increase  sparsity and reduce the memory usage.
Our results suggest that solely using the sparse formulation allows for the extension of existing deep learning architectures to 3D ULM, while preserving the performance robustness at high concentration. However, more complex modifications to the architecture had a less important impact on the memory usage while degrading the performance. 

\subsection{Reducing memory  usage of existing methods in 2D}
The sparse formulation offers a relatively simple approach to divide by nearly two-fold the memory requirement of an existing architecture in 2D, while outperforming conventional ULM. 
Indeed, the sparse formulation reconstructed quantitatively more precise angiograms for high concentration (5MB/FOV and 10 MB/FOV). Dense deep learning approaches consistently outperformed the sparse formulation. Leveraging larger and more complex datasets could help mitigate the performance degradation, while still benefiting from the computation time and memory improvements. Combining the shorter acquisition time with reduced constraints on computation, the proposed sparse formulation could be relevant in near real-time processing of ULM acquisition. 
The robustness of the sparse formulation to additive noise suggests that the information loss introduced by the \textit{dense-to-sparse} operation and the resulting drop in performance are primarily due to features that are sensitive to noise.

\subsection{Scaling to 3D imaging}
In 3D, the sparse formulation allowed for training, with less than 11 GB of memory, an architecture that would require close to $700$GB of memory in dense formulation.  
As expected, the sparse formulation impact on memory was greater in 3D than 2D. Experimentally, the ratio between the sparse and the dense formulation was multiplied by $\delta_{exp} = 2.6 \times 10^{-2}$ when extending from 2D to 3D whereas the theoretical value from Eq. \ref{eq:scaling_factor} was  $\delta = 3.1 \times 10^{-2}$. The smaller ratio could be explained by the aforementioned experimental factors such as temporal context considerations, as well as the higher threshold for the \textit{dense-to-sparse} strategy in 3D.
The performance gap between the sparse formulation of Deep-stULM and the conventional ULM was larger in 3D, suggesting an even higher potential for deep learning based approach in 3D ULM. 
On the one hand, the drop of performance of conventional ULM might be caused by the important side lobes of the simulated PSF in 3D. 
Indeed, these sidelobes create false detection which are supposed to be filtered during the tracking step based on the track length.
However, in high concentration, candidate detections in successive frame are multiple causing the filtering on track length to be less effective and contribute to 3D conventional ULM map with a high number of false positives. 
On the other hand, it is also possible that the additional dimension increases the capability of the deep learning approach to distinguish crossing trajectories, leading to better performance relatively to conventional ULM.
The results in $3$D shows that akin to the dense approach in 2D, sparse models can accurately reconstruct angiograms at concentration where conventional ULM is failing.
When translated in vivo, such performance in high concentration would allow for reduced acquisition time.
In 3D, where dataset tends to be larger, reducing acquisition time is even more crucial as it also reduces the storage needed and the associated transfer time.

\subsection{Further reductions of memory and performance trade-off}

\subsubsection{\textit{Dense-to-sparse} strategies} The parameter studies showed that top-k and thresholding \textit{dense-to-sparse} strategies performed similarly, and a better trade-off could be achieved with deep-learning based \textit{dense-to-sparse} operations. However, this approach lacked flexibility in the trade-off between sparsity achieved and performance, which translated to very restrictive filtering and poor overall performance. Top-k and threshold \textit{dense-to-sparse} strategies still filtered out some microbubbles signal, presumably altering the localization performance. Local misalignment between the correlation map peak intensity and the localization of the microbubble caused all the proposed \textit{dense-to-sparse} strategies to fail. Such mismatches are expected to occur when microbubbles are closer than the size of the PSF used for correlation. Removing the correlation step and directly using beamformed IQ could improve the performance of \textit{dense-to-sparse} strategies. Additionally, increasing the number of pixels considered in the sparse formulation reduced the performance gap between sparse and dense formulation of Deep-stULM, which also suggested suboptimal \textit{dense-to-sparse} operations.
The sparse formulation might benefit from more sophisticated \textit{dense-to-sparse} strategies to further improve the proposed trade-off between performance and memory requirements in training.

\subsubsection{Architecture modifications} It appears that architecture modifications such as deep-supervision, cascaded learning and pruning had an important impact on the performance with smaller gain in memory than that then sole sparse formulation. Cascaded learning and deep-supervision negative impact on the performance could be explained by the fact that microbubble detection at coarse resolution is different from localization at super-resolution and therefore intermediate losses enforce the learning of less relevant latent representations.  
Considering the additional complexity in training and architecture constraint going with this modifications, it is less clear that they present an interesting trade-off between memory usage and performance. 

\subsection{Limitations and perspectives}
The proposed method demonstrates that sparse tensor neural networks can extend the benefits of deep learning based approach in ULM from 2D to 3D imaging by improving the scaling law of memory costs with dimensionality and resolution. However, some limitations should be mentioned and could be addressed by future studies. 
\subsubsection{Problem formulation}
This study is mostly focused on measuring the impact of using a sparse formulation in deep learning for ULM. For this reason, it does not tackle some key challenges in framing the learning problem.
On the one hand, it is important to mention that even though the Dice loss has been proposed and successfully applied in the previous study \cite{mileckiDeepLearningFramework2021}, it is unstable and lacks smoothness when working with temporal projection of trajectories.
In addition, the final representation of the prediction, being a projection of the microbubble trajectories on a grid, does not offer the same liberty for downstream analysis as the individual detections provided by conventional ULM. A formulation tackling these issues has been recently proposed \cite{shinContextAwareDeepLearning2023} based on the DECODE method \cite{speiserDeepLearningEnables2021} and could be worth investigating.
On the other hand, the encoding of the real and imaginary parts of the input signal as channels lacks the proper arithmetic of complex numbers. Using a complex value neural network \cite{trabelsiDeepComplexNetworks2018} could allow a better representation of the signal with less overfitting and better overall performance. Such networks have shown interesting potential when dealing with ultrasound data \cite{luComplexConvolutionalNeural2022, xingPhaseAberrationCorrection2024}.
\subsubsection{Validity of the \textit{in silico} model}
The proposed approach reaches the level of performance of the state-of-the-art dense model \textit{in silico} under varying concentrations of microbubbles. It would be interesting to evaluate the validity of the conclusions on real data as it has not been tested \textit{in vivo}.  
As current training datasets for ULM are limited in their diversity and realism, they did not allow for direct domain transfer to \textit{in vivo} of both dense and proposed methods. Consequently, \textit{in vivo} applications required carefully tuned pre-processing steps, which were not consistently successful. Due to the preprocessing inconsistency between training and testing, it is not clear that better learning ability on the simulated dataset would lead to improved performance \textit{in vivo}. Furthermore, as the main benefit of the proposed method is an improvement of the scaling law of the memory complexity, it is reasonable to assume that it is still valid \textit{in vivo}.

\section{Conclusion}
In this study, we studied the potential of sparse formulation when applying deep learning in 3D ULM. We also proposed further optimization of memory efficiency through pruning of the 3D ULM deep learning model. The proposed Sparse Deep-stULM method successfully improve the scaling of memory requirements of deep learning based approaches, addressing the challenge of their extension in 3D. While it comes at a small cost of performance in 2D, the use of deep-learning in 3D ULM seems to be even more beneficial than in 2D. To the best of our knowledge, it is the first application of deep learning for 3D ULM, and it could pave the way for further approaches both \textit{in silico} and \textit{in vivo}. Further applications of such models could allow for improved architectures capable of fitting to more diverse datasets, yielding better results both in 2D and 3D ULM or DynULM.

\printbibliography

@article{bar-zionSUSHISparsityBasedUltrasound2018,
  title = {{{SUSHI}}: {{Sparsity-Based Ultrasound Super-Resolution Hemodynamic Imaging}}},
  shorttitle = {{{SUSHI}}},
  author = {{Bar-Zion}, A. and Solomon, O. and {Tremblay-Darveau}, C. and Adam, D. and Eldar, Y. C.},
  year = {2018},
  month = dec,
  journal = {IEEE Transactions on Ultrasonics, Ferroelectrics, and Frequency Control},
  volume = {65},
  number = {12},
  pages = {2365--2380},
  issn = {1525-8955},
  doi = {10.1109/TUFFC.2018.2873380}
}

@article{belgharbiAnatomicallyRealisticSimulation2023,
  title = {An {{Anatomically Realistic Simulation Framework}} for {{3D Ultrasound Localization Microscopy}}},
  author = {Belgharbi, Hatim and Por{\'e}e, Jonathan and Damseh, Rafat and Perrot, Vincent and Milecki, L{\'e}o and {Delafontaine-Martel}, Patrick and Lesage, Fr{\'e}d{\'e}ric and Provost, Jean},
  year = {2023},
  journal = {IEEE Open Journal of Ultrasonics, Ferroelectrics, and Frequency Control},
  volume = {3},
  pages = {1--13},
  issn = {2694-0884},
  doi = {10.1109/OJUFFC.2023.3235766}
}

@inproceedings{paszkePyTorchImperativeStyle2019,
  title = {{{PyTorch}}: {{An Imperative Style}}, {{High-Performance Deep Learning Library}}},
  shorttitle = {{{PyTorch}}},
  booktitle = {Advances in {{Neural Information Processing Systems}}},
  author = {Paszke, Adam and Gross, Sam and Massa, Francisco and Lerer, Adam and Bradbury, James and Chanan, Gregory and Killeen, Trevor and Lin, Zeming and Gimelshein, Natalia and Antiga, Luca and Desmaison, Alban and Kopf, Andreas and Yang, Edward and DeVito, Zachary and Raison, Martin and Tejani, Alykhan and Chilamkurthy, Sasank and Steiner, Benoit and Fang, Lu and Bai, Junjie and Chintala, Soumith},
  year = {2019},
  volume = {32},
  publisher = {Curran Associates, Inc.},
  urldate = {2023-11-07}
}

@article{bodardUltrasoundLocalizationMicroscopy2023,
  title = {Ultrasound Localization Microscopy of the Human Kidney Allograft on a Clinical Ultrasound Scanner},
  author = {Bodard, Sylvain and Denis, Louise and Hingot, Vincent and Chavignon, Arthur and H{\'e}l{\'e}non, Olivier and Anglicheau, Dany and Couture, Olivier and Correas, Jean-Michel},
  year = {2023},
  month = may,
  journal = {Kidney International},
  volume = {103},
  number = {5},
  pages = {930--935},
  publisher = {Elsevier},
  issn = {0085-2538},
  doi = {10.1016/j.kint.2023.01.027},
  urldate = {2024-05-28},
  langid = {english},
  pmid = {36841476}
}

@article{bourquinQuantitativePulsatilityMeasurements2024,
  title = {Quantitative Pulsatility Measurements Using {{3D}} Dynamic Ultrasound Localization Microscopy},
  author = {Bourquin, Chlo{\'e} and Por{\'e}e, Jonathan and Rauby, Brice and Perrot, Vincent and Ghigo, Nin and Belgharbi, Hatim and B{\'e}langer, Samuel and {Ramos-Palacios}, Gerardo and Cortes, Nelson and Ladret, Hugo and Ikan, Lamyae and Casanova, Christian and Lesage, Fr{\'e}d{\'e}ric and Provost, Jean},
  year = {2024},
  month = feb,
  journal = {Physics in Medicine \& Biology},
  volume = {69},
  number = {4},
  pages = {045017},
  issn = {0031-9155, 1361-6560},
  doi = {10.1088/1361-6560/ad1b68},
  urldate = {2024-05-21},
  langid = {english}
}

@article{bourquinVivoPulsatilityMeasurement2022,
  title = {In {{Vivo Pulsatility Measurement}} of {{Cerebral Microcirculation}} in {{Rodents Using Dynamic Ultrasound Localization Microscopy}}},
  author = {Bourquin, Chlo{\'e} and Por{\'e}e, Jonathan and Lesage, Fr{\'e}d{\'e}ric and Provost, Jean},
  year = {2022},
  month = apr,
  journal = {IEEE Transactions on Medical Imaging},
  volume = {41},
  number = {4},
  pages = {782--792},
  issn = {1558-254X},
  doi = {10.1109/TMI.2021.3123912}
}

@article{chavignon3DTranscranialUltrasound2021,
  title = {{{3D Transcranial Ultrasound Localization Microscopy}} in the {{Rat Brain}} with a {{Multiplexed Matrix Probe}}},
  author = {Chavignon, Arthur and Heiles, Baptiste and Hingot, Vincent and Orset, Cyrille and Vivien, Denis and Couture, Olivier},
  year = {2021},
  month = dec,
  journal = {IEEE transactions on bio-medical engineering},
  volume = {PP},
  issn = {1558-2531},
  doi = {10.1109/TBME.2021.3137265},
  langid = {english},
  pmid = {34932470}
}

@article{chavignon3DTranscranialUltrasound2022,
  title = {{{3D}} Transcranial Ultrasound Localization Microscopy for Discrimination between Ischemic and Hemorrhagic Stroke in Early Phase},
  author = {Chavignon, Arthur and Hingot, Vincent and Orset, Cyrille and Vivien, Denis and Couture, Olivier},
  year = {2022},
  month = aug,
  journal = {Scientific Reports},
  volume = {12},
  number = {1},
  pages = {14607},
  publisher = {Nature Publishing Group},
  issn = {2045-2322},
  doi = {10.1038/s41598-022-18025-x},
  urldate = {2024-05-28},
  copyright = {2022 The Author(s)},
  langid = {english}
}

@article{chenDeepLearningBasedMicrobubble2022,
  title = {Deep {{Learning-Based Microbubble Localization}} for {{Ultrasound Localization Microscopy}}},
  author = {Chen, Xi and Lowerison, Matthew R. and Dong, Zhijie and Han, Aiguo and Song, Pengfei},
  year = {2022},
  month = apr,
  journal = {IEEE Transactions on Ultrasonics, Ferroelectrics, and Frequency Control},
  volume = {69},
  number = {4},
  pages = {1312--1325},
  issn = {1525-8955},
  doi = {10.1109/TUFFC.2022.3152225}
}

@article{chenLocalizationFreeSuperresolution2023,
  title = {Localization Free Super-Resolution Microbubble Velocimetry Using a Long Short-Term Memory Neural Network},
  author = {Chen, Xi and Lowerison, Matthew R. and Dong, Zhijie and Sekaran, Nathiya Vaithiyalingam Chandra and Llano, Daniel A. and Song, Pengfei},
  year = {2023},
  journal = {IEEE Transactions on Medical Imaging},
  pages = {1--1},
  issn = {1558-254X},
  doi = {10.1109/TMI.2023.3251197}
}

@inproceedings{choy4DSpatioTemporalConvNets2019,
  title = {{{4D Spatio-Temporal ConvNets}}: {{Minkowski Convolutional Neural Networks}}},
  shorttitle = {{{4D Spatio-Temporal ConvNets}}},
  booktitle = {2019 {{IEEE}}/{{CVF Conference}} on {{Computer Vision}} and {{Pattern Recognition}} ({{CVPR}})},
  author = {Choy, Christopher and Gwak, JunYoung and Savarese, Silvio},
  year = {2019},
  month = jun,
  pages = {3070--3079},
  publisher = {IEEE},
  address = {Long Beach, CA, USA},
  doi = {10.1109/CVPR.2019.00319},
  urldate = {2021-12-15},
  isbn = {978-1-72813-293-8},
  langid = {english}
}

@article{christensen-jeffriesVivoAcousticSuperResolution2015,
  title = {In {{Vivo Acoustic Super-Resolution}} and {{Super-Resolved Velocity Mapping Using Microbubbles}}},
  author = {{Christensen-Jeffries}, K. and Browning, R. J. and Tang, M. and Dunsby, C. and Eckersley, R. J.},
  year = {2015},
  month = feb,
  journal = {IEEE Transactions on Medical Imaging},
  volume = {34},
  number = {2},
  pages = {433--440},
  issn = {1558-254X},
  doi = {10.1109/TMI.2014.2359650}
}

@article{cormierDynamicMyocardialUltrasound2021,
  title = {Dynamic {{Myocardial Ultrasound Localization Angiography}}},
  author = {Cormier, Philippe and Por{\'e}e, Jonathan and Bourquin, Chlo{\'e} and Provost, Jean},
  year = {2021},
  journal = {IEEE Transactions on Medical Imaging},
  pages = {1--1},
  issn = {1558-254X},
  doi = {10.1109/TMI.2021.3086115}
}

@article{demeneTranscranialUltrafastUltrasound2021,
  title = {Transcranial Ultrafast Ultrasound Localization Microscopy of Brain Vasculature in Patients},
  author = {Demen{\'e}, Charlie and Robin, Justine and Dizeux, Alexandre and Heiles, Baptiste and Pernot, Mathieu and Tanter, Mickael and Perren, Fabienne},
  year = {2021},
  month = mar,
  journal = {Nature biomedical engineering},
  volume = {5},
  number = {3},
  pages = {219--228},
  issn = {2157-846X},
  doi = {10.1038/s41551-021-00697-x},
  urldate = {2022-08-11},
  pmcid = {PMC7610356},
  pmid = {33723412}
}

@article{demeulenaereVivoWholeBrain2022,
  title = {In Vivo Whole Brain Microvascular Imaging in Mice Using Transcranial {{3D Ultrasound Localization Microscopy}}},
  author = {Demeulenaere, Oscar and Bertolo, Adrien and Pezet, Sophie and {Ialy-Radio}, Nathalie and Osmanski, Bruno and Papadacci, Cl{\'e}ment and Tanter, Mickael and Deffieux, Thomas and Pernot, Mathieu},
  year = {2022},
  month = may,
  journal = {eBioMedicine},
  volume = {79},
  publisher = {Elsevier},
  issn = {2352-3964},
  doi = {10.1016/j.ebiom.2022.103995},
  urldate = {2022-05-05},
  langid = {english},
  pmid = {35460988}
}

@article{denisSensingUltrasoundLocalization2023,
  title = {Sensing Ultrasound Localization Microscopy for the Visualization of Glomeruli in Living Rats and Humans},
  author = {Denis, Louise and Bodard, Sylvain and Hingot, Vincent and Chavignon, Arthur and Battaglia, Jacques and Renault, Gilles and Lager, Franck and Aissani, Abderrahmane and H{\'e}l{\'e}non, Olivier and Correas, Jean-Michel and Couture, Olivier},
  year = {2023},
  month = apr,
  journal = {eBioMedicine},
  volume = {91},
  pages = {104578},
  issn = {2352-3964},
  doi = {10.1016/j.ebiom.2023.104578},
  urldate = {2024-06-03},
  pmcid = {PMC10149190},
  pmid = {37086650}
}

@article{erricoUltrafastUltrasoundLocalization2015,
  title = {Ultrafast Ultrasound Localization Microscopy for Deep Super-Resolution Vascular Imaging},
  author = {Errico, Claudia and Pierre, Juliette and Pezet, Sophie and Desailly, Yann and Lenkei, Zsolt and Couture, Olivier and Tanter, Mickael},
  year = {2015},
  month = nov,
  journal = {Nature},
  volume = {527},
  number = {7579},
  pages = {499--502},
  publisher = {Nature Publishing Group},
  issn = {1476-4687},
  doi = {10.1038/nature16066},
  urldate = {2023-03-10},
  copyright = {2015 Nature Publishing Group, a division of Macmillan Publishers Limited. All Rights Reserved.},
  langid = {english}
}

@article{garciaSIMUSOpensourceSimulator2022,
  title = {{{SIMUS}}: {{An}} Open-Source Simulator for Medical Ultrasound Imaging. {{Part I}}: {{Theory}} \& Examples},
  shorttitle = {{{SIMUS}}},
  author = {Garcia, Damien},
  year = {2022},
  month = may,
  journal = {Computer Methods and Programs in Biomedicine},
  volume = {218},
  pages = {106726},
  issn = {0169-2607},
  doi = {10.1016/j.cmpb.2022.106726},
  urldate = {2023-06-08},
  langid = {english}
}

@article{goudotAssessmentTakayasuArteritis2023,
  title = {Assessment of {{Takayasu}}'s Arteritis Activity by Ultrasound Localization Microscopy},
  author = {Goudot, Guillaume and Jimenez, Anatole and Mohamedi, Nassim and Sitruk, Jonas and Khider, Lina and Mortelette, H{\'e}l{\`e}ne and Papadacci, Cl{\'e}ment and Hyafil, Fabien and Tanter, Micka{\"e}l and Messas, Emmanuel and Pernot, Mathieu and Mirault, Tristan},
  year = {2023},
  month = apr,
  journal = {eBioMedicine},
  volume = {90},
  publisher = {Elsevier},
  issn = {2352-3964},
  doi = {10.1016/j.ebiom.2023.104502},
  urldate = {2024-05-28},
  langid = {english},
  pmid = {36893585}
}

@incollection{gwakGenerativeSparseDetection2020,
  title = {Generative {{Sparse Detection Networks}} for {{3D Single-Shot Object Detection}}},
  booktitle = {Computer {{Vision}} -- {{ECCV}} 2020},
  author = {Gwak, JunYoung and Choy, Christopher and Savarese, Silvio},
  editor = {Vedaldi, Andrea and Bischof, Horst and Brox, Thomas and Frahm, Jan-Michael},
  year = {2020},
  volume = {12349},
  pages = {297--313},
  publisher = {Springer International Publishing},
  address = {Cham},
  doi = {10.1007/978-3-030-58548-8_18},
  urldate = {2023-10-12},
  isbn = {978-3-030-58547-1 978-3-030-58548-8},
  langid = {english}
}

@article{hansen-shearerUltrafast3DTranscutaneous2024,
  title = {Ultrafast 3-{{D Transcutaneous Super Resolution Ultrasound Using Row-Column Array Specific Coherence-Based Beamforming}} and {{Rolling Acoustic Sub-aperture Processing}}: {{In Vitro}}, in {{Rabbit}} and in {{Human Study}}},
  shorttitle = {Ultrafast 3-{{D Transcutaneous Super Resolution Ultrasound Using Row-Column Array Specific Coherence-Based Beamforming}} and {{Rolling Acoustic Sub-aperture Processing}}},
  author = {{Hansen-Shearer}, Joseph and Yan, Jipeng and Lerendegui, Marcelo and Huang, Biao and Toulemonde, Matthieu and Riemer, Kai and Tan, Qingyuan and Tonko, Johanna and Weinberg, Peter D. and Dunsby, Chris and Tang, Meng-Xing},
  year = {2024},
  month = jul,
  journal = {Ultrasound in Medicine and Biology},
  volume = {50},
  number = {7},
  pages = {1045--1057},
  publisher = {Elsevier},
  issn = {0301-5629, 1879-291X},
  doi = {10.1016/j.ultrasmedbio.2024.03.020},
  urldate = {2024-07-03},
  langid = {english},
  pmid = {38702285}
}

@article{harputTwoStageMotionCorrection2018,
  title = {Two-{{Stage Motion Correction}} for {{Super-Resolution Ultrasound Imaging}} in {{Human Lower Limb}}},
  author = {Harput, Sevan and {Christensen-Jeffries}, Kirsten and Brown, Jemma and Li, Yuanwei and Williams, Katherine J. and Davies, Alun H. and Eckersley, Robert J. and Dunsby, Christopher and Tang, Meng-Xing},
  year = {2018},
  month = may,
  journal = {IEEE Transactions on Ultrasonics, Ferroelectrics, and Frequency Control},
  volume = {65},
  number = {5},
  pages = {803--814},
  issn = {1525-8955},
  doi = {10.1109/TUFFC.2018.2824846},
  urldate = {2024-05-28}
}

@article{heilesPerformanceBenchmarkingMicrobubblelocalization2022,
  title = {Performance Benchmarking of Microbubble-Localization Algorithms for Ultrasound Localization Microscopy},
  author = {Heiles, Baptiste and Chavignon, Arthur and Hingot, Vincent and Lopez, Pauline and Teston, Eliott and Couture, Olivier},
  year = {2022},
  month = may,
  journal = {Nature Biomedical Engineering},
  volume = {6},
  number = {5},
  pages = {605--616},
  publisher = {Nature Publishing Group},
  issn = {2157-846X},
  doi = {10.1038/s41551-021-00824-8},
  urldate = {2023-10-10},
  copyright = {2021 The Author(s), under exclusive licence to Springer Nature Limited},
  langid = {english}
}

@article{heilesUltrafast3DUltrasound2019,
  title = {Ultrafast {{3D Ultrasound Localization Microscopy Using}} a 32{\texttimes}32 {{Matrix Array}}},
  author = {Heiles, Baptiste and Correia, Mafalda and Hingot, Vincent and Pernot, Mathieu and Provost, Jean and Tanter, Mickael and Couture, Olivier},
  year = {2019},
  month = sep,
  journal = {IEEE Transactions on Medical Imaging},
  volume = {38},
  number = {9},
  pages = {2005--2015},
  issn = {1558-254X},
  doi = {10.1109/TMI.2018.2890358}
}

@article{heilesVolumetricUltrasoundLocalization2022,
  title = {Volumetric {{Ultrasound Localization Microscopy}} of the {{Whole Rat Brain Microvasculature}}},
  author = {Heiles, Baptiste and Chavignon, Arthur and Bergel, Antoine and Hingot, Vincent and Serroune, Hicham and Maresca, David and Pezet, Sophie and Pernot, Mathieu and Tanter, Mickael and Couture, Olivier},
  year = {2022},
  journal = {IEEE Open Journal of Ultrasonics, Ferroelectrics, and Frequency Control},
  volume = {2},
  pages = {261--282},
  issn = {2694-0884},
  doi = {10.1109/OJUFFC.2022.3214185},
  urldate = {2024-07-16}
}

@article{hingotMicrovascularFlowDictates2019,
  title = {Microvascular Flow Dictates the Compromise between Spatial Resolution and Acquisition Time in {{Ultrasound Localization Microscopy}}},
  author = {Hingot, Vincent and Errico, Claudia and Heiles, Baptiste and Rahal, Line and Tanter, Mickael and Couture, Olivier},
  year = {2019},
  month = feb,
  journal = {Scientific Reports},
  volume = {9},
  number = {1},
  pages = {2456},
  issn = {2045-2322},
  doi = {10.1038/s41598-018-38349-x},
  langid = {english},
  pmcid = {PMC6385220},
  pmid = {30792398}
}

@article{huangShortAcquisitionTime2020,
  title = {Short {{Acquisition Time Super-Resolution Ultrasound Microvessel Imaging}} via {{Microbubble Separation}}},
  author = {Huang, Chengwu and Lowerison, Matthew R. and Trzasko, Joshua D. and Manduca, Armando and Bresler, Yoram and Tang, Shanshan and Gong, Ping and Lok, U.-Wai and Song, Pengfei and Chen, Shigao},
  year = {2020},
  month = apr,
  journal = {Scientific Reports},
  volume = {10},
  number = {1},
  pages = {6007},
  publisher = {Nature Publishing Group},
  issn = {2045-2322},
  doi = {10.1038/s41598-020-62898-9},
  urldate = {2023-07-17},
  copyright = {2020 The Author(s)},
  langid = {english}
}

@article{huangSuperresolutionUltrasoundLocalization2021,
  title = {Super-Resolution Ultrasound Localization Microscopy Based on a High Frame-Rate Clinical Ultrasound Scanner: An in-Human Feasibility Study},
  shorttitle = {Super-Resolution Ultrasound Localization Microscopy Based on a High Frame-Rate Clinical Ultrasound Scanner},
  author = {Huang, Chengwu and Zhang, Wei and Gong, Ping and Lok, U.-Wai and Tang, Shanshan and Yin, Tinghui and Zhang, Xirui and Zhu, Lei and Sang, Maodong and Song, Pengfei and Zheng, Rongqin and Chen, Shigao},
  year = {2021},
  month = apr,
  journal = {Physics in Medicine \& Biology},
  volume = {66},
  number = {8},
  pages = {08NT01},
  publisher = {IOP Publishing},
  issn = {0031-9155},
  doi = {10.1088/1361-6560/abef45},
  urldate = {2024-05-28},
  langid = {english}
}

@article{jensenAnatomicFunctionalImaging2022,
  title = {Anatomic and {{Functional Imaging Using Row}}--{{Column Arrays}}},
  author = {Jensen, J{\o}rgen Arendt and Schou, Mikkel and J{\o}rgensen, Lasse Thurmann and Tomov, Borislav G. and Stuart, Matthias Bo and Traberg, Marie Sand and Taghavi, Iman and {\O}ygaard, Sigrid Hueseb{\o} and Ommen, Martin Lind and Steenberg, Kitty and Thomsen, Erik Vilain and Panduro, Nathalie Sarup and Nielsen, Michael Bachmann and S{\o}rensen, Charlotte Mehlin},
  year = {2022},
  month = oct,
  journal = {IEEE Transactions on Ultrasonics, Ferroelectrics, and Frequency Control},
  volume = {69},
  number = {10},
  pages = {2722--2738},
  issn = {1525-8955},
  doi = {10.1109/TUFFC.2022.3191391},
  urldate = {2024-07-03}
}

@misc{kingmaAdamMethodStochastic2017,
  title = {Adam: {{A Method}} for {{Stochastic Optimization}}},
  shorttitle = {Adam},
  author = {Kingma, Diederik P. and Ba, Jimmy},
  year = {2017},
  month = jan,
  number = {arXiv:1412.6980},
  eprint = {1412.6980},
  primaryclass = {cs},
  publisher = {arXiv},
  doi = {10.48550/arXiv.1412.6980},
  urldate = {2023-09-18},
  archiveprefix = {arXiv}
}

@article{kuhnHungarianMethodAssignment1955,
  title = {The {{Hungarian}} Method for the Assignment Problem},
  author = {Kuhn, H. W.},
  year = {1955},
  journal = {Naval Research Logistics Quarterly},
  volume = {2},
  number = {1-2},
  pages = {83--97},
  issn = {1931-9193},
  doi = {10.1002/nav.3800020109},
  urldate = {2024-01-22},
  copyright = {Copyright {\copyright} 1955 Wiley Periodicals, Inc., A Wiley Company},
  langid = {english}
}

@article{leconteTrackingPriorLocalization2024,
  title = {A {{Tracking}} Prior to {{Localization}} Workflow for {{Ultrasound Localization Microscopy}}},
  author = {Leconte, Alexis and Por{\'e}e, Jonathan and Rauby, Brice and Wu, Alice and Ghigo, Nin and Xing, Paul and LEE, Stephen and Bourquin, Chlo{\'e} and {Ramos-Palacios}, Gerardo and Sadikot, Abbas F. and Provost, Jean},
  year = {2024},
  journal = {IEEE Transactions on Medical Imaging},
  pages = {1--1},
  issn = {1558-254X},
  doi = {10.1109/TMI.2024.3456676},
  urldate = {2024-09-23}
}

@inproceedings{leeDeeplySupervisedNets2015,
  title = {Deeply-{{Supervised Nets}}},
  booktitle = {Proceedings of the {{Eighteenth International Conference}} on {{Artificial Intelligence}} and {{Statistics}}},
  author = {Lee, Chen-Yu and Xie, Saining and Gallagher, Patrick and Zhang, Zhengyou and Tu, Zhuowen},
  year = {2015},
  month = feb,
  pages = {562--570},
  publisher = {PMLR},
  issn = {1938-7228},
  urldate = {2023-06-19},
  langid = {english}
}

@article{liSuperresolutionUltrasoundLocalization2024,
  title = {Super-Resolution Ultrasound Localization Microscopy for the Non-Invasive Imaging of Human Testicular Microcirculation and Its Differential Diagnosis Role in Male Infertility},
  author = {Li, Maoyao and Chen, Lei and Yan, Jipeng and Jayasena, Channa Nalin and Liu, Zhangshun and Li, Jia and Li, Ao and Zhu, Jiang and Wang, Ronghui and Li, Jianchun and Zhang, Chaoxue and Guo, Jingyi and Zhao, Yuwu and Feng, Chao and Tang, Mengxing and Zheng, Yuanyi},
  year = {2024},
  journal = {VIEW},
  volume = {5},
  number = {2},
  pages = {20230093},
  issn = {2688-268X},
  doi = {10.1002/VIW.20230093},
  urldate = {2024-07-02},
  langid = {english}
}

@article{liuDeepLearningUltrasound2020,
  title = {Deep {{Learning}} for {{Ultrasound Localization Microscopy}}},
  author = {Liu, Xin and Zhou, Tianyang and Lu, Mengyang and Yang, Yi and He, Qiong and Luo, Jianwen},
  year = {2020},
  month = oct,
  journal = {IEEE Transactions on Medical Imaging},
  volume = {39},
  number = {10},
  pages = {3064--3078},
  issn = {1558-254X},
  doi = {10.1109/TMI.2020.2986781},
  urldate = {2024-05-01}
}

@article{lokThreeDimensionalUltrasoundLocalization2022,
  title = {Three-{{Dimensional Ultrasound Localization Microscopy}} with {{Bipartite Graph-Based Microbubble Pairing}} and {{Kalman-Filtering-Based Tracking}} on a 256-{{Channel Verasonics Ultrasound System}} with a 32{\texttimes}32 {{Matrix Array}}},
  author = {Lok, U-Wai and Huang, Chengwu and Trzasko, Joshua D. and Kim, Yohan and Lucien, Fabrice and Tang, Shanshan and Gong, Ping and Song, Pengfei and Chen, Shigao},
  year = {2022},
  month = dec,
  journal = {Journal of Medical and Biological Engineering},
  volume = {42},
  number = {6},
  pages = {767--779},
  issn = {2199-4757},
  doi = {10.1007/s40846-022-00755-y},
  urldate = {2022-12-22},
  langid = {english}
}

@article{lowerisonSuperResolutionUltrasoundReveals2024,
  title = {Super-{{Resolution Ultrasound Reveals Cerebrovascular Impairment}} in a {{Mouse Model}} of {{Alzheimer}}'s {{Disease}}},
  author = {Lowerison, Matthew R. and Sekaran, Nathiya Vaithiyalingam Chandra and Dong, Zhijie and Chen, Xi and You, Qi and Llano, Daniel A. and Song, Pengfei},
  year = {2024},
  month = feb,
  journal = {Journal of Neuroscience},
  volume = {44},
  number = {9},
  publisher = {Society for Neuroscience},
  issn = {0270-6474, 1529-2401},
  doi = {10.1523/JNEUROSCI.1251-23.2024},
  urldate = {2024-05-28},
  chapter = {Research Articles},
  copyright = {Copyright {\copyright} 2024 Lowerison et al.. This is an open-access article distributed under the terms of the Creative Commons Attribution 4.0 International license, which permits unrestricted use, distribution and reproduction in any medium provided that the original work is properly attributed.},
  langid = {english},
  pmid = {38253533}
}

@article{luComplexConvolutionalNeural2022,
  title = {Complex {{Convolutional Neural Networks}} for {{Ultrafast Ultrasound Imaging Reconstruction From In-Phase}}/{{Quadrature Signal}}},
  author = {Lu, Jingfeng and Millioz, Fabien and Garcia, Damien and Salles, Sebastien and Ye, Dong and Friboulet, Denis},
  year = {2022},
  month = feb,
  journal = {IEEE Transactions on Ultrasonics, Ferroelectrics, and Frequency Control},
  volume = {69},
  number = {2},
  pages = {592--603},
  issn = {0885-3010, 1525-8955},
  doi = {10.1109/TUFFC.2021.3127916},
  urldate = {2023-10-19},
  langid = {english}
}

@article{marquezDeepCascadeLearning2018,
  title = {Deep {{Cascade Learning}}},
  author = {Marquez, Enrique S. and Hare, Jonathon S. and Niranjan, Mahesan},
  year = {2018},
  month = nov,
  journal = {IEEE Transactions on Neural Networks and Learning Systems},
  volume = {29},
  number = {11},
  pages = {5475--5485},
  issn = {2162-2388},
  doi = {10.1109/TNNLS.2018.2805098}
}

@article{mileckiDeepLearningFramework2021,
  title = {A {{Deep Learning Framework}} for {{Spatiotemporal Ultrasound Localization Microscopy}}},
  author = {Milecki, L{\'e}o and Por{\'e}e, Jonathan and Belgharbi, Hatim and Bourquin, Chlo{\'e} and Damseh, Rafat and {Delafontaine-Martel}, Patrick and Lesage, Fr{\'e}d{\'e}ric and Gasse, Maxime and Provost, Jean},
  year = {2021},
  month = may,
  journal = {IEEE Transactions on Medical Imaging},
  volume = {40},
  number = {5},
  pages = {1428--1437},
  issn = {1558-254X},
  doi = {10.1109/TMI.2021.3056951}
}

@article{opacicMotionModelUltrasound2018,
  title = {Motion Model Ultrasound Localization Microscopy for Preclinical and Clinical Multiparametric Tumor Characterization},
  author = {Opacic, Tatjana and Dencks, Stefanie and Theek, Benjamin and Piepenbrock, Marion and Ackermann, Dimitri and Rix, Anne and Lammers, Twan and Stickeler, Elmar and Delorme, Stefan and Schmitz, Georg and Kiessling, Fabian},
  year = {2018},
  month = apr,
  journal = {Nature Communications},
  volume = {9},
  number = {1},
  pages = {1527},
  publisher = {Nature Publishing Group},
  issn = {2041-1723},
  doi = {10.1038/s41467-018-03973-8},
  urldate = {2024-05-28},
  copyright = {2018 The Author(s)},
  langid = {english}
}

@article{porteUltrasoundLocalizationMicroscopy2024,
  title = {Ultrasound {{Localization Microscopy}} for {{Breast Cancer Imaging}} in {{Patients}}: {{Protocol Optimization}} and {{Comparison}} with {{Shear Wave Elastography}}},
  shorttitle = {Ultrasound {{Localization Microscopy}} for {{Breast Cancer Imaging}} in {{Patients}}},
  author = {Porte, C{\'e}line and Lisson, Thomas and Kohlen, Matthias and {von Maltzahn}, Finn and Dencks, Stefanie and {von Stillfried}, Saskia and Piepenbrock, Marion and Rix, Anne and Dasgupta, Anshuman and Koczera, Patrick and Boor, Peter and Stickeler, Elmar and Schmitz, Georg and Kiessling, Fabian},
  year = {2024},
  month = jan,
  journal = {Ultrasound in Medicine \& Biology},
  volume = {50},
  number = {1},
  pages = {57--66},
  issn = {0301-5629},
  doi = {10.1016/j.ultrasmedbio.2023.09.001},
  urldate = {2024-07-02}
}

@article{renaudinFunctionalUltrasoundLocalization2022,
  title = {Functional Ultrasound Localization Microscopy Reveals Brain-Wide Neurovascular Activity on a Microscopic Scale},
  author = {Renaudin, No{\'e}mi and Demen{\'e}, Charlie and Dizeux, Alexandre and {Ialy-Radio}, Nathalie and Pezet, Sophie and Tanter, Mickael},
  year = {2022},
  month = aug,
  journal = {Nature Methods},
  volume = {19},
  number = {8},
  pages = {1004--1012},
  publisher = {Nature Publishing Group},
  issn = {1548-7105},
  doi = {10.1038/s41592-022-01549-5},
  urldate = {2023-10-10},
  copyright = {2022 The Author(s)},
  langid = {english}
}

@misc{shinContextAwareDeepLearning2023,
  title = {Context-{{Aware Deep Learning Enables High-Efficacy Localization}} of {{High Concentration Microbubbles}} for {{Super-Resolution Ultrasound Localization Microscopy}}},
  author = {Shin, YiRang and Lowerison, Matthew R. and Wang, Yike and Chen, Xi and You, Qi and Dong, Zhijie and Anastasio, Mark A. and Song, Pengfei},
  year = {2023},
  month = apr,
  primaryclass = {New Results},
  pages = {2023.04.21.536599},
  publisher = {bioRxiv},
  doi = {10.1101/2023.04.21.536599},
  urldate = {2023-06-07},
  archiveprefix = {bioRxiv},
  chapter = {New Results},
  copyright = {{\copyright} 2023, Posted by Cold Spring Harbor Laboratory. This pre-print is available under a Creative Commons License (Attribution-NonCommercial-NoDerivs 4.0 International), CC BY-NC-ND 4.0, as described at http://creativecommons.org/licenses/by-nc-nd/4.0/},
  langid = {english}
}

@article{solomonExploitingFlowDynamics2019,
  title = {Exploiting {{Flow Dynamics}} for {{Superresolution}} in {{Contrast-Enhanced Ultrasound}}},
  author = {Solomon, Oren and {van Sloun}, Ruud J. G. and Wijkstra, Hessel and Mischi, Massimo and Eldar, Yonina C.},
  year = {2019},
  month = oct,
  journal = {IEEE Transactions on Ultrasonics, Ferroelectrics, and Frequency Control},
  volume = {66},
  number = {10},
  pages = {1573--1586},
  issn = {1525-8955},
  doi = {10.1109/TUFFC.2019.2926062},
  urldate = {2024-05-28}
}

@article{songEffectsSpatialSampling2018,
  title = {On the {{Effects}} of {{Spatial Sampling Quantization}} in {{Super-Resolution Ultrasound Microvessel Imaging}}},
  author = {Song, Pengfei and Manduca, Armando and Trzasko, Joshua D. and Daigle, Ronald E. and Chen, Shigao},
  year = {2018},
  month = dec,
  journal = {IEEE Transactions on Ultrasonics, Ferroelectrics, and Frequency Control},
  volume = {65},
  number = {12},
  pages = {2264--2276},
  issn = {1525-8955},
  doi = {10.1109/TUFFC.2018.2832600},
  urldate = {2024-04-18}
}

@article{speiserDeepLearningEnables2021,
  title = {Deep Learning Enables Fast and Dense Single-Molecule Localization with High Accuracy},
  author = {Speiser, Artur and M{\"u}ller, Lucas-Raphael and Hoess, Philipp and Matti, Ulf and Obara, Christopher J. and Legant, Wesley R. and Kreshuk, Anna and Macke, Jakob H. and Ries, Jonas and Turaga, Srinivas C.},
  year = {2021},
  month = sep,
  journal = {Nature Methods},
  volume = {18},
  number = {9},
  pages = {1082--1090},
  publisher = {Nature Publishing Group},
  issn = {1548-7105},
  doi = {10.1038/s41592-021-01236-x},
  urldate = {2023-05-12},
  copyright = {2021 The Author(s), under exclusive licence to Springer Nature America, Inc.},
  langid = {english}
}

@article{taghaviUltrasoundSuperresolutionImaging2022,
  title = {Ultrasound Super-Resolution Imaging with a Hierarchical {{Kalman}} Tracker},
  author = {Taghavi, Iman and Andersen, Sofie Bech and Hoyos, Carlos Armando Villag{\'o}mez and Schou, Mikkel and Gran, Fredrik and Hansen, Kristoffer Lindskov and Nielsen, Michael Bachmann and S{\o}rensen, Charlotte Mehlin and Stuart, Matthias Bo and Jensen, J{\o}rgen Arendt},
  year = {2022},
  month = may,
  journal = {Ultrasonics},
  volume = {122},
  pages = {106695},
  issn = {0041-624X},
  doi = {10.1016/j.ultras.2022.106695},
  urldate = {2022-07-04},
  langid = {english}
}

@article{tangKalmanFilterBasedMicrobubble2020,
  title = {Kalman {{Filter-Based Microbubble Tracking}} for {{Robust Super-Resolution Ultrasound Microvessel Imaging}}},
  author = {Tang, Shanshan and Song, Pengfei and Trzasko, Joshua D. and Lowerison, Matthew and Huang, Chengwu and Gong, Ping and Lok, U-Wai and Manduca, Armando and Chen, Shigao},
  year = {2020},
  month = sep,
  journal = {IEEE Transactions on Ultrasonics, Ferroelectrics, and Frequency Control},
  volume = {67},
  number = {9},
  pages = {1738--1751},
  issn = {1525-8955},
  doi = {10.1109/TUFFC.2020.2984384}
}

@inproceedings{trabelsiDeepComplexNetworks2018,
  title = {Deep {{Complex Networks}}},
  booktitle = {International {{Conference}} on {{Learning Representations}}},
  author = {Trabelsi, Chiheb and Bilaniuk, Olexa and Zhang, Ying and Serdyuk, Dmitriy and Subramanian, Sandeep and Santos, Joao Felipe and Mehri, Soroush and Rostamzadeh, Negar and Bengio, Yoshua and Pal, Christopher J.},
  year = {2018},
  month = feb,
  urldate = {2021-12-20},
  langid = {english}
}

@article{vanslounSuperResolutionUltrasoundLocalization2021,
  title = {Super-{{Resolution Ultrasound Localization Microscopy Through Deep Learning}}},
  author = {{van Sloun}, Ruud J. G. and Solomon, Oren and Bruce, Matthew and Khaing, Zin Z. and Wijkstra, Hessel and Eldar, Yonina C. and Mischi, Massimo},
  year = {2021},
  month = mar,
  journal = {IEEE Transactions on Medical Imaging},
  volume = {40},
  number = {3},
  pages = {829--839},
  issn = {1558-254X},
  doi = {10.1109/TMI.2020.3037790}
}

@misc{wu3DTranscranialDynamic2024,
  title = {{{3D}} Transcranial {{Dynamic Ultrasound Localization Microscopy}} in the Mouse Brain Using a {{Row-Column Array}}},
  author = {Wu, Alice and Por{\'e}e, Jonathan and {Ramos-Palacios}, Gerardo and Bourquin, Chlo{\'e} and Ghigo, Nin and Leconte, Alexis and Xing, Paul and Sadikot, Abbas F. and Chass{\'e}, Micha{\"e}l and Provost, Jean},
  year = {2024},
  month = jun,
  number = {arXiv:2406.01746},
  eprint = {2406.01746},
  primaryclass = {physics},
  publisher = {arXiv},
  doi = {10.48550/arXiv.2406.01746},
  urldate = {2024-07-03},
  archiveprefix = {arXiv}
}

@article{xingPhaseAberrationCorrection2024,
  title = {Phase {{Aberration Correction}} for {{In Vivo Ultrasound Localization Microscopy Using}} a {{Spatiotemporal Complex-Valued Neural Network}}},
  author = {Xing, Paul and Por{\'e}e, Jonathan and Rauby, Brice and Malescot, Antoine and Martineau, Eric and Perrot, Vincent and Rungta, Ravi L. and Provost, Jean},
  year = {2024},
  month = feb,
  journal = {IEEE Transactions on Medical Imaging},
  volume = {43},
  number = {2},
  pages = {662--673},
  issn = {1558-254X},
  doi = {10.1109/TMI.2023.3316995},
  urldate = {2024-05-16}
}

@article{zhuSuperResolutionUltrasoundLocalization2022,
  title = {Super-{{Resolution Ultrasound Localization Microscopy}} of {{Microvascular Structure}} and {{Flow}} for {{Distinguishing Metastatic Lymph Nodes}} -- {{An Initial Human Study}}},
  author = {Zhu, Jiaqi and Zhang, Chao and {Christensen-Jeffries}, Kirsten and Zhang, Ge and Harput, Sevan and Dunsby, Christopher and Huang, Pintong and Tang, Meng-Xing},
  year = {2022},
  month = dec,
  journal = {Ultraschall in der Medizin - European Journal of Ultrasound},
  volume = {43},
  number = {6},
  pages = {592--598},
  publisher = {Georg Thieme Verlag KG},
  issn = {0172-4614, 1438-8782},
  doi = {10.1055/a-1917-0016},
  urldate = {2024-05-28},
  copyright = {Georg Thieme Verlag KG R{\"u}digerstra{\ss}e 14, 70469 Stuttgart, Germany},
  langid = {english}
}
\end{document}